\documentclass[11pt]{article}

\PassOptionsToPackage{hyphens}{url} % 允许 URL 自动换行

\usepackage{jheppub} % 会自动加载 hyperref，不要重复加载 hyperref！

\usepackage[T1]{fontenc}
\usepackage{amsmath, amssymb, mathtools, amsthm, amsfonts, amsbsy}
\usepackage{physics}
\usepackage{bm, bbold, bbm}
\usepackage{slashed}
\usepackage{comment}
\usepackage{enumerate, exscale}
\usepackage{float}
\usepackage{dcolumn}
\usepackage{times}
\usepackage{tabularx}
\usepackage{euscript}
\usepackage{wasysym}
\usepackage{graphicx, psfrag, grffile}
\usepackage{subfigure}
\usepackage{algorithm}
\usepackage{algpseudocode}
\usepackage{tikz}

\usepackage{xcolor}
\definecolor{ForestGreen}{RGB}{34, 139, 34}

\usepackage{url}
\usepackage{etoolbox}
\makeatletter
\patchcmd{\@biblabel}{\hfill}{\relax}{}{}
\makeatother
\newcommand{\omits}[1]{}

\def \i{\mathrm{i}}

\def\bc{\begin{center}}
\def\nno{\nonumber}
\def\ec{\end{center}}
\def\be{\begin{eqnarray}}
\def\ee{\end{eqnarray}}

\title{\boldmath Hyperfine Structure of Quantum Entanglement}

\affiliation[a]{School of Physics, Sun Yat-sen University, Guangzhou, 510275, China}
\affiliation[b]{School of Physics and Astronomy, Sun Yat-sen University, Guangzhou, 510275, China}

\author[a]{Liang-Hong Mo,}
\author[a]{Yao Zhou,}
\author[b,*]{Jia-Rui Sun }
\author[a,*]{and Peng Ye}
\emailAdd{sunjiarui@mail.sysu.edu.cn}
\emailAdd{yepeng5@mail.sysu.edu.cn}

\footnotetext[1]{Corresponding authors.}

\abstract{Quantum entanglement, crucial for understanding quantum many-body systems and quantum gravity, is commonly assessed through various measures such as von Neumann entropy, mutual information, and entanglement contour, each with its inherent advantages and limitations. In this work, we introduce the hyperfine structure of entanglement, which decomposes entanglement contours known as the fine structure into particle-number cumulants. This measure exhibits a set of universal properties with its significance in quantum information science. We apply it across diverse contexts: in Fermi gases, establishing connections to mutual information and interacting conformal field theory; in AdS$_3$/CFT$_2$ correspondence, unveiling finer subregion-subregion duality; and in Chern insulators, distinguishing between different quantum phases, especially topological gapped state and trivial gapped state. Our findings suggest experimental accessibility, offering fresh insights into quantum entanglement across physical systems.}

\begin{document}
\maketitle
\flushbottom
 \section{Introduction}

Quantum entanglement, a cornerstone of quantum information, is crucial for understanding the behavior of quantum many-body systems and the nature of gravity~\cite{2009RvMP...81..865H,amicoEntanglement2008,eisertColloquium2010,laflorencieQuantum2016,Rangamani_2017,almheiriEntropy2021,faulknerSnowmass2022,2024NatCo..15.1601L} . Over recent decades, entanglement has propelled advancements in quantum computation~\cite{2018Quant...2...79P}, uncovered novel quantum phases~\cite{2006PhRvL..96k0405L,2008PhRvL.101a0504L,2012PhRvL.108s6402Q}, and shed light on quantum gravity phenomena~\cite{2006PhRvL..96r1602R,2013JHEP...08..090L,2011JHEP...05..036C}.
Quantitatively, various entanglement measures, such as von Neumann entropy (entanglement entropy), R\'{e}nyi entropy, mutual information, and negativity, have been introduced to quantify the entanglement structure~\cite{2009RvMP...81..865H} and provide an intuitive picture of how entanglement distributes. However, these measures are insufficient to fully characterize quantum entanglement, as each has its strengths and weaknesses in capturing different aspects of quantum entanglement. For example, entanglement entropy is defined for pure state entanglement, whereas for mixed states, negativity is often used. However, negativity can be zero even if the state is entangled~\cite{1996PhLA..223....1H}. Mutual information can detect mixed state entanglement but struggles to distinguish between quantum entanglement and classical correlations, a distinction where entanglement entropy excels. What's more, as single numbers, they also fail to reveal much information about the eigen properties of the density matrix of the subregion $A$ (denoted as $\hat{\rho}_A$), i.e., entanglement spectrum and entanglement wave function (Schmidt vectors)~\cite{2008PhRvL.101a0504L, 2009PhRvL.102j0502K, 2010PhRvB..82a2405S}.

Recently, Chen and Vidal put forward an entanglement quantity called ``entanglement contour''~\cite{2014JSMTE..10..011C} that can be regarded as the \textit{fine structure of entanglement} distribution in real space. More precisely, the entanglement entropy $S(A)$ between a subregion $A$ and its complement $\bar A$ is decomposed as the sum of entanglement contour $s(i)$ over all sites, expressed as $S(A) = \sum_{i \in A} s(i)$. Since then, the contour function $s(i)$ has been explored in various contexts: fractal lattice~\cite{2023arXiv231101199Z}, Gaussian states~\cite{2014JSMTE..10..011C}, conformal field theory (CFT)~\cite{2016JSMTE..12.3103C,2018JSMTE..04.3105T}, geometric construction in the AdS/CFT correspondence, and partial entanglement entropy derived by the additive linear combination of subset entanglement entropy~\cite{2018PhRvD..98j6004W}. The extension to the $n$-th order \textit{R\'{e}nyi entropy} $S_n(A)$, defined as $S_n(A)=\frac{1}{1-n}\ln\tr\hat \rho_A^n$, has also been explored by introducing the $n$-th order \textit{R\'{e}nyi contour} $s_{n}(i)$\cite{2014JSMTE..10..011C,2016JSMTE..12.3103C,2017JPhA...50E4001C,2020PhRvR...2b3170W}, as below:
 \begin{equation*}
{   \small    \begin{array}{cccccc}
          &  S(A)\quad\quad\quad & \Longrightarrow & s(i) \quad\quad\quad&  \Longrightarrow &h_{1;k}(i) \quad\quad\\
           & \uparrow n=1  & & \uparrow\small{n=1} && \uparrow \small{n=1}\\
           &   S_n(A)( \tilde S_n(A)) &  \Longrightarrow & s_n(i)( \tilde s_n(i)) &  \Longrightarrow &h_{n;k}(i) (\tilde h_{n;k}(i) )

        \end{array}
 }\end{equation*}

In the present paper, we significantly expand the zoo of entanglement quantities and uncover a deeper understanding of quantum entanglement by introducing the \textit{hyperfine structure of entanglement}. Specifically, we precisely decompose the fine structure of entanglement $s_{n}(i)$, into contributions from particle-number cumulants, i.e., $s_n(i)=\sum_{k=1}^\infty h_{n;k}(i)$ defined in Eq.~(\ref{RenyiContour}). We rigorously prove that the hyperfine structure, represented by $h_{n;k}(i)$, is an exotic entanglement quantity with several universal properties, including additivity, normalization, exchange symmetry, and local unitary invariance, making the hyperfine structure physical and meaningful just like other quantum informative measures. We then apply this quantity to the following three distinct physical contexts, demonstrating plentiful findings. \textit{First}, in a Fermi gas, we find that $s_{n}(i) \approx h_{n;2}(i)$, with all $h_{n;k}$ for $k > 2$ being suppressed. Using this simplified relation, we establish the connection between $h_{n;2}$ and mutual information, showing that $h_{n;2}$ negatively contribute to the mutual information.  We also generalize these results into generally interacting (1+1)D CFT. \textit{Second}, from a holographic perspective, the hyperfine structure of entanglement unveils a finer subregion-subregion duality in the bulk reconstruction. Specifically, we demonstrate that the hyperfine structure of holographic \textit{refined} R\'{e}nyi entropy is dual to the density-density correlator in the boundary subregion. Surprisingly, for a given boundary subregion, the bulk extremal surfaces associated with refined R\'{e}nyi entropy can extend beyond the entanglement wedge determined by entanglement entropy. It indicates that the holographic dual of refined R\'{e}nyi entropy and its hyperfine structure can be utilized to construct more information outside the entanglement wedge. \textit{Third}, by investigating a prototypical (2+1)D lattice model, we numerically observe significant distinctions in the hyperfine structure across parameter regions, such as trivial mass gap, topological gap (supporting edge states and nonzero bulk Chern number), Dirac critical point, and Fermi surface. In addition, This hyperfine structure, based on local measurement, is experimentally accessible, we first introduce a promising experimental protocol based on quantum point contact to measure the hyperfine structure.

This paper is organized as follows. In section \ref{sec:free fermion hyperfine}, we briefly review the entanglement entropy and its fine structure (i.e., the entanglement contour) of free fermions, then we introduce the definitions and properties of the hyperfine structure of entanglement for free fermions. After that, we give the hyperfine decomposition of R\'{e}nyi entropy for free Fermi gas in $d+1$ dimensions. In section \ref{sec:Holography} we investigate the holographic dual description for the hyperfine structure of entanglement by considering the refined R\'{e}nyi entropy. In section \ref{Chern insulator}, we utilize the hyperfine structure of entanglement to distinguish different parameter regions in (2+1)D Chern insulator. In section \ref{experiment}, we propose an experiment protocol using quantum point contact to measure the hyperfine structure of entanglement. Finally, we conclude by discussing several future directions based on these findings in section \ref{conclusion}.

%%%%%%%

%%%%%%%%%%%%%%%%%%%%%%%%%%%%%%%%%%%%%%%%%%%%%%%%%%%%%%%%%%%%%%%%%%%%%%%%
\section{The hyperfine structure of entanglement for free fermion}
\label{sec:free fermion hyperfine}
%%%%%%%%%%%%%%%%%%%%%%%%%%%%%%%%%%%%%%%%%%%%%%%%%%%%%%%%%%%%%%%%%%%%%%%%
\subsection{Entanglement and its fine structure of free fermion}
\label{subsec:FreeFermionFineStructure}
%%%%%%%%%%%%%%%%%%%%%%%%%%%%%%%%%%%%%%%%%%%%%%%%%%%%%%%%%%%%%%%%%%%%%%%%
We consider a generic free-fermion lattice Hamiltonian of the quadratic form
\begin{equation}
    \hat H=\sum_{i,j}\hat c^{\dagger}_{i}\,\mathcal{H}_{ij}\,\hat c_{j},
    \label{eq:QuadHam}
\end{equation}
where $\hat c^{\dagger}_{j}$ ($\hat c_{j}$) creates (annihilates) a fermion at site $j$, and $\mathcal{H}$ denotes the single-particle Hamiltonian matrix. Let $|G\rangle$ denote the many-body ground state of $\hat H$, with the corresponding density matrix $\hat \rho = |G\rangle \langle G|$.

To study the entanglement structure, we partition the system into a spatial subregion $A$ and its complement $\bar A$, and consider the reduced density matrix of $A$ obtained by tracing out $\bar A$:
\begin{equation}
    \hat\rho_{A}= \tr_{\bar A}(\hat \rho) = \frac{1}{Z}\,e^{-\hat K_{A}}, \qquad
    \hat K_{A} = \sum_{i,j\in A}\hat c^{\dagger}_{i}\,H^{A}_{ij}\,\hat c_{j},
    \label{eq:ModHam}
\end{equation}
where $Z = \Tr\, e^{-\hat K_{A}}$ ensures normalization, and $\hat K_A$ is known as the entanglement Hamiltonian.

For Gaussian states such as free fermion ground states, the entanglement Hamiltonian is entirely determined by the equal-time correlation matrix $M_{ij} = \langle G|\hat c^{\dagger}_{i} \hat c_{j}|G\rangle$ restricted to region $A$. According to Peschel's formula~\cite{2003JPhA...36L.205P}, the matrix $H^A$ appearing in Eq.~(\ref{eq:ModHam}) is given by
\begin{equation}
    H^{A} = \ln\!\bigl[M^{-1} - \mathbf{1}\bigr].
\end{equation}
Diagonalizing $M$ yields eigenvalues $\{\xi_k\} \in (0,1)$, which constitute the entanglement spectrum and fully determine all R\'{e}nyi entropies.

In particular, the von Neumann entropy of region $A$ reads
\begin{equation}
    S_1 = -\sum_{k} \left[ \xi_k \ln \xi_k + (1 - \xi_k) \ln (1 - \xi_k) \right],
\end{equation}
while the R\'{e}nyi entropy of order $n > 0$ is
\begin{equation}\label{renenr}
    S_n = \frac{1}{1 - n} \sum_k \ln \left[ \xi_k^n + (1 - \xi_k)^n \right].
\end{equation}

To resolve the spatial structure of entanglement within region $A$, Chen and Vidal introduced the entanglement contour $s(j)$, a site-resolved quantity whose sum reproduces the von Neumann entropy~\cite{2014JSMTE..10..011C}. This concept can be generalized to R\'{e}nyi entropies via the R\'{e}nyi contour $s_n(j)$. Given the projector $\hat P_j = |j\rangle \langle j|$ onto site $j$, the entanglement contour and R\'{e}nyi contour functions can be respectively expressed as
\begin{align}
    s(j) &= -\Tr \left\{ \hat P_j \left[ M \ln M + (\mathbf{1} - M)\ln(\mathbf{1} - M) \right] \right\}, \\
    s_n(j) &= \frac{1}{1 - n} \Tr \left\{ \hat P_j \ln\left[ M^n + (\mathbf{1} - M)^n \right] \right\},
    \label{eq:RenyiContourDef}
\end{align}
which satisfies the sum rule
\begin{equation}
    S_n = \sum_{j \in A} s_n(j).
    \label{eq:ContourSumRule}
\end{equation}
These local quantities describe the fine structure of entanglement and provide the basis for the hyperfine decomposition $s_n(j) = \sum_{k\ge1} h_{n;k}(j)$ introduced in the next subsection.

%%%%%%%%%%%%%%%%%%%%%%%%%%%%%%%%%%%%%%%%%%%%%%%%%%%%%%%%%%%%%%%%%%%%%%%%
\subsection{Definition and properties of the hyperfine structure}
\label{subsec:HyperfineDef}
%%%%%%%%%%%%%%%%%%%%%%%%%%%%%%%%%%%%%%%%%%%%%%%%%%%%%%%%%%%%%%%%%%%%%%%%
In the following, we will introduce the hyperfine structure of entanglement by decomposing the fine structure $s_n(j)$, i.e., the R\'enyi entanglement contour. Denoting the particle-number operator in $A$ as $\hat N_A\equiv\sum_{i\in A}\hat c_i^{\dagger}\hat c_i$ and its maximal expectation value, or equivalently, the sites of subregion $A$ as $\mathcal{N}_A$. For charge-conserving free fermions, the R\'enyi entanglement contour $s_n(j)$ can be further resolved into contributions from particle-number fluctuations at all orders. This leads to the definition of the \textit{hyperfine structure of entanglement}, a decomposition of $s_n(j)$ into a sum over local cumulant densities:
\begin{equation}
    s_{n}(j)=\sum_{k=1}^{\infty} h_{n;k}(j)
            =\sum_{k=1}^{\infty} \beta_{k}(n)\,C_{k}(j),
    \label{RenyiContour}
\end{equation}
where each term $h_{n;k}(j)$ encodes the $k$-th-order contribution to the R\'enyi contour, and the coefficients $\beta_k(n)$ depend only on the R\'enyi index $n$. These coefficients are given by
\begin{equation}
    \beta_{k}(n)
      =\frac{2}{n-1}\,\frac{1}{k!}
       \left(\frac{2\pi i}{n}\right)^{k}
       \zeta\!\left(-k,\tfrac{n+1}{2}\right),
\end{equation}
where $\zeta$ denotes the Hurwitz zeta function~\cite{olver2010nist}. Notably, $\beta_k(n)$ vanishes for all odd $k$, so only even cumulants contribute. For example, one finds $\beta_2(2) = \pi^2/4$ and $\beta_4(2) = -\pi^4/192$.

The local cumulant density $C_k(j)$ is defined through the generating function of particle-number fluctuations:
\begin{equation}
  C_{k}(j)\equiv
  \left[(-i\partial_{\lambda})^{k-1}
        \frac{\langle e^{i\lambda\hat N_{A}}\hat n_{j}\rangle}
             {\langle e^{i\lambda\hat N_{A}}\rangle}\right]_{\lambda=0},
  \label{densitycorr}
\end{equation}
where $\chi(\lambda,\mathcal{N} _A)\equiv\langle e^{i\lambda  \hat N_A}\rangle$ is the cumulant generating function, $\hat n_j$ is the number operator at site $j$, $\hat N_A$ is the total number operator on region $A$, and $\lambda \in \mathbb{R}$. This object captures non-local correlations between site $j$ and the rest of $A$. For instance, the first nontrivial contribution is
\begin{equation}
   C_{2}(j)=\sum_{i\in A}
           \left[\langle\hat n_{i}\hat n_{j}\rangle
                -\langle\hat n_{i}\rangle\langle\hat n_{j}\rangle\right],
\end{equation}
which encodes the two-point density-density correlation between $j$ and all other sites in $A$. Besides, the summation of $C_k(j)$ over subregion $A$ is given by the particle-number cumulant $C_k\equiv\sum_{j\in A}C_k(j)$ and one can easily check that  $C_k=(-i\partial_{\lambda})^{k}\ln  \chi(\lambda,\mathcal{N}_A)\big|_{\lambda=0}$.

Since Eq.~\eqref{RenyiContour} is a linear relation, knowledge of the R\'enyi contour up to some cutoff $k_{\mathrm{max}}$ allows for a reconstruction of the corresponding cumulants $\{C_k(j)\}$, and vice versa.
\begin{proof}
Using the projection operator $\hat{P}_l$, we can obtain the following useful conclusion
\begin{align}
   \sum_m \hat P_j\langle  \exp(i\lambda \hat N_A)\hat c_m^{\dagger}\hat c_m\rangle\hat P_j=\langle  \exp(i\lambda \hat N_A)\hat c_j^{\dagger}\hat c_j\rangle\label{dms},
\end{align}
which we will demonstrate later. According to the Ref.~\cite{2012PhRvB..85c5409S}, R\'{e}nyi entropy in Eq.~(\ref{renenr}) can be written as the summation of particle-number cumulants as follows
\begin{align}
    S_n  &=\frac{1}
    {1-n}\sum_{k=1}^{\infty}\beta_k(n)C_k\nonumber\\
    &=\frac{1}
    {1-n}\sum_{k=1}^{\infty}\beta_k(n)(-i\partial_{\lambda})^k\ln \chi(\lambda)|_{\lambda=0}\nonumber\\
     &=\frac{1}{1-n}\sum_{k=1}^{\infty}\beta_k(n)(-i\partial_{\lambda})^{k-1}\frac{\langle  \exp(i\lambda \hat N_A)\hat N_A\rangle}{\langle \exp(i\lambda \hat N_A)\rangle}\bigg{|}_{\lambda=0}\label{two}.
\end{align}
We can also define the cumulant dissemination of $S_n$ as $S_n=\sum_k^{\infty} h_{n;k}$ with $h_{n;k}\equiv\frac{1}
    {1-n}\beta_k(n)C_k$.
By applying the projection operator on both sides of Eq.~(\ref{two})  and referring to the Eq.~(\ref{dms}), we have
\begin{align}
 \hat P_j S_n \hat P_j
     &=\frac{1}{1-n}\sum_{k=1}^{\infty}\beta_k(n)\sum_m(-i\partial_{\lambda})^{k-1}\hat P_j\frac{\langle  \exp(i\lambda \hat N_A)\hat c_m^{\dagger}\hat c_m\rangle}{\langle \exp(i\lambda \hat N_A)\rangle}\hat P_j\bigg{|}_{\lambda=0}\nonumber\\
       &=\frac{1}{1-n}\sum_{k=1}^{\infty}\beta_k(n)(-i\partial_{\lambda})^{k-1}\frac{\langle  \exp(i\lambda \hat N_A)\hat n_j\rangle}{\langle \exp(i\lambda \hat N_A)\rangle}\bigg{|}_{\lambda=0}
\end{align}
Therefore, we demonstrate that the R\'{e}nyi contour defined in Eq.~(\ref{eq:RenyiContourDef}) can be written as
\begin{align}
    s_n(j)  =\sum_{k}h_{n;k}(j),
\end{align}
where we define the hyperfine structure of entanglement as
\begin{align}
    h_{n;k}(j)=\frac{1}{1-n}\sum_{k=1}^{\infty}\beta_k(n)(-i\partial_{\lambda})^{k-1}\frac{\langle  \exp(i\lambda \hat N_A)\hat n_j\rangle}{\langle \exp(i\lambda \hat N_A)\rangle}\bigg{|}_{\lambda=0}.
\end{align}
\end{proof}
\medskip\noindent
In the following, we will demonstrate the Eq.~(\ref{dms}).
\begin{proof}
 The proof primarily relies on the equation $\langle e^{i\lambda(\sum_{i=1}^{\mathcal{N}_A}c_i^{\dagger}c_i)}c_l^{\dagger}c_l\rangle=e^{i\lambda}\langle e^{i\lambda(\sum_{i=1,i\neq l}^{\mathcal{N}_A}c_i^{\dagger}c_i)}c_l^{\dagger}c_l\rangle$, which we will demonstrate first.
\begin{align}
    \sum_l  \left\langle e^{i\lambda(\sum_{i=1}^{\mathcal{N}_A}\hat c_i^{\dagger}\hat c_i)}\hat c_l^{\dagger}\hat c_l\right\rangle
     &=\left\langle \left (\prod_{i=1}^{\mathcal{N}_A}\left [1+   (e^{i\lambda}-1)\hat c_i^{\dagger}\hat c_i\right ]\right)\hat c_l^{\dagger}\hat c_l\right\rangle \nonumber\\
     &=\left\langle \left(\prod_{i=1,i\neq l}^{\mathcal{N}_A}\left [1+(e^{i\lambda}-1)\hat c_i^{\dagger}\hat c_i\right ]\right)\left [1+(e^{i\lambda}-1)\hat c_l^{\dagger}\hat c_l\right ]\hat c_l^{\dagger}\hat c_l\right\rangle\nonumber\\
      % &=\left\langle \prod_{i=1,i\neq l}^{\mathcal{N}_A}\left [1+(e^{i\lambda}-1)\hat c_i^{\dagger}\hat c_i\right ]\left [\hat c_l^{\dagger}\hat c_l+(e^{i\lambda}-1)\hat c_l^{\dagger}\hat c_l\hat c_l^{\dagger}\hat c_l\right ]\right\rangle\nonumber\\
       &=\left\langle \left(\prod_{i=1,i\neq l}^{\mathcal{N}_A}\left [1+(e^{i\lambda}-1)\hat c_i^{\dagger}\hat c_i\right ]\right)\left [\hat c_l^{\dagger}\hat c_l+(e^{i\lambda}-1)(\hat c_l^{\dagger}\hat c_l-\hat c_l^{\dagger}\hat c_l^{\dagger}\hat c_l\hat c_l)\right ]\right\rangle\nonumber\\
        &=\left\langle\left ( \prod_{i=1,i\neq l}^{\mathcal{N}_A}\left [1+(e^{i\lambda}-1)\hat c_i^{\dagger}\hat c_i\right ]\right)\left [\hat c_l^{\dagger}\hat c_l+(e^{i\lambda}-1)\hat c_l^{\dagger}\hat c_l\right ]\right\rangle\nonumber\\
        &=e^{i\lambda}\left\langle \left (\prod_{i=1,i\neq l}^{\mathcal{N}_A}\left [1+   (e^{i\lambda}-1)\hat c_i^{\dagger}\hat c_i\right ]\right)\hat c_l^{\dagger}\hat c_l\right\rangle \nonumber\\
        &=e^{i\lambda} \left\langle e^{i\lambda(\sum_{i=1,i\neq l}^{\mathcal{N}_A}\hat c_i^{\dagger}\hat c_i)}\hat c_l^{\dagger}\hat c_l\right\rangle\label{dms2}.
\end{align}
(1). When the subregion $A$ only contains one site $j$, that is $\hat N_A= \hat n_j$,
\begin{align}
  \sum_m  \hat P_j\left\langle  \exp(i\lambda \hat N_A)\hat c_m^{\dagger}\hat c_m\right\rangle\hat P_j&=\hat P_j\left\langle \left [1+(e^{i\lambda}-1)\hat c_j^{\dagger}\hat c_j\right ]\hat c_j^{\dagger}\hat c_j\right\rangle\hat P_j\nonumber\\
    &=\hat P_j\left\langle e^{i\lambda}\hat c_j^{\dagger}c_j\right\rangle\hat P_j\nonumber\\
    &=\left\langle  \exp(i\lambda \hat N_A)\hat c_j^{\dagger}\hat c_j\right\rangle.
\end{align}
(2). When the subregion $A$ contains two sites $i,j$, the process is similar.

\begin{align}
 &\sum_l\hat P_j\left\langle  \exp(i\lambda \hat N_A)\hat c_l^{\dagger}\hat c_l\right\rangle\hat P_j\nonumber\\
     =& \sum_l\hat P_j \left\langle \left [1+(e^{i\lambda}-1)\hat c_i^{\dagger}\hat c_i\right ]\left [1+(e^{i\lambda}-1)\hat c_j^{\dagger}\hat c_j\right ]\hat c_l^{\dagger}\hat c_l\right\rangle \hat P_j\nonumber\\
    =&\sum_l\hat P_j\left\langle \left [1+   (e^{i\lambda}-1)\hat c_{\bar l}^{\dagger}\hat c_{\bar l}\right ]e^{i\lambda}\hat c_l^{\dagger}\hat c_l\right\rangle\hat P_j\nonumber\\
    =&\sum_le^{i\lambda}\hat P_j\left [\left\langle \hat c_l^{\dagger}\hat c_l\right\rangle+   (e^{i\lambda}-1)\left (\left \langle \hat c_l^{\dagger}\hat c_l\right\rangle \left\langle \hat c_{\bar l}^{\dagger}\hat c_{\bar l}\right\rangle-\left\langle \hat c_l^{\dagger}\hat c_{\bar l}\right\rangle\left \langle \hat c_{\bar l}^{\dagger}\hat c_l\right\rangle\right ) \right ]\hat P_j\nonumber\\
    =&e^{i\lambda}\left [\left\langle \hat c_j^{\dagger}\hat c_j\right\rangle+   (e^{i\lambda}-1)\left (\left\langle \hat c_j^{\dagger}\hat c_j\right\rangle \left\langle \hat c_{i}^{\dagger}\hat c_{i}\right\rangle-\left\langle \hat c_j^{\dagger}\hat c_{i}\right\rangle\left\langle \hat c_{i}^{\dagger}\hat c_j\right\rangle \right ) \right ]\nonumber\\
    =&\left\langle \left [1+   (e^{i\lambda}-1)\hat c_{i}^{\dagger}\hat c_{i}\right ]e^{i\lambda}\hat c_j^{\dagger}\hat c_j\right\rangle\nonumber\\
    =&\left\langle \left [1+(e^{i\lambda}-1)c_i^{\dagger}c_i\right ]\left [1+(e^{i\lambda}-1)\hat c_j^{\dagger}\hat c_j\right ]\hat c_j^{\dagger}\hat c_j\right\rangle\nonumber\\
   =&\left\langle  \exp(i\lambda \hat N_A)\hat c_j^{\dagger}\hat c_j\right\rangle,
\end{align}
where  $\bar l$ is the site inside $A$ but excludes the site $l$.

(3).
The results can be easily extended to cases with any $\mathcal{N}_A$ by recognizing the central role played by the Wick theorem:

\begin{align}
 &\sum_l\hat P_j\left \langle  \exp(i\lambda \hat N_A)\hat c_l^{\dagger}\hat c_l\right \rangle\hat P_j\nonumber\\
  =&\sum_l\hat P_j\left \langle\left (\prod_{i=1,i\neq l}^{\mathcal{N}_A} [1+   (e^{i\lambda}-1)c_{i}^{\dagger}c_{i}]\right )e^{i\lambda}c_l^{\dagger}c_l\right \rangle\hat P_j\nonumber\\
 % =&\sum_le^{i\lambda}\hat P_j\left[\left \langle c_l^{\dagger}c_l\right \rangle+ (e^{i\lambda}-1)\left (\sum_{m\neq l}(\left \langle c_l^{\dagger}c_l\right \rangle\left \langle c_m^{\dagger}c_m\right \rangle-\left \langle c_l^{\dagger}c_m\right \rangle\left \langle c_m^{\dagger}c_l\right \rangle)\left \langle ...\right \rangle+\sum_{p\neq l,q\neq l}\left \langle c_l^{\dagger}c_p\right \rangle\left \langle c_q^{\dagger}c_l\right \rangle\left \langle ...\right \rangle\right )\right]\hat P_j\nonumber\\
 =&e^{i\lambda}\left [\left \langle c_j^{\dagger}c_j\right \rangle+ (e^{i\lambda}-1)\left (\sum_{m\neq j}\left(\left \langle c_j^{\dagger}c_j\right \rangle\left \langle c_m^{\dagger}c_m\right \rangle-\left \langle c_j^{\dagger}c_m\right \rangle\left \langle c_m^{\dagger}c_j\right \rangle\right)\left \langle ...\right \rangle+\sum_{p\neq l,q\neq l}\left \langle c_j^{\dagger}c_p\right \rangle\left \langle c_q^{\dagger}c_j\right \rangle\left \langle ...\right \rangle\right )\right ]\hat P_j\nonumber\\
  =&\left \langle\left (\prod_{i=1,i\neq j}^{\mathcal{N}_A} [1+   (e^{i\lambda}-1)c_{i}^{\dagger}c_{i}]\right )e^{i\lambda}c_j^{\dagger}c_j\right \rangle,\nonumber\\
  =&\left\langle  \exp(i\lambda \hat N_A)\hat c_j^{\dagger}\hat c_j\right\rangle,
 \end{align}
where $\langle ...\rangle$ represents the coefficient part obtained using the Wick's theorem, excluding the contribution from $\hat c_l^{\dagger},\hat c_l$ in the third row and $\hat c_j^{\dagger},\hat c_j$ in the fourth row.
\end{proof}

%%%%%%%%%%%%%%%%%%%%%%%%%%%%%%%%%%%%%%%%%%%%%%%%%%%%%%%%%%%%%%%%%%%%%%%%
\subsection{Information reconstruction from the fine structure $s_{n}(j)$ }\label{Ap:complete}
%%%%%%%%%%%%%%%%%%%%%%%%%%%%%%%%%%%%%%%%%%%%%%%%%%%%%%%%%%%%%%%%%%%%%%%%
From the above subsections, we know that the complete set of $h_{n;k}(j)$ with even $k$ can be utilized to obtain the complete set of $s_n(j)$. In the following, we will demonstrate that the fine structure $s_n(j)$ can be used to reconstruct the sets of $C_k(j)$ when a cutoff is applied. Furthermore, based on the fine structure, we can derive the R\'{e}nyi entropy, which allows us to reconstruct the entanglement spectra.

%%%%%%%%%%%%%%%%%%%%%%%%%%%%%%%%%%%%%%%%%%%%%%%%%%%%%%%%%%%%%%%%%%%%%%%%
\subsubsection{Particle-number correlation}
%%%%%%%%%%%%%%%%%%%%%%%%%%%%%%%%%%%%%%%%%%%%%%%%%%%%%%%%%%%%%%%%%%%%%%%%
Using R\'{e}nyi Contour as the building block, we can reconstruct the information of the particle-number correlation. To illustrate this, consider a free-fermion system with $\hat N_A$ fermions in subsystem A. Now, let's introduce  a \textit{cumulant coefficient matrix} denoted as $B$
\begin{align}
    B=\begin{pmatrix}
 \beta_2(1) &   \beta_4(1) &...  & \beta_{2\mathcal{N}_A}(1)  \nonumber\\
 \beta_2(2) &   \beta_4(2) &...  & \beta_{2\mathcal{N}_A}(2) \nonumber\\
 \vdots  &  & \ddots  & \nonumber\\
\beta_2(\mathcal{N}_A) &   \beta_4(\mathcal{N}_A) &...  & \beta_{\mathcal{N}_A}(\mathcal{N}_A)
\end{pmatrix},\label{pmatrix}
\end{align}
where $\beta_k(n)$ is non-zero only when $k$ is even and $\beta_k(n)=\frac{2}{n-1}\frac{1}{k!}\left(\frac{2\pi i}{n}\right)^k\zeta\left(-k,\frac{n+1}{2}\right)$, then Eq.~(1) of the main text can be written as
\begin{align}
    [s_1(j),
 s_2(j) ,
... ,
s_{\mathcal{N}_A}(j)]^T=B[C_2(j),
C_4(j) ,
... ,
C_{2\mathcal{N}_A}(j)]^T.
\end{align} By solving this linear algebra Eq.~(\ref{pmatrix})  inversely, we can reconstruct the even-rank particle-number fluctuation cumulants $C_k$ with the cut off $k_c=2\mathcal{N}_A.$ Since $C_k$ are the coefficient of the  Taylor expansion  $\chi(\lambda,\mathcal{N}_A)=\sum_{k=0}^{\infty} C_k\frac{\lambda^k}{k!}$,  centered around $\lambda=0$, the dominant $C_k$ terms correspond to smaller values of $k$, such as the former $k_c$ terms.

%%%%%%%%%%%%%%%%%%%%%%%%%%%%%%%%%%%%%%%%%%%%%%%%%%%%%%%%%%%%%%%%%%%%%%%%
\subsubsection{Entanglement spectrum}
%%%%%%%%%%%%%%%%%%%%%%%%%%%%%%%%%%%%%%%%%%%%%%%%%%%%%%%%%%%%%%%%%%%%%%%%
Besides the set of particle-number cumulant, R\'{e}nyi contour can also be utilized to reconstruct the R\'{e}nyi entropy and entanglement spectrum.  The summation of R\'{e}nyi contour over all sites yields the R\'{e}nyi entropy, from which the entanglement spectrum can be derived.

Let $T_n$ be the trace of the $n$-th power of the reduced density matrix~\cite{2012PhRvB..85c5409S}:
\begin{align}
T_n = \tr(\hat{\rho}_A^n) = e^{(1-n)S_n}.
\end{align}
Let us define \(T_1 = 1\) for a normalized density matrix and introduce a \(D \times D\) matrix as follows:
\begin{align}
U = \begin{pmatrix}
  1 & 1 & 0 & \dots  & \\
  T_2 & 1 & 2 & 0 & \dots  \\
  \vdots & \ddots & \ddots & \ddots & 0 \\
  T_{D-1} & T_{D-2} & \dots & T_2 & D-1 \\
  T_D & T_{D-1} & \dots & T_2 & 1
\end{pmatrix}.
\end{align}

Let's further denote the matrix obtained by taking the first \(n \times n\) submatrix of \(U\) as \(U_n\). Then, we can construct the following polynomial:
\begin{align}
P(x) = \sum_{n=0}^D \frac{(-1)^n}{n!} (\det U_n) x^{D-n},
\end{align}
where $\det U_0 = 1$. This polynomial is actually the characteristic polynomial of the density matrix $\hat{\rho}_A$: $P(x) = \det(xI - \hat{\rho}_A)$. Therefore, the roots of $P(x)$ are the entanglement spectrum.

%%%%%%%%%%%%%%%%%%%%%%%%%%%%%%%%%%%%%%%%%%%%%%%%%%%%%%%%%%%%%%%%%%%%%%
\subsection{Properties of the hyperfine structure}\label{property}
%%%%%%%%%%%%%%%%%%%%%%%%%%%%%%%%%%%%%%%%%%%%%%%%%%%%%%%%%%%%%%%%%%%%%%
The hyperfine components $h_{n;k}(j)$ exhibit several universal and desirable properties, which promote them to genuine quantum-informational measures (see Appendix \ref{Ap:property} for proofs):
\begin{enumerate}
  \item \textbf{Additivity.}
        For disjoint sites $i, j \in A$, one has
        \[
          h_{n;k}(i) + h_{n;k}(j) = h_{n;k}(i \cup j).
        \]

  \item \textbf{Normalization.}
        The spatial sum of $h_{n;k}(j)$ reproduces the $k$-th particle-number cumulant,
        \[
          C_k = \sum_{j\in A} \frac{h_{n;k}(j)}{\beta_k(n)},
        \]
        where $C_k = (-i\partial_\lambda)^k \ln \chi(\lambda, \hat N_A) \big|_{\lambda=0}$ and
        $\chi(\lambda, \hat N_A) = \langle e^{i\lambda \hat N_A} \rangle$.

  \item \textbf{Exchange symmetry.}
        If a unitary $\hat T$ satisfies $\hat T \hat \rho_A \hat T^\dagger = \hat \rho_A$ and exchanges sites $i \leftrightarrow j$, then
        \[
        h_{n;k}(i) = h_{n;k}(j).
        \]

  \item \textbf{Local-unitary invariance.}
        Let $\hat U^X$ be a unitary supported only on a subset $X \subseteq A$. Under the transformation $|\psi\rangle \mapsto \hat U^X |\psi\rangle$, the quantities $h_{n;k}(j)$ remain invariant for all $j \in X$.
\end{enumerate}

We emphasize that, unlike the total R\'enyi contour $s_n(j)$, which is strictly non-negative, both $h_{n;k}(j)$ and $C_k(j)$ can take positive or negative values.

%%%%%%%%%%%%%%%%%%%%%%%%%%%%%%%%%%%%%%%%%%%%%%%%%%%%%%%%%%%%%%%%%%%%%%%%
\subsection{Hyperfine structure of R\'{e}nyi entropy}
\label{subsec:FermiGas}
%%%%%%%%%%%%%%%%%%%%%%%%%%%%%%%%%%%%%%%%%%%%%%%%%%%%%%%%%%%%%%%%%%%%%%%%
\subsubsection{Free fermi gas case}
%%%%%%%%%%%%%%%%%%%%%%%%%%%%%%%%%%%%%%%%%%%%%%%%%%%%%%%%%%%%%%%%%%%%%%%%
Next, we will illustrate the hyperfine decomposition of R\'{e}nyi entropy in the context of the continuum, non-interacting Fermi gas in $(d\!+\!1)$ spacetime dimensions. In this setting, all hyperfine components $h_{n;k}$ with $k > 2$ are parametrically suppressed in the infrared (IR) limit, so that Eq.~\eqref{RenyiContour} simplifies for large subsystem size $k_F R \to \infty$ as
\begin{equation}
    \frac{s_{n}(x)}{h_{n;2}(x)} = 1 + o(1),
    \label{eq:sn_over_hn2}
\end{equation}
where $k_F$ is the Fermi momentum, and $R$ is the characteristic linear size of region $A$.
\begin{proof}

According to the following exact relation between particle fluctuation and R\'{e}nyi entropy~in Fermi gas~\cite{2012EL.....9820003C}
\begin{align}
  \frac{S_n}{C_2}&= \frac{(1+n^{-1})\pi^2}{6} +o(1),
 \end{align}
we notice that the contribution from higher cumulant ($C_{k>2}$)  become negligible when compared to $C_2$. Since the elements of projection operator is finite, by applying the projection operator,
\begin{align}
\hat P_j  S_n  \hat P_j=\frac{(1+n^{-1})\pi^2}{6}\hat P_j  C_2  \hat P_j+o(1)\left(\frac{(1+n^{-1})\pi^2}{6}\hat P_j  C_2  \hat P_j\right),
\end{align}
the second part ($\sum_k h_{n;k>2}$) is also of a magnitude less than $h_{n;2}$.  Hence,
\begin{align}
    \frac{s_n(x)}{h_{n;2}(x)}= 1+o(1).
\end{align}

\end{proof}

The dominant contribution to the R\'enyi contour is thus given by the second-order hyperfine density
\begin{equation}
    h_{n;2}(x) = \beta_{2}(n)
        \int_{y \in A} \!\! dy\,
        \Bigl[\langle \hat n(x)\hat n(y) \rangle
              - \langle \hat n(x) \rangle \langle \hat n(y) \rangle \Bigr],
    \label{eq:hn2_def}
\end{equation}
where $\hat n(x)$ is the local number operator and $\beta_2(n) = \pi^2/4$ for $n=2$.

Since integrating $h_{n;2}(x)$ over region $A$ yields the second particle-number cumulant, using the relationship between $h_{n;2}$ and the particle-number cumulant density, we can further derive an important property that relates $h_{n;2}(j)$ or $s_n(j)$ to the mutual information $I_2(A,B)$ between regions $A$ and $B$. This relationship is mathematically expressed as:
\begin{align}
   I_2(A_1,A_2) = S(A_1) + S(A_2) - S(A_1\cup A_2=A) = -2 \sum_{i \in A_1, j \in A_2} [\langle \hat n_i \hat n_j \rangle - \langle \hat n_i \rangle \langle \hat n_j \rangle].
\end{align}
We derive that:
\begin{align}
    S_A - h_{1;2}(i\in A_1) &= \sum_{i \in A_1, j \in A_1} [\langle \hat n_i \hat n_j \rangle - \langle \hat n_i \rangle \langle \hat n_j \rangle] - \sum_{i \in A_1, j \in A} [\langle \hat n_i \hat n_j \rangle - \langle \hat n_i \rangle \langle \hat n_j \rangle] \\
    &= -\sum_{i \in A_1, j \in A_2} [\langle \hat n_i \hat n_j \rangle - \langle \hat n_i \rangle \langle \hat n_j \rangle].
\end{align}
Therefore, the mutual information between regions $A_1$ and $A_2$ inside $A$ can be reformulated as:
\begin{align}
I_2(A_1, A_2)&= 2(S(A_1) - h_{1;2}(i\in A_1)),
\end{align}
which can also be written in a continuous form as
\begin{align}
I_2(A_1, A_2)&= 2(S(A_1) - \int_{x\in {A_1}}h_{1;2}(x)dx),
\end{align}
where  $h_{1;2}(x)$ is the hyperfine structure of region $A_1\cup A_2=A$.

%%%%%%%%%%%%%%%%%%%%%%%%%%%%%%%%%%%%%%%%%%%%%%%%%%%%%%%%%%%%%%%%%%%%%%%%
\subsubsection{(1+1)D CFT case}
\label{2d cft case}
%%%%%%%%%%%%%%%%%%%%%%%%%%%%%%%%%%%%%%%%%%%%%%%%%%%%%%%%%%%%%%%%%%%%%%%%
To make this concrete, we present an example to calculate the hyperfine structure in (1+1)D CFT, which will be useful in the section of holographic duality. In (1+1)D CFT, the leading term of the hyperfine structure $h_{n;2}$ in region A is given by
\begin{align}
    h_{n;2}(x)
    &=\frac{(1+n^{-1})\pi^2}{6}\int dy[\langle \hat n(x) \hat n(y)\rangle-\langle \hat n(x)\rangle\langle \hat n(y)\rangle]\label{eq:renyi222}.
\end{align}
Specifically, we consider the Euclidean action of the Dirac fermion as follows:
\begin{align}
    S=\frac{1}{2} \int d^2x \bar \psi\gamma_{\mu}\partial^{\mu}\psi,
\end{align}where $\psi=(\psi_R,\psi_L)^T.$ The Dirac fermions  $\psi_R$ and $\psi_L$ have conformal weights $(1/2,0)$ and $(0,1/2)$, respectively.  Their corresponding correlation functions are
\begin{align}
    \langle \psi_R^{\dagger}(z)\psi_R(\omega)\rangle=-\frac{1}{2\pi }\frac{1}{z-\omega},\quad \langle \psi_L^{\dagger}(\bar z)\psi_L(\bar \omega)\rangle=-\frac{1}{2\pi }\frac{1}{\bar z-\bar \omega}.
\end{align}

The particle-number operator in this field is
\begin{align}
    \hat n(x)=\psi_R^{\dagger}(x)\psi_R(x)+\psi_L^{\dagger}(x)\psi_L(x).
\end{align}
Inserting the above expression inside Eq.~(\ref{eq:renyi222}), we have
\begin{align}
    \langle \hat n(x)\hat n(y)\rangle=&\langle \psi_R^{\dagger}(x)\psi_R(x)\psi_R^{\dagger}(y)\psi_R(y)\rangle+\langle \psi_L^{\dagger}(x)\psi_L(x)\psi_L^{\dagger}(y)\psi_L(y)\rangle\nonumber\\
    &+\langle \psi_R^{\dagger}(x)\psi_R(x)\psi_L^{\dagger}(y)\psi_L^{\dagger}(y)\rangle+\langle \psi_L^{\dagger}(x)\psi_L(x)\psi_R^{\dagger}(y)\psi_R^{\dagger}(y)\rangle,
\end{align}
and
\begin{align}
    \langle \hat n(x)\rangle\langle \hat n(y)\rangle=&\langle \psi_R^{\dagger}(x)\psi_R(x)\rangle\langle \psi_R^{\dagger}(y)\psi_R(y)\rangle+\langle \psi_R^{\dagger}(x)\psi_R(x)\rangle\langle \psi_L^{\dagger}(y)\psi_L(y)\rangle\nonumber\\
&  + \langle \psi_L^{\dagger}(x)\psi_L(x)\rangle\langle \psi_R^{\dagger}(y)\psi_R(y)\rangle+\langle \psi_L^{\dagger}(x)\psi_L(x)\rangle\langle \psi_L^{\dagger}(y)\psi_L(y)\rangle.
\end{align}
Based on Wick's theorem and the chiral symmetry condition \(\langle \psi_R^{\dagger}\psi_L \rangle = \langle \psi_L^{\dagger}\psi_R \rangle = 0\), we have
\begin{align}
    \langle \hat n(x)\hat n(y)\rangle-\langle \hat n(x)\rangle\langle \hat n(y)\rangle=&-\langle \psi_R^{\dagger}(x)\psi_R(y)\rangle\langle \psi_R^{\dagger}(y)\psi_R(x)\rangle-\langle \psi_L^{\dagger}(x)\psi_L(y)\rangle\langle \psi_L^{\dagger}(y)\psi_L(y)\rangle\nonumber\\
    =&\frac{1}{2\pi^2 }\frac{1}{(x-y)^2}.
\end{align}
Therefore, the leading hyperfine structure $h_{n;2}$ can be written as
\begin{align}
     h_{n;2}(x)&=\frac{(1+n^{-1})\pi^2}{6}\int_{-R+\epsilon}^{R-\epsilon} \frac{1}{2\pi^2 }\frac{1}{(x-y)^2}dy\nonumber\\
     &=\frac{(1+n^{-1})}{12}(\frac{1}{R-x}+\frac{1}{R+x}).
\end{align}

By integrating $s_{n}(x)$ over $A$, we obtain the R\'{e}nyi entropy as
\begin{align}
    S_n=\int_{-R+\epsilon}^{R-\epsilon} dx s_n(x)
\approx\frac{\left(1+n^{-1}\right)}{6}\ln\frac{2R}{\epsilon}.
\end{align}
Comparing the known equality $S_n=[\left(1+n^{-1}\right)c/6]\ln\frac{2R}{\epsilon}$,  we end up with the value of the central charge $c=1$, which agrees with the known fact of Dirac fermions models. Importantly, while $S_n$ retains sensitivity to the UV cutoff $\epsilon$, the hyperfine density $h_{n;2}(x)$ is independent of $\epsilon$, emphasizing that it captures purely infrared information. This highlights the utility of the hyperfine decomposition in isolating universal, scale-invariant features of entanglement structure in free fermionic systems.

Here, we observe that the contour function in a Fermi gas, given by $s_{n} \approx h_{n;2} = \frac{c}{12} (1 + n^{-1}) \left(\frac{1}{R-x} + \frac{1}{R+x}\right)$, aligns with the contour function with conformal symmetry as defined in Ref.~\cite{2016JSMTE..12.3103C}. Remarkably, this expression holds true regardless of the presence of interaction or the specifics of the Hamiltonian.

Consequently, it is natural to ask whether the relationship between the contour function and the density of the particle-number cumulant still persists in general interacting (1+1)D CFT. In the following, we will demonstrate that this relationship does indeed hold, up to a constant. Specifically, we find that
\begin{align}
    \frac{s_{n}(x)}{C_2(x)}=\text{constant}+o(1).
\end{align}
\begin{proof}
According to the definition of the density of particle-number cumulant , we have the following relation
\begin{align}
    C_2(i)&\equiv\sum_{j\in A}[\langle \hat n_i \hat n_j \rangle - \langle \hat n_i \rangle \langle \hat n_j \rangle].
\end{align}
By dividing the subsystem $A$ into $A_1, A_2, A_3$. It is easy to check the linear combination relation between the density of particle-number cumulant contributed from region $A_2$ and the particle-number cumulant
\begin{align}
    C_2(A_2)=\frac{1}{2}\big(C_2(A_1\cup A_2)+C_2(A_2\cup A_3)-C_2(A_1)-C_2(A_3)\big)\label{eq:ALC}.
\end{align}
The left side of the equation is
\begin{align}
      C_2(A_2)=\sum_{i\in A_2}\sum_{j\in A}[\langle \hat n_i \hat n_j \rangle - \langle \hat n_i \rangle \langle \hat n_j \rangle],
\end{align}
which is the same as the right side of the equation.

The continuous form of Eq.~(\ref{eq:ALC})  in 1+1d CFT is
   \begin{align}
   C_2(x)=\frac{1}{2}\left(\frac{\partial   C_2(x_1,x)}{\partial x}-\frac{\partial   C_2(x,x_2)}{\partial x}\right)\label{continuous},
    \end{align}
where $x_1,x_2$ are the are the coordinates of the boundary of region A, and $C(x_1,x)$ is the particle-number fluctuation for the interval with boundary at $x_1, x$.

In (1+1)D CFT, the particle-number fluctuation is ~\cite{2010PhRvB..82a2405S}
\begin{align}
  C_2(A)=\frac{g}{\pi^2}\ln(\frac{2R}{\varepsilon})\label{eq:def},
\end{align}
where the prefactor $g$ can be fixed by considering the conserved charge and $\varepsilon$ is the cutoff.

Based on Eq.~(\ref{continuous}) and Eq.~(\ref{eq:def}), we have
\begin{align}
    C_2(x)=\frac{g}{2\pi^2}\left(\frac{1}{R-x}+\frac{1}{R+x}\right).
\end{align}

Therefore, the ratio between $s_2(x)$ and $c_2(x)$ is a constant
\begin{align}
    \frac{s_n(x)}{C_2(x)}=\frac{\frac{c}{12}(1+n^{-1})(\frac{1}{R-x}+\frac{1}{R+x})}{\frac{g}{2\pi^2}(\frac{1}{R-x}+\frac{1}{R+x})}=\frac{\pi^2 c(1+n^{-1})}{6g}.
\end{align}

This relation can be generalized to $h_{n;2}$ in CFT directly.
\end{proof}

%%%%%%%%%%%%%%%%%%%%%%%%%%%%%%%%%%%%%%%%%%%%%%%%%%%%%%%%%%%%%%%%%%%%%%%%
\section{Holographic duality of the hyperfine structure of entanglement}
\label{sec:Holography}
%%%%%%%%%%%%%%%%%%%%%%%%%%%%%%%%%%%%%%%%%%%%%%%%%%%%%%%%%%%%%%%%%%%%%%%%
From the AdS/CFT correspondence, the entanglement entropy of a boundary subregion $A$ can be determined by the extremal surface (homologous to the boundary of $A$) in bulk AdS spacetime, which is called the Ryu-Takayanagi (RT) surface $\mathcal{E}_A$~\cite{2006PhRvL..96r1602R,Hubeny_2007} in Fig.~\ref{nHRT}(c). It is natural to ask what the holographic duality of the hyperfine structure is.

%%%%%%%%%%%%%%%%%%%%%%%%%%%%%%%%%%%%%%%%%%%%%%%%%%%%%%%%%%%%%%%%%%%%%%%%
\subsection{Refined R\'{e}nyi entropy and its hyperfine structure}
\label{refined}
%%%%%%%%%%%%%%%%%%%%%%%%%%%%%%%%%%%%%%%%%%%%%%%%%%%%%%%%%%%%%%%%%%%%%%%%
In the following, we will answer this question by considering the refined R\'{e}nyi entropy defined as $ \tilde S_n \equiv n^2\partial_n\left(\frac{(n-1)}{n}S_n\right)$, which is dual to the cosmic brane in the bulk AdS spacetime~\cite{2016NatCo...712472D}. The relation between particle-number fluctuation and R\'{e}nyi contour in Eq.~(\ref{eq:sn_over_hn2}) can be rewritten for the $n$-th refined R\'{e}nyi contour as
\begin{align}
\!\! \! \! \tilde s_n(x)\!\approx \!\tilde h_{n;2}(x)\!\!=\!\!\frac{\pi^2}{3n} \!\int_{y\in A}\!\! dy[\langle \hat n(x) \hat n(y)\rangle\!-\!\langle \hat n(x)\rangle \!\langle \hat n(y)\rangle].\!\label{Eq:refinedrenyiholo}
\end{align}
It is crucial to recognize that although Eq.~(\ref{Eq:refinedrenyiholo}) is initially derived from the non-interacting Fermi gases model, it remains valid for generally interacting CFTs up to a specific constant coefficient, as proved in \ref{2d cft case}. Thus, Eq.~(\ref{Eq:refinedrenyiholo}) remains applicable in the large central charge limit, which has a dual description in terms of classical bulk gravity~\cite{2000PhR...323..183A}.

We will first show that the refined R\'{e}nyi entropy represents the entanglement entropy of a new density matrix, $\tilde{\rho}_A = \frac{(\hat{\rho}_A)^n}{\tr(\hat{\rho}_A)^n}$. Subsequently, we will derive Eq. (\ref{Eq:refinedrenyiholo}).

\begin{proof}

Considering a pure state $|\psi\rangle$ in Hilbert space $\mathcal{H}=\mathcal{H}_A\otimes \mathcal{H}_B$,  the Schmidt decomposition is as follows,

\begin{align}
|\psi\rangle=\sum_{q}a_q|\psi_A^q\rangle\otimes |\psi_B^q\rangle,
\end{align}
where $\{|\psi_A^q\rangle\}$ is an orthogonal basis for $\mathcal{H}_A$ and $\{|\psi_B^q\rangle\}$ is  an orthogonal basis for $\mathcal{H}_B$ and the Schmidt numbers sum up to one,  $\sum_q a_q^2=1$.
The reduced density matrix of region A is
\begin{align}
\hat \rho_A
&=\sum_q a_q^2(|\psi_A^q\rangle\langle \psi_A^q|).
\end{align}
The refined R\'{e}nyi contour $\tilde S_n(\hat \rho_A)$ can be written as the von Neumann entropy of a new density matrix $\tilde{\rho}_A=\frac{\hat \rho_A^n}{\tr \hat \rho_A^n}$,
\begin{align}
\tilde S_n(\hat \rho_A) &= -n^2\partial_n(\frac{1}{n}\ln \tr \hat \rho_A^n) \nonumber\\
&= \ln\tr\hat \rho_A^n - n\partial_n\ln\tr\hat \rho_A^n \nonumber\\
&= \ln\tr\hat \rho_A^n - \frac{\tr (\hat \rho_A^n\ln\hat \rho_A^n)}{\tr(\hat \rho^n_A)} \nonumber\\
&= -\tr\left(\frac{\hat\rho_A^n}{\tr \hat\rho_A^n}\right)\ln \left(\frac{\hat\rho_A^n}{\tr \hat\rho_A^n}\right) \nonumber\\
&= S\left(\frac{\hat \rho_A^n}{\alpha_n}\right)\label{newrho}
\end{align}
where we have denoted $\alpha_n=\tr \hat \rho_A^n$.

\end{proof}

Here we further discuss the entanglement Hamiltonian of the $n$-replica reduced density matrix $\tilde{\rho}_A=\frac{(\hat \rho_A)^n}{\tr (\hat \rho_A)^n}$. The original density density matrix is
\begin{align}
    \hat \rho_A&=\sum_p a_p^2 |\psi_A^p\rangle\langle\psi_A^p|\nonumber\\
    &=\frac{e^{- \hat K_A}}{Z},
\end{align}where  the corresponding  entanglement Hamiltonian is
\begin{align}
\hat K_A=\sum_p- \ln a_p^2 |\psi_A^p\rangle\langle\psi_A^p|,
\end{align}and $Z$ is the normalization constant. The density matrix is a Gibbs state with temperature $T=1$. The $n$-replica density matrix is defined as
\begin{align}
    \tilde\rho_A^{(n)}&\equiv\sum_p \frac{a_p^{2n} }{\alpha_n}|\psi_A^p\rangle\langle\psi_A^p|\nonumber\\
    &=\frac{e^{- \hat K_A^{(n)}}}{Z^{(n)}},
\end{align}
where $\alpha_n=\sum_q a_q^{2n}$  and
\begin{align}
\hat K_A^{(n)}
&=n\hat K_A\label{KAn}.
\end{align}
Here, $Z^{(n)}$ represents a new normalization constant, and the temperature of the Gibbs state is now set to $T = \frac{1}{n}$. It is important to emphasize that the two entanglement Hamiltonians commute, $[\hat{K}_A^{(n)}, \hat{K}_A] = 0$, yet their spectra differ.

Next, we will present two different ways to derive the Eq.~(\ref{Eq:refinedrenyiholo}). Firstly, note that Eq.~(\ref{Eq:refinedrenyiholo}) can be easily determined from the definition of the refined R\'{e}nyi contour. Secondly, we can use the analytical form of entanglement Hamiltonian $\hat K_A$ and $\hat K_A^{(n)}$ to verify it. In the Fermi gas, the entanglement Hamiltonian takes an analytical form resembling the following tridiagonal matrix, as described in Ref.~\cite{2017JPhA...50B4003E}:
\begin{align}
  K = \begin{pmatrix}
  d_1 & t_1 &  &  & \nonumber\\
  t_1 & d_2 & t_2 &  & \nonumber\\
  & t_2 & d_3 &  & \nonumber\\
  & & & \ddots & t_{\mathcal{N}_A-1} \nonumber\\
  & & & t_{\mathcal{N}_A-1} & d_{\mathcal{N}_A}
\end{pmatrix}, \label{modularham}
\end{align}
with the matrix elements defined as:
\begin{align}
    t_i = \frac{i}{\mathcal{N}_A} \left(1 - \frac{i}{\mathcal{N}_A}\right), \quad d_i = -2 \cos q_F \frac{2i-1}{2\mathcal{N}_A} \left(1 - \frac{2i-1}{2\mathcal{N}_A}\right),
\end{align}
where $q_F = \frac{\pi}{2}$ represents the Fermi momentum at the critical point.
Based on Eq.~\ref{KAn} and Eq.~\ref{modularham} as well as exact diagonalization, it is easy to confirm that the refined R\'{e}nyi contour $\tilde s_n(x)$ derived from $nT$ and the R\'{e}nyi contour derived from $T$ satisfy the relationship:
\begin{align}
    \tilde s_n(x) = \frac{1}{n} s_1(x), \label{refinedrenyia}
\end{align}
which aligns with Eq.~(\ref{Eq:refinedrenyiholo}). Equivalently, we have
\begin{align}
    \tilde S_n=\frac{S}{n}.
\end{align}
and
\begin{align}
    \tilde{S}_n &\equiv n^2 \partial_n \left(\frac{n-1}{n} S_n\right)\\\nno
   \omits{ &=-n^2 \partial_n  \left(\frac{n-1}{n}\frac{1}{1-n}(n-\frac{1}{n})\log L\right)\frac{c}{6}\\\nno
     &=-n^2 \partial_n  \left((1-\frac{1}{n^2})\log L\right)\\\nno
      &=-n^2 \partial_n  \left((1-\frac{1}{n^2})\right)\log L\\\nno }
      &=\frac{c}{3n}\log L,
\end{align}
in which
\begin{align}
S_n = \left(1+ \frac{1}{n}\right)\frac{c}{6}\log L.
\end{align}

%%%%%%%%%%%%%%%%%%%%%%%%%%%%%%%%%%%%%%%%%%%%%%%%%%%%%%%%%%%%%%%%%%%%%%%%
\subsection{Rindler transformation in AdS$_3$/CFT$_2$ correspondence}
\label{Ap:Rindler}
%%%%%%%%%%%%%%%%%%%%%%%%%%%%%%%%%%%%%%%%%%%%%%%%%%%%%%%%%%%%%%%%%%%%%%%%
\begin{figure}[t] %
\centering
\includegraphics[width=0.7\linewidth]{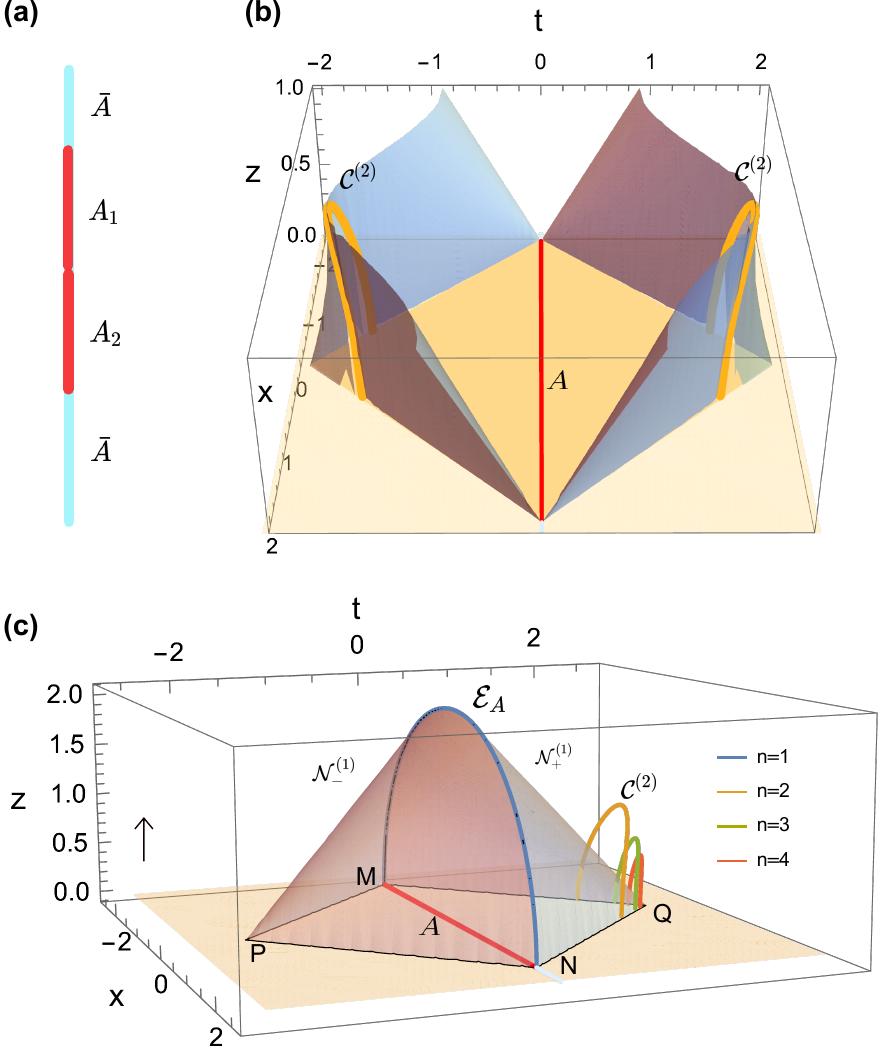}
\caption{ (a) A tripartition of the system to calculate the mutual information $I_2(A_1,A_2)$, where $A_1$ and $A_2$ can be disjoint.  In (c) (b),  the two transparent surfaces are the null hypersurfaces $\mathcal{N}^{(1)}_{\pm} (\mathcal{N}^{(2)}_{\pm})$, intersecting at the RT surface $\mathcal{E}_A$ (the extremal surface $\mathcal{C}^{(2)}$).
Discontinuities occur on the null hypersurfaces $\mathcal{N}^{(n\neq 1)}_{\pm}$ due to conical defect and the backreaction from $\mathcal{C}^{(n)}$. These singularities vanish when $n=1$.
For $n \geq 2$, the intersections of  $\mathcal{N}^{(n)}_{\pm}$ form  $\mathcal{C}^{(n)}$, depicted as yellow, green, and orange curves for $n=2, 3, 4$ respectively.
Clusters of $\mathcal{C}^{(n)}$ appear on both sides of $\mathcal{E}_A$, though for clarity, only the clusters on the right are shown in (c).
We set $R= 2$.}
\label{nHRT}
\end{figure}

\omits{
\begin{figure}[t]
\centering
\includegraphics[width=0.7\linewidth]{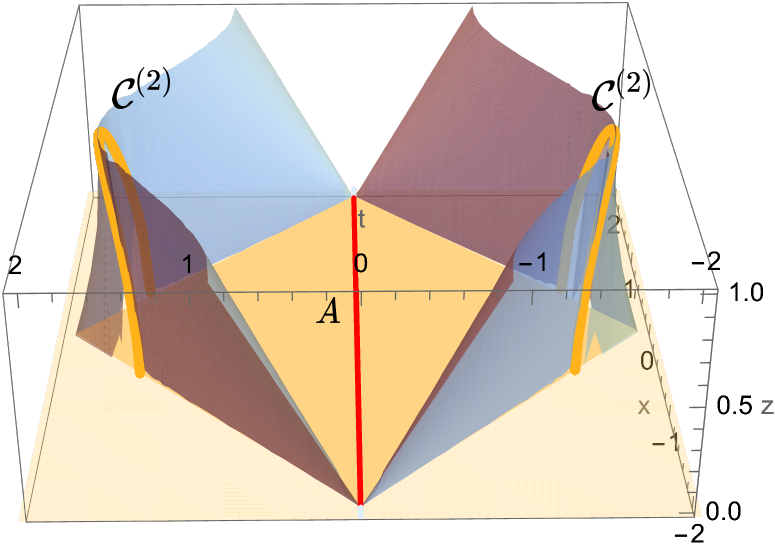}
 \caption{The red line represents subregion A. The two transparent surfaces are $\mathcal{N}^{(n)}_{\pm}$ with $n=2$, and their intersections are highlighted by yellow lines, which correspond to the extremal surfaces for $n=2$. These extremal surfaces, denoted $\mathcal{C}^{(2)}$, are anchored at and traverse along the entangling surface. They also exhibit tension and conical defects that cause backreactions to the bulk geometry, such as the discontinuity in $\mathcal{N}^{(n)}_{\pm}$.
We set $l_u = l_v = 2$ in the figure. }
            \label{n=2}
\end{figure}
}

To elucidate the holographic duality of hyperfine structure $h_{n;2}$ in the AdS$_3$ spacetime, we employ the Rindler transformation~\cite{2011JHEP...05..036C}. The fundamental idea of the Rindler transformation is to map the entanglement entropy of a region to the thermal entropy of another region, which can then be associated with the horizon entropy of a hyperbolic black hole. It turns out that the horizon corresponds to the null hypersurfaces $\mathcal{N}^{(1)}_{\pm}$ in the original AdS spacetime, as illustrated in Fig.~\ref{nHRT}(c). The intersection of these null hypersurfaces is the extremal surface $\mathcal{E}_A$.  It can be shown that the refined R\'{e}nyi entropy of $\hat{\rho}_A$ is equivalent to the entanglement entropy of the $n$-replica reduced density matrix $\tilde{\rho}_A \equiv \frac{(\hat{\rho}_A)^n}{\mathrm{Tr} (\hat{\rho}_A)^n}$. Therefore, the intersection of the $n$-dependent null hypersurfaces $\mathcal{N}^{(n)}_{\pm}$ is the holographic dual of $\tilde{S}_n$ in Fig.~\ref{nHRT}(b) and (c). Notably, within the entanglement wedge enclosed by the boundary $PMQN$ and  $\mathcal{E}_A$, the bulk operators can be reconstructed from the boundary CFT, which is the celebrated subregion-subregion duality in holographic duality~\cite{2014JHEP...12..162H,2016PhRvL.117b1601D,2020JHEP...12..083L}.

Specifically, consider a single interval \(A = \{(x, t = 0)\,|\,x \in [-R, R]\}\) in the vacuum of a CFT$_2$. In Poincar\'{e} coordinates $(x, t, z)$ with metric
\begin{equation}\label{poincare ads3}
  ds^{2} = \frac{1}{z^{2}}\left(dx^{2} + dz^{2} - dt^{2}\right),
\end{equation}
the interval endpoints lie at $(x_1, t_1)=(-R, 0)$ and $(x_2, t_2)=(R, 0)$. Further introducing light-cone coordinates $u = (x + t)/2$, $v = (x - t)/2$ and the radial coordinate $r = 2/z^{2}$, the metric (\ref{poincare ads3}) becomes
\begin{equation}
  ds^{2} = 2r\,du\,dv + \frac{dr^{2}}{4r^{2}},
  \label{eq:poincare_metric}
\end{equation}
and the two boundaries of the interval $A$ are $(u_1,v_1)=(-\frac{l_u}{2},-\frac{l_v}{2})$ and $(u_2,v_2)=(\frac{l_u}{2},\frac{l_v}{2})$, with $R=\frac{l_u+l_v}{2}$. The causal development of $A$ defines a diamond region bounded by $|u|, |v| \le R/2$. A Rindler transformation maps this region to a hyperbolic black hole, whose horizon area yields the entanglement entropy $S_1(A)$~\cite{2011JHEP...05..036C}.

To access refined R\'enyi entropies, one introduces an $n$-dependent Rindler-like metric~\cite{2016NatCo...712472D}:
\begin{equation}\label{eq:nRindler_metric}
  ds^{2} = \rho^{2} d\tilde y^{2} + (\rho^{2} - \rho_h^{2}) d\tilde\tau^{2}
         + \frac{d\rho^{2}}{\rho^{2} - \rho_h^{2}},
\end{equation}
where $\rho_h = \frac{1}{n}$.

\omits{
After coordinate redefinitions
\be
\tilde r + 1 = 2\rho^{2},\quad
\tilde u = \frac{\tilde y + \tau}{2},\quad
\tilde v = \frac{\tilde y - \tau}{2}, \quad (\tau = i\tilde\tau),
\ee
and rescalings
\be
u^{*} = \rho_h \tilde u,\quad
v^{*} = \rho_h \tilde v,\quad
r^{*} = \frac{\tilde r + 1}{\rho_h^{2}} - 1,
\ee
one finally  obtain the canonical ``Rindler-AdS'' form:
\begin{align}
    ds^{2} = du^{*2} + 2r^{*}du^{*}dv^{*} + dv^{*2}
      + \frac{dr^{*2}}{4(r^{*2} - 1)}.
\end{align}
The surface $r^{*} = 1$ maps to two null hypersurfaces $\mathcal{N}_{\pm}^{(n)}$ in the original coordinates. Solving $r^{*} = 1$ gives
\begin{align}
 \mathcal{N}_{+}^{(n)}:
 \quad r &= \frac{-2n^{2}}{l_{u}^{2}(n^{2} - 2) + 4n^{2}uv
           + 2l_{u} \sqrt{l_{u}^{2}(1 - n^{2}) - 4n^{2}uv + n^{4}(u + v)^{2}}}, \\
 \mathcal{N}_{-}^{(n)}:
 \quad r &= \frac{-2n^{2}}{l_{u}^{2}(n^{2} - 2) + 4n^{2}uv
           - 2l_{u} \sqrt{l_{u}^{2}(1 - n^{2}) - 4n^{2}uv + n^{4}(u + v)^{2}}}.
 \label{eq:null_surfaces}
\end{align}
Their intersection defines a codimension-two surface $\mathcal{C}^{(n)}$—the bulk dual of refined R\'enyi entropy. In $(x, t, z)$ coordinates, this surface is given parametrically by
\begin{align}
 x_{n}^{+}(u) &= \frac{2n^{2}u - 2\sqrt{-n^{4}(1 - n^{2})(1 - u^{2})}}{n^{4}}, \nonumber\\
 t_{n}^{+}(u) &= 2\left(u - \frac{u}{n^{2}}
               + \frac{\sqrt{-n^{4}(1 - n^{2})(1 - u^{2})}}{n^{4}}\right), \nonumber\\
 z_{n}^{+}(u) &= \left[l_{u}^{2}\left(1 - \frac{2}{n^{2}}\right)
               + \frac{4u \left(n^{2}(2 - n^{2})u
               + \sqrt{n^{4}(1 - n^{2})(l_{u}^{2} - 4u^{2})}\right)}{n^{4}}
               \right]^{1/2},
 \label{eq:xp_tp_zp}
\end{align}
with analogous expressions for the second branch $\mathcal{N}_{-}^{(n)}$. For $n = 1$, these reduce to the usual RT surface. For $n > 1$, the two branches tilt in opposite time directions and meet at a conical defect on the $t = 0$ slice.
}

Then we will utilize the Rindler transformation to derive the extremal surface associated with this metric, which corresponds to the refined R\'{e}nyi entropy. Before proceeding further, we include some remarks regarding the application of this metric. As shown in Ref.~\cite{2011JHEP...12..047H}, the entanglement entropy is equivalent to the thermal entropy, or the horizon entropy, of the dual black hole at a temperature $T_0 = \frac{1}{2\pi R}$.
\begin{align}
  &  S= S_{\text{thermal}}(T_0).
\end{align}
The R\'{e}nyi entropy can also be derived from the thermal entropy through
\begin{align}
     S_n= \frac{n}{n-1}\frac{1}{T_0} \int_{T_0/n}^{T_0}S_{\text{thermal}}(T)dT.
\end{align}
Note that the refined R\'{e}nyi entropy is nothing but the thermal entropy at temperature $T_n=T_0/n$, namely,
\begin{align}
    \tilde S_n&=n^2\frac{d}{dn}\left ( \frac{n-1}{n}S_n\right )\nno\\
    &=-\frac{n^2}{T_0}S_{\text{thermal}}(\frac{T_0}{n}) \frac{d}{dn}(T_0/n)\nno\\
    &=S_{\text{thermal}}(T_0/n).
\end{align}
Therefore, we can use Rindler transformation to derive the refined R\'{e}nyi entropy.
% \[
% \frac{d}{dt}\int^{a(t)}_{b(t)} f(x)dx=f(a(t))a'(t)-f(b(t))b'(t)
% \]

The  $n$-dependent Rindler-like metric (\ref{eq:nRindler_metric}) can also be written as
\begin{align}
ds^2=&\frac{d\tilde r^2}{4(\tilde r+1)(\tilde r+1-2\rho_h^2)}
+\rho_h^2d\tilde u^2+\rho_h^2d\tilde v^2+2\left(\frac{\tilde r+1}{\rho_h^2}-1\right)\rho_h^2d\tilde u d\tilde v,
\end{align}
by setting
\begin{align}
\tilde r+1=2\rho^2,\quad \tilde u=\frac{\tilde y+\tau}{2},\quad \tilde v=\frac{\tilde y- \tau}{2},\quad \tau =i\tilde \tau.
\end{align}
The above metric can be further transformed into the Rindler $\mathrm {AdS_3}$
\begin{align}
ds^2=du^{*2}+2r^{*}  du^{*}dv^{*}  +dv^{*2}+\frac{dr^{*2}}{4(r^{*2}-1)},
\end{align}
using
\begin{align}
u^*=\rho_h \tilde u,v^*=\rho_h \tilde v,r^*=\frac{\tilde r+1}{\rho_h^2}-1.
\end{align}
Therefore, the Rindler transformation from coordinate $(u,v,r)$ to $(u^*,v^*,r^*)$ is
\begin{align}
&u^*=\frac{\rho_h}{4}\ln (\frac{4(rv(l_u+2u)+1)^2-l_v^22 r^2(l_u+2u)^2}{4(rv(l_u-2u)-1)^2-l_v^2 r^2(l_u-2u)^2}),\nonumber\\
&v^*=\frac{\rho_h}{4}\ln(\frac{l_u^2r^2(l_v+2v)^2-4(l_vru+2ruv+1)^2}{l_u^2 r^2(l_v-2v)^2-4(-l_v ru+2ruv+1)^2}),\nonumber\\
&(r^*+1)\rho_h^2-1=\frac{r^2(l_u^2(l_v^2-4v^2)-4l_v^2 u^2)+4(2ruv+1)^2}{4l_ul_v r}.
\end{align}

By setting $r^*=1$,  the horizon of the $n$-Rindler AdS$_3$ maps to the two null hypersurfaces $\mathcal{N}_{\pm}^{(n)}$ in the original space $(u,v,r)$
\begin{align}
&\mathcal{N}^{(n)}_+:
\quad r=\frac{-2 n^2}{l_u^2(n^2-2)+4 n^2 u v+2l_u \sqrt{l_u^2(1-n^2)-4 n^2 uv+n^4(u+v)^2)}},\\
&\mathcal{N}^{(n)}_-:
\quad r=\frac{-2 n^2}{l_u^2(n^2-2)+4 n^2 u v-2l_u \sqrt{l_u^2(1-n^2)-4 n^2 uv+n^4(u+v)^2)}}.
\end{align}

The horizon of the $n$-Rindler $\mathrm {AdS_3}$ in Poincar\'{e} coordinates can be written as
\begin{align}
x_n^+&=    \frac{2 n^2 u - 2 \sqrt{-n^4 (1 - n^2) (1 - u^2)}}{n^4},\\
t_n^+&=2 \left(u - \frac{u}{n^2} + \frac{\sqrt{-n^4 (1 - n^2) (1 - u^2)}}{n^4}\right),\\
z_n^+&=\sqrt{l_u^2 \left(1 - \frac{2}{n^2}\right) + \frac{4 u \left(n^2 (2 - n^2) u + \sqrt{n^4 (1 - n^2) (l_u^2 - 4 u^2)}\right)}{n^4}},
\end{align}
and
\begin{align}
    x_n^-&=\frac{2 n^2 u + \sqrt{n^4 (1 - n^2) (l_u^2 - 4 u^2)}}{n^4},\\
t_n^-&=-\frac{2 n^2 u - 2 n^4 u + \sqrt{n^4 (1 - n^2) (l_u^2 - 4 u^2)}}{n^4},\\
z_n^-&=\sqrt{l_u^2 \left(1 - \frac{2}{n^2}\right) + 4 u^2 - \frac{4 u \left(2 n^2 u + \sqrt{n^4 (1 - n^2) (l_u^2 - 4 u^2)}\right)}{n^4}},
\end{align}
where we use $l_u=l_v$.
\begin{figure}[t]
\centering
\includegraphics[width=0.5\linewidth]{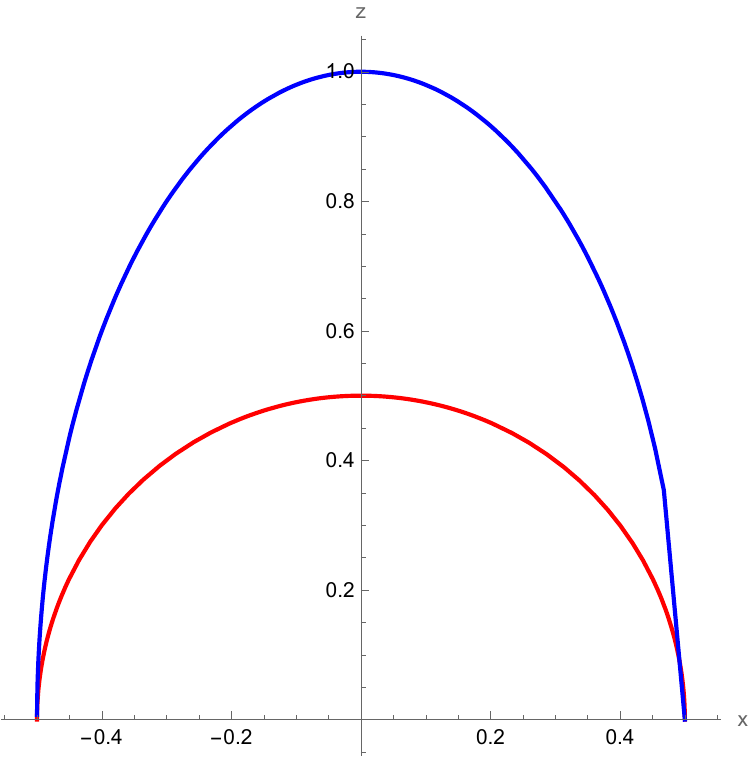}
\caption{The holography duality of refined R\'{e}nyi entropy lies outside the entanglement wedge. Here we take the time slice $t_n=\frac{1}{n^2}$, at which the extremal surface anchored on. The blue line denotes the the projection of $\mathcal{C}^{(2)}$ on the time slice and the red line denotes the time slice of the entanglement wedge.}
\label{Fig:outside}
\end{figure}
As we can see in the Fig.~\ref{nHRT}(c), the intersection of the null hypersurfaces $\mathcal{N}^{(1)}_{\pm}$ provides the RT surface $\mathcal{E}_A$. The entanglement wedge is defined as the  bulk domain of dependence bounded by $A$ and the RT surface $\mathcal{E}_A$. \omits{Within the entanglement wedge, the bulk operators can be reconstructed from the boundary CFT, which is the well-known subregion-subregion duality in holographic duality~\cite{2014JHEP...12..162H}.}

The intersection of $\mathcal{N}^{(n)}_{\pm}$ provides the extremal surface $\tilde{S}_n$ within the original coordinates $(x,t,z)$, as depicted in Fig.~\ref{nHRT}(b) .  The extremal surfaces, denoted as $\mathcal{C}^{(n)}$, for $n > 1$ differ markedly from the $n=1$ RT surface. It is noteworthy that the extremal surfaces $\mathcal{C}^{(n)}$ are inclined and exhibit a conical defect, manifesting as discontinuous regions in the null hypersurface $\mathcal{N}_{\pm}^{(n)}$. However, this singularity disappears when $n=1$.

Interestingly, note that $\mathcal{C}^{(n)}$ extends beyond the entanglement wedge, indicating that the hyperfine structure provides a finer description for the subregion-subregion duality beyond the entanglement wedge. It is noteworthy that $\mathcal{C}^{(n)}$ with $n > 1$ differ significantly from the $n=1$ RT surface $\mathcal{E}_A$. First, $\mathcal{C}^{(n)}$ are inclined in the time direction and exhibit a conical defect, manifesting as discontinuous regions in $\mathcal{N}_{\pm}^{(n)}$. Second, $\mathcal{C}^{(n)}$ cannot be connected to $\mathcal{E}_A$ through time evolution, as unitary time evolution does not alter the entanglement spectrum of a density matrix.

Moreover, as is shown in Fig.~\ref{Fig:contourplane}, with the help of the hyperfine structure we introduced, we can find out a finer description for the subregion-subregion duality in entanglement wedge and R\'enyi entanglement wedge. We can use the planes that link the boundary sites with the extremal surfaces $\mathcal{C}^{(n)}$ to slice the extremal surface  according to the contour distribution. Each segment derived from the extremal surface represents the holographic dual of the hyperfine structure. Thus, the holographic description of R\'enyi entropy is refined into the duality between the boundary density-density correlator and the segment of bulk extremal surface. This finer correspondence indicates that a bulk local operator can be reconstructed by CFT operators in a segment of the subregion on the boundary.

\begin{figure}[t]
\centering
\includegraphics[width=0.7\linewidth]{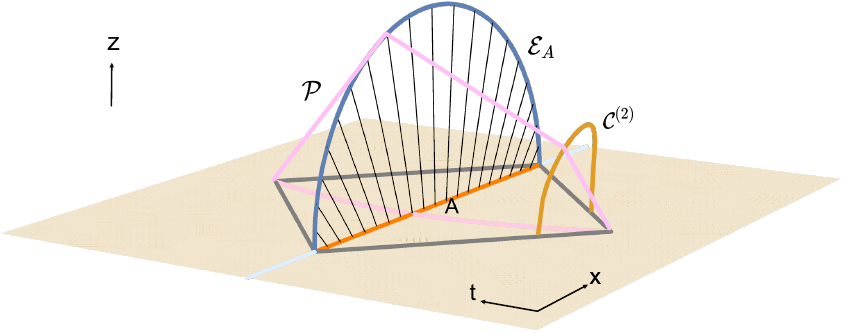}
\caption{Slice the extremal surfaces according to the distribution of the hyperfine structure using the pink plane.}
\label{Fig:contourplane}
\end{figure}
%%%%%%%%%%%%%%%%%%%%%%%%%%%%%%%%%%%%%%%%%%%%%%%%%%%%%%%%%%%%%%%%%%%%%%%%
\subsection{Discussions on the R\'{e}nyi entanglement wedge}
%%%%%%%%%%%%%%%%%%%%%%%%%%%%%%%%%%%%%%%%%%%%%%%%%%%%%%%%%%%%%%%%%%%%%%%%
Based on observations in the above subsection, we can further define the R\'{e}nyi entanglement wedge enclosed by $\mathcal{C}^{(n)}$, which extends beyond the entanglement wedge. One may wonder whether this contradicts the well-known entanglement wedge reconstruction, which states that the largest region for bulk reconstruction is typically the entanglement wedge. To verify the bulk reconstruction property, we further analyze the distribution of boundary and bulk modular flows associated with subregion $A$.
%%%%%%%%%%%%%%%%%%%%%%%%%%%%%%%%%%%%%%%%%%%%%%%%%%%%%%%%%%%%%%%%%%%%%%%%
\omits{\subsubsection{Modular flow on the boundary and bulk}}
%%%%%%%%%%%%%%%%%%%%%%%%%%%%%%%%%%%%%%%%%%%%%%%%%%%%%%%%%%%%%%%%%%%%%%%%

The modular Hamiltonian is the generator along the thermal circle $k_t\equiv \tilde\beta^i\partial_{\tilde x^i}$ in Rindler space. In order to map it to the original space, we need to solve the following differential equations
\begin{align}
\partial_u&=(\partial_u \tilde u)\partial_{\tilde u}+(\partial_u \tilde v)\partial_{\tilde v}+(\partial_u \tilde r)\partial_{\tilde r},\\
\partial_v&=(\partial_v \tilde u)\partial_{\tilde u}+(\partial_v \tilde v)\partial_{\tilde v}+(\partial_v \tilde r)\partial_{\tilde r},\\
\partial_r&=(\partial_r \tilde u)\partial_{\tilde u}+(\partial_r \tilde v)\partial_{\tilde v}+(\partial_r \tilde r)\partial_{\tilde r}.
\end{align}
For the general \( n \) case, these equations become:
\begin{align}
\partial_u&=(\partial_u u^*)\partial_{ u^*}+(\partial_u v^*)\partial_{\tilde v}+(\partial_u r^*)\partial_{r^*},\\
\partial_v&=(\partial_v u^*)\partial_{u^*}+(\partial_v  v^*)\partial_{ v^*}+(\partial_v r^*)\partial_{ r^*},\\
\partial_r&=(\partial_r u^*)\partial_{u^*}+(\partial_r v^*)\partial_{v^*}+(\partial_r r^*)\partial_{r^*}.
\end{align}
By plugging the bulk Rindler transformations (from \((u, v, r)\) to \((u^*, v^*, r^*)\)) into the above differential equations and solving them, we obtain the modular flow in both the boundary and bulk. In Rindler AdS$_3$, the generator of the Hamiltonian is simply the translation along the thermal circle. Mapping this to the original space yields the modular flow in the boundary and bulk.

The modular flow in the bulk is given by:
\begin{align}
k_t^{\text{bulk}} &= \beta^*_{u^*} \partial_{u^*} + \beta^*_{v^*} \partial_{v^*} \nno\\
&= -\pi \partial_{u^*} + \pi \partial_{v^*} \nno\\
&= -n\pi \partial_{\tilde{u}} + n\pi \partial_{\tilde{v}} \nno\\
&= n\left(\frac{2\pi u^2}{l_u} - \frac{\pi l_u}{2} + \frac{\pi}{l_v r}\right) \partial_u + 4n\pi r \left(\frac{v}{l_v} - \frac{u}{l_u}\right) \partial_r \nno\\
&\quad + \frac{n}{2} \pi \left(-\frac{2}{l_u r} - \frac{4v^2}{l_v} + l_v\right) \partial_v.
\end{align}
Solving the system of differential equations:
\begin{align}
\frac{\partial u(s)}{\partial s} &= n\left(\frac{2\pi u^2}{l_u} - \frac{\pi l_u}{2} + \frac{\pi}{l_v r}\right), \\
\frac{\partial v(s)}{\partial s} &= \frac{n}{2} \pi \left(-\frac{2}{l_u r} - \frac{4v^2}{l_v} + l_v\right), \\
\frac{\partial r(s)}{\partial s} &= 4n\pi r \left(\frac{v}{l_v} - \frac{u}{l_u}\right),
\end{align}
we obtain the modular flow in the bulk. By reparameterizing time as \( s \rightarrow ns \), these equations reduce to the modular flow of the \( n = 1 \) case. This implies that although the R\'{e}nyi entanglement wedge extends beyond the entanglement wedge, the largest bulk region in which local operators can be reconstructed using the modular flow is still the entanglement wedge. However, the extension beyond the entanglement wedge can be used to reconstruct other information about the bulk spacetime, such as the metric.

The modular flow on the boundary is given by:
\begin{align}
k_t = \left(\frac{2n\pi u^2}{l_u} - \frac{n\pi l_u}{2}\right) \partial_u + \frac{n}{2} \pi \left(-\frac{4v^2}{l_v} + l_v\right) \partial_v.
\end{align}
Solving the differential equations \( \left(\frac{\partial u(s)}{\partial s}, \frac{\partial v(s)}{\partial s}\right) = k_t \), we obtain the boundary solution:
\begin{align}
u(s) &= -\frac{1}{2} l_u \tanh(n\pi s - 2 l_u k), \\
v(s) &= \frac{1}{2} l_v \tanh(n\pi s + 2 l_v k),
\end{align}
which scales the time $s$ by a factor of $n$.

In summary, we verified that the boundary and bulk modular flows in the R\'{e}nyi entanglement wedge remain consistent with those in the entanglement wedge, confirming that the largest bulk region in which local operators can be reconstructed by the modular flow in the bulk is still the entanglement wedge. However, the fact that the R\'{e}nyi entanglement wedge extends beyond the entanglement wedge suggests that the refined R\'{e}nyi entropy captures additional information about the bulk geometry that is not encoded in the entanglement wedge alone, e.g., the bulk metric. Similar result has been found in Ref.~\cite{2020PhRvD.101f6011B} in which a lower dimensional bulk minimal surface (with boundary within a given stripe like subregion $A$) can extend beyond the higher dimensional bulk minimal surface (i.e. extend beyond the entanglement wedge) of $A$. While in our paper, the R\'{e}nyi entanglement wedge has the same dimension with the entanglement wedge, which is a novel finding. The R\'{e}nyi entanglement wedge might provide a new framework for reconstructing additional information in the bulk~\cite{2024ScPP...16..144A,2021JHEP...04..062A,2020PhRvD.101f6011B,2024arXiv240804016B}, a direction we leave for future work.

%%%%%%%%%%%%%%%%%%%%%%%%%%%%%%%%%%%%%%%%%%%%%%%%%%%%%%%%%%%%%%%%%%%%%%%%
\section{Hyperfine structure of entanglement in Chern insulators}
\label{Chern insulator}
%%%%%%%%%%%%%%%%%%%%%%%%%%%%%%%%%%%%%%%%%%%%%%%%%%%%%%%%%%%%%%%%%%%%%%%%
It has been shown in subsection~\ref{subsec:FermiGas} that in Fermi gas systems, higher ranks of the hyperfine structure are suppressed. However, in more general cases, we observe a variety of parameter regions where $h_{n;k>2}$ gives nontrivial contributions to both the fine structure and entropy. In this section, we will explore the behavior of all ranks of the hyperfine structure in a class of lattice fermion models--the Chern insulators. Initially, we will examine the distribution function of $h_{n;k>2}$ across different parameter regions. Subsequently, we will demonstrate the emergent scaling law of normalized $h_{n;k}(m;k_x=0,\pi)$ in the presence of topological edge states.
%%%%%%%%%%%%%%%%%%%%%%%%%%%%%%%%%%%%%%%%%%%%%%%%%%%%%%%%%%%%%%%%%%%%%%%%
\subsection{$s_n$ and $h_{n;k}$ in various parameter regions}
\label{Ap:scaling}
%%%%%%%%%%%%%%%%%%%%%%%%%%%%%%%%%%%%%%%%%%%%%%%%%%%%%%%%%%%%%%%%%%%%%%%%
Considering the model of $(2+1)$D Chern insulator with Hamiltonian~\cite{2006PhRvB..74h5308Q,2011RvMP...83.1057Q, 2016RvMP...88c5005C}
\begin{align}
 \hat{\mathcal{H}}=\sum_{\mathbf{k}}\hat c_{ \mathbf{k}}^{\dagger}H(\mathbf{k})\hat c_{ \mathbf{k} },
\end{align}
where
\begin{align}
H(\mathbf{k})=(m+\cos  {k}_x+\cos  {k}_y)\sigma_z+\lambda(\sin  {k}_x\sigma_x+\sin  {k}_y\sigma_y)-\mu \mathbf{I},
\end{align}
and $\sigma_i$ represent the Pauli matrices and  $\mathbf{I}$  is the $2\times 2$ identity matrix. This Hamiltonian comprises a Zeeman splitting term induced by the mass parameter  $m$ and a spin-orbit coupling term with amplitude $\lambda$. The energy gap closes at $ m=\pm 2$, forming a Dirac point respectively at $ \textbf{k}=(0,0)$ and $ \textbf{k}=(\pi,\pi)$. For $m=0$, the energy gaps close, creating two Dirac points with vanishing density-of-states: one at $ \textbf{k}=(0,\pi)$ and the other at $ \textbf{k}=(\pi,0)$. In the following, we consider four distinct scenarios: (i) trivial mass gap; (ii) topological mass gap (supporting nonzero Chern number and robust edge states); (iii) critical Dirac point; (iv) Fermi surface.

\begin{figure*}[t]
 \centering
 \includegraphics[width=1\linewidth]{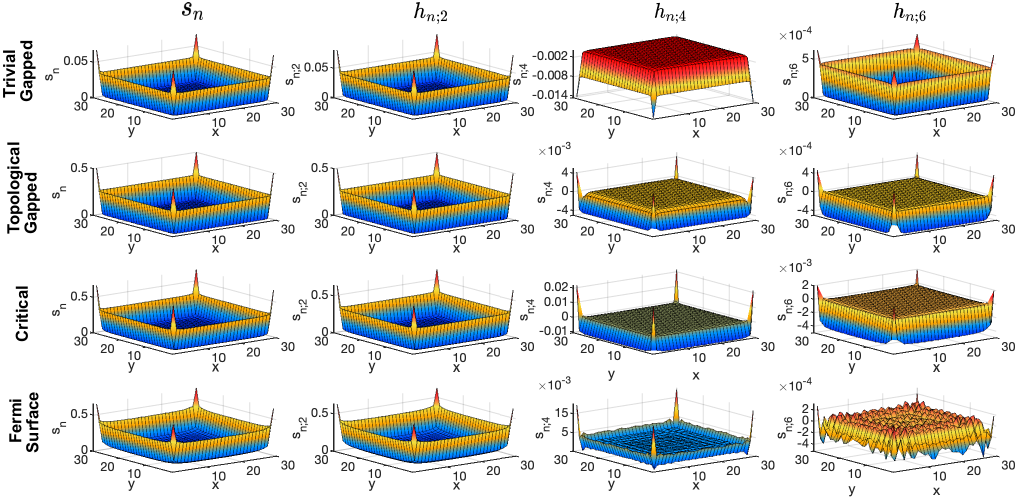}
 \caption{Site-resolved distributions of the R\'enyi contour
  $s_{n}(j)$ and its hyperfine components $h_{n;2}(j)$,
  $h_{n;4}(j)$, $h_{n;6}(j)$ for a $60\times60$ square lattice (periodic
  boundaries) with subregion $A$ the central $30\times30$ block.
  From top to bottom: (i) trivial insulator ($m\!=\!3,\mu\!=\!0$),
  (ii) Chern insulator ($m\!=\!1,\mu\!=\!0$), (iii) Dirac critical
  point ($m\!=\!0,\mu\!=\!0$), and (iv) metallic phase with a Fermi
  surface ($m\!=\!0,\mu\!=\!2/3$).  All panels use $n=2$ and
  hopping amplitude $\lambda=1$.  The hyperfine hierarchy clearly
  distinguishes topological, critical, and Fermi-surface regimes.}
 \label{allrankrenyi}
\end{figure*}
%Among them, the entanglement entropies of the  former three cases support the  area law at the scaling limit, while, the last one supports super-area law based on the Widom conjecture {\color{red}[XXX]} of asymptotic behavior of Toeplitz matrices.

Fig.~\ref{allrankrenyi} displays the distributions of $s_n(j), h_{n;2}(j), h_{n;4}(j),$ and $h_{n;6}(j)$ for various types of parameters. The four rows correspond to cases of trivial mass gap, topological mass gap, critical Dirac point, and Fermi surface. The four columns represent different terms of the hyperfine structure. Notably, the behavior of $h_{n;k>6}$ resembles that of $h_{n;4}(j)$ and $h_{n;6}(j)$, thus we focus on the first three non-zero terms $h_{n;2,4,6}(j)$.

Examining the first two columns within each row, we observe that the distribution and magnitude of $s_n(j)$ and $h_{n;2}(j)$ are very similar, indicating that $h_{n;2}$ is dominant in the series $h_{n;k}$. This aligns with previous findings~\cite{2015PhRvB..92k5129F} showing a numerical resemblance between the entanglement contour and the density of $C_2$. In addition to this observation, we find more interesting features as follows.

% Comparing the distribution of $h_{n;2}(j)$ across the three system types, we note that in cases of mass gap (either trivial or topological) and critical point, $h_{n;2}(j)$ decays more rapidly from the boundary of $A$ to the center of $A$ compared to the Fermi surface case, suggesting different scaling laws between mass gap/critical systems and Fermi surface systems. \textbf{XXXX[Here, add texts and connect to new SM section]}

Comparing the distribution of $h_{n;2}(j)$ across the three system types, we note that in cases of mass gap (either trivial or topological) and critical points, $h_{n;2}(j)$ decays more rapidly from the boundary of $A$ to the center of $A$ compared to the Fermi surface case. To further explore the distribution function of $h_{n;2}$ across various parameter regions, we present a cross-sectional analysis of the hyperfine structure $h_{n;2}$ in Fig.~\ref{Fig:scal}. For clarity, we set $n = 2$. As shown in Fig.~\ref{Fig:scal}(a), it is apparent that in both trivially gapped and topologically gapped scenarios, $h_{n;2}$ exhibit an exponential decrease. In contrast, at critical points and along Fermi surfaces, $h_{n;2}$ decreases following a power-law decay, which is notably slower than the exponential decline.

Furthermore, it is noteworthy that the power-law exponents at critical points and Fermi surfaces are distinct. The decay of $h_{n;2}$ at critical points is faster than that at Fermi surfaces. Despite the similarities in scaling behavior of entanglement entropy between critical points and trivially gapped cases, the distribution functions of the hyperfine structure $h_{n;2}$ exhibit marked differences.

 \begin{figure}[t]
            \centering
            \subfigure[]{\includegraphics[width=0.45\linewidth]{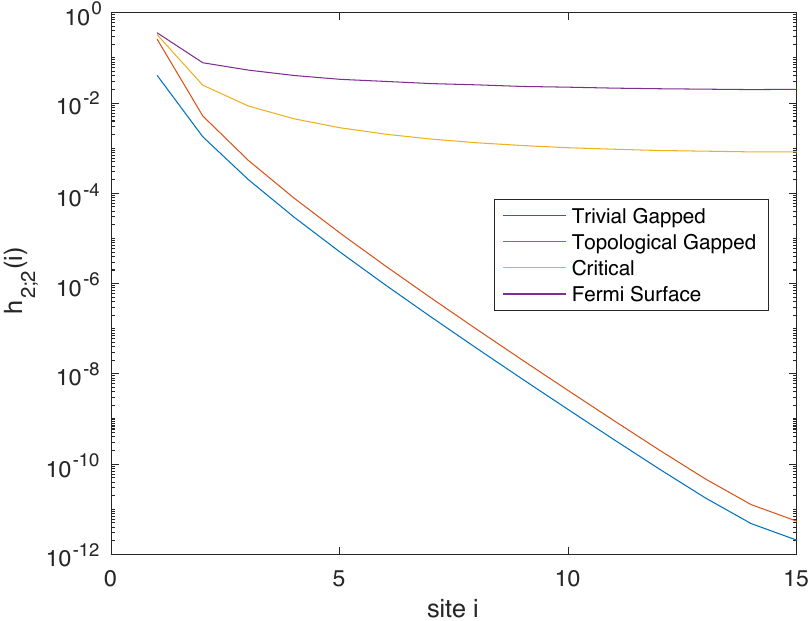}}
            \subfigure[]{\includegraphics[width=0.45\linewidth]{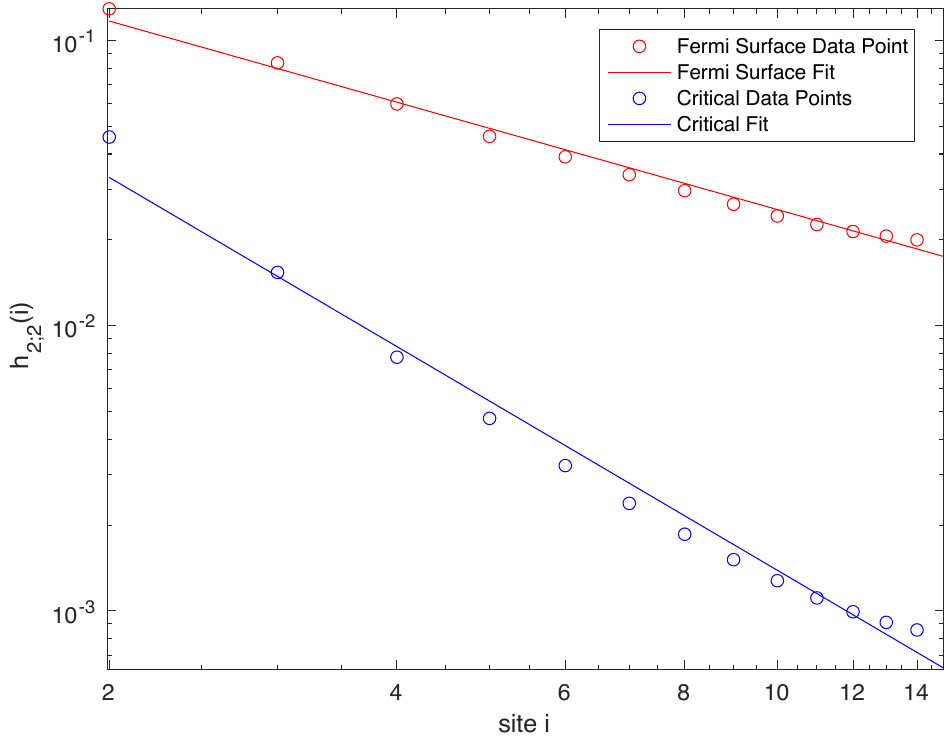}}
            \caption{The cross section of the hyperfine structure $h_{2;2}(x)$ in the four parameter regions. From Fig.(a) we can find that in the cases of both trivial gapped and topological gapped, $h_{2;2}(x)$ decrease exponentially from the boundary to the bulk. From both (a)  and (b) we can see that the $h_{2;2}(x)$ of critical and Fermi surface decrease in a power law. }
            \label{Fig:scal}
      \end{figure}

As Eq.~(\ref{RenyiContour}) allows for the analysis of contributions from all values of $k$, we investigate the hyperfine structure $h_{n;k}(j)$ for larger $k$ ($k > 2$), as depicted in the third and fourth columns, where negative values of $h_{n;k}(j)$ are observed. In the trivial mass gap scenario, for any given $n$ and $k$, the values of $h_{n;k}(j)$ across the entire region $A$ maintain a consistent sign at all sites $j$, with corner and hinge sites sharing the same sign. However, in both the topological mass gap and critical Dirac point cases, the sign of $h_{n;k}(j)$ at corner sites may differ from that at hinge sites. In this way, we can  distinguish  the case of  trivial mass gap from topological mass gap as well as critical point.
%%%%%%%%%%%%%%%%%%%%%%%%%%%%%%%%%%%%%%%%%%%%%%%%%%%%%%%%%%%%%%%%%%%%%%%%
\subsection{Edge states of Chern insulator}
\label{Ap:edge}
%%%%%%%%%%%%%%%%%%%%%%%%%%%%%%%%%%%%%%%%%%%%%%%%%%%%%%%%%%%%%%%%%%%%%%%%
\begin{figure}[t] %
\centering
\includegraphics[width=\linewidth]{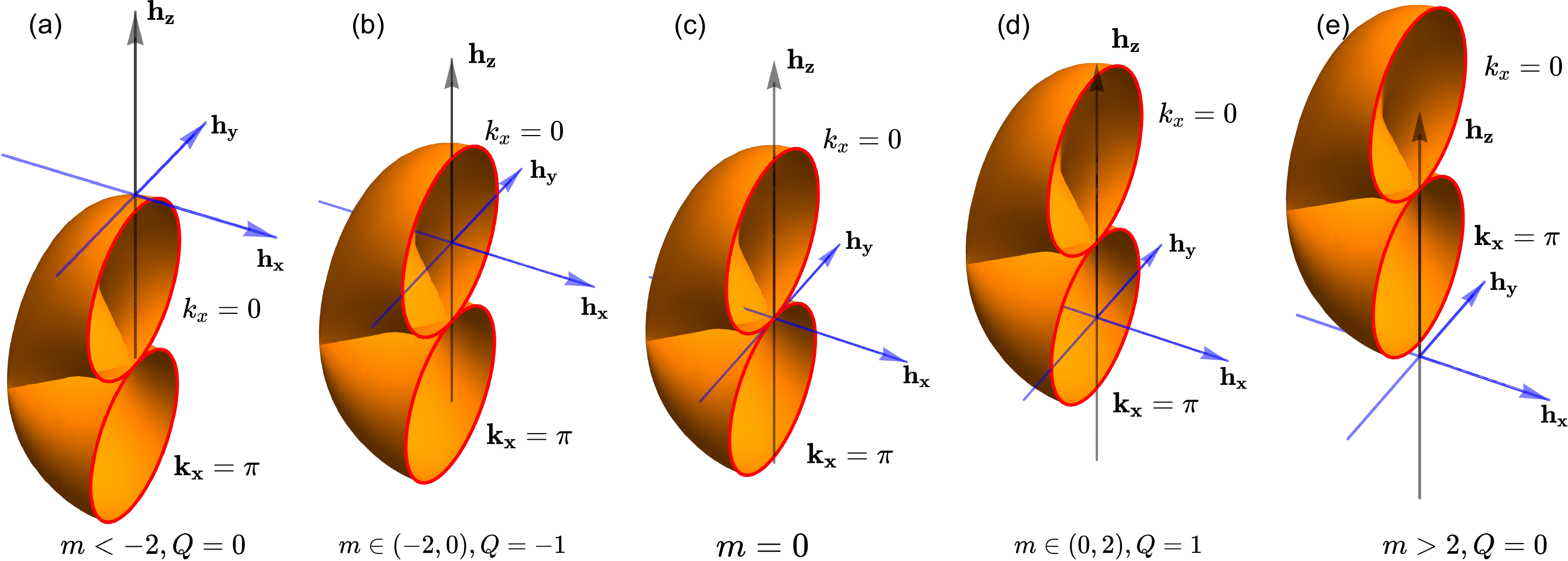}
\caption{The torus $\mathbf{h}(\mathbf{k}) $ of the Chern insulator for different values of m. In (a) ,(c) and (e), the torus do not contain the origin hence Chern number Q=0. In (b), the red circle with $k_x=0$ encloses the origin, meaning that there are zero-energy edge states at this $k_x$.  At $k_x=\pi$, the red loop lies in the plane of the origin without containing it, indicating no edge state at $k_x=\pi$.  (d) is similar to (b) but has zero-energy edge state at $k_x=\pi$.}
\label{Fig:bbc}
\end{figure}

\begin{figure}[t] %
\centering
\includegraphics[width=1\linewidth]{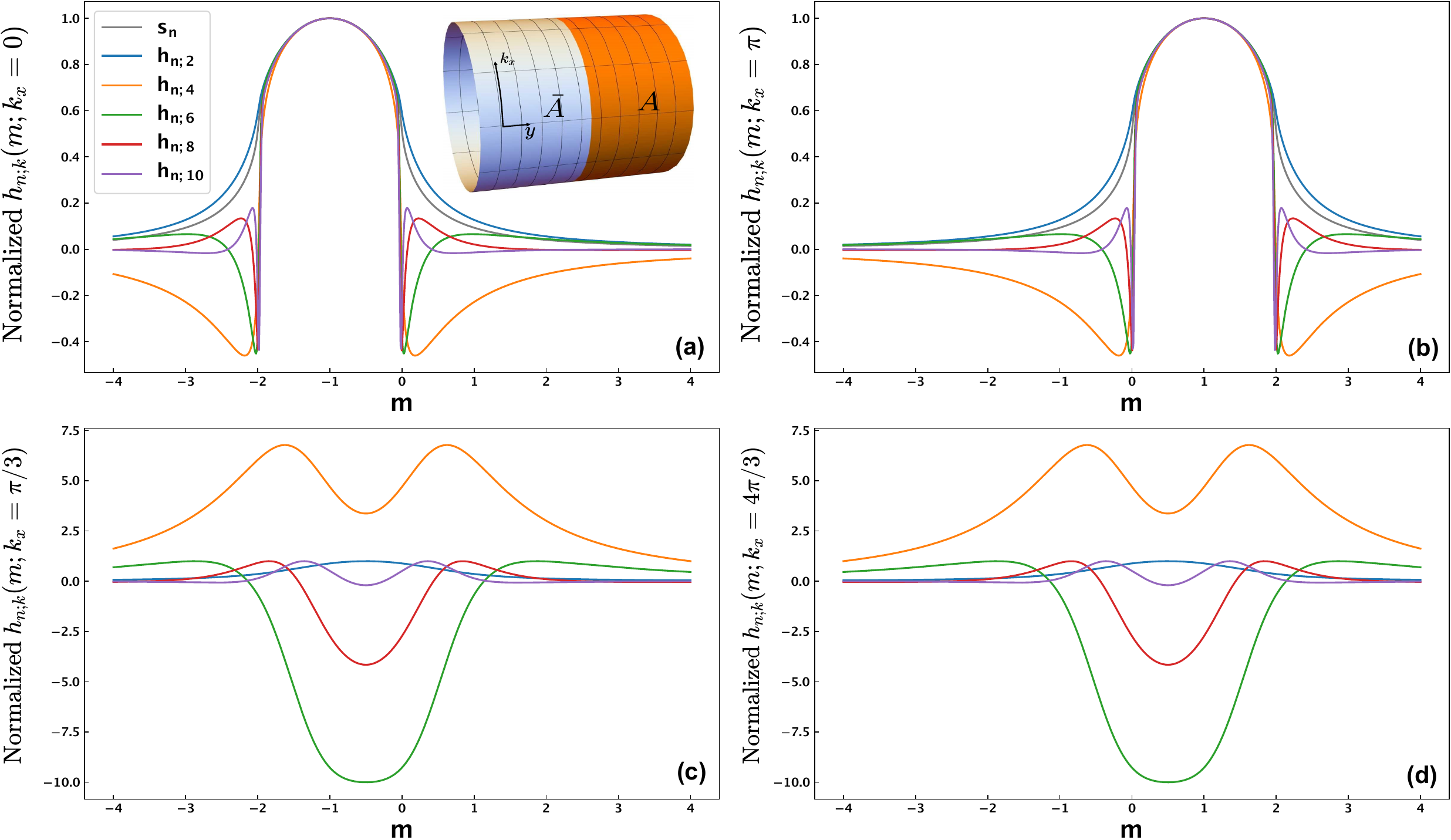}
\caption{Normalized $h_{n;k}(m; k_x)$ as a function of $m$: (a) $k_x=0$, (b) $k_x=\pi$, (c) $k_x=\pi/3$, and (d) $k_x=4\pi/3$. We set $n=2$ and $\lambda=1$ for generality.
}
\label{fig:edgestate}
\end{figure}

From the distribution of hyperfine structure in Fig.~\ref{allrankrenyi}, the behavior of topological phase is similar to the critical point. To further detect the topological edge modes, we will demonstrate the emergent scaling law of normalized $h_{n;k}(m;k_x=0,\pi)$ in the presence of topological edge states. Initially, we establish the link between the bulk Chern number and the zero-energy modes in the Chern insulator model, denoted as  $\hat{\mathcal{H}}=\sum_{\mathbf{k}}\hat c_{ \mathbf{k}}^{\dagger}h(\mathbf{k})\cdot \mathbf{\sigma} \hat c_{  \mathbf{k} }$ with $ h(\mathbf{k})=(\sin  {k}_x,\sin  {k}_y,m+\cos  {k}_x+\cos  {k}_y),$where $\sigma$ is the vector of Pauli matrices. By considering an open boundary along the $y$-axis and a periodic boundary along the $x$-axis, we classify different quantum states using the quantum number $k_x$. We demonstrate that within the topological interval $m\in(-2,0)$, the zero-energy edge mode corresponds to $k_x=0$, whereas for $m\in(0,2)$, it aligns with $k_x=\pi$.
Here we only present a torus argument as shown in Fig.~\ref{Fig:bbc} with the rigorous proof demonstrated in Ref.~\cite{2011PhRvB..83l5109M}. The critical loops of fixed $k_x$ are highlighted  in red. In Fig.~\ref{Fig:bbc}(b) and (d), since the Chern number $Q\neq 0$,  the torus wraps the origin and it is always possible to find two loops that are coplanar with the origin, one of which encloses the origin and the other does not.  Specifically, at $k_x=\pi$ in Fig.~\ref{Fig:bbc}(d), the loop that is coplanar with and encircles the origin signifies the presence of zero-energy edge states at this $k_y$ value.

According to Refs.~\cite{2009arXiv0909.3119T,2010PhRvL.104m0502F}, it has been identified that in specific topological zones, zero-energy edge states at $k_x=0$ for $m \in (-2,0)$ and at $k_x=\pi$ for $m \in (0,2)$ correspond to entanglement modes with a fractional value of $1/2$.

In the topological region where $m \in (-2,0)$, we derive the following expression:
\begin{align}
   h_{n;k}(m;k_x=0) &= \beta_k(n) C_k(m;j \in \partial A, k_x=0) \nonumber \\
   &\approx \beta_k(n) C_k(m; k_x=0) \nonumber \\
   &= \beta_k(n) (-i\partial_\lambda)^k \ln \det \left(1 + M(m; k_x=0) \exp(i\lambda)\right)\bigg|_{\lambda=0} \nonumber \\
   &= \beta_k(n) \operatorname{tr}\left[(-i\partial_\lambda)^k \ln \left(1 + M(m; k_x=0) \exp(i\lambda)\right)\right]\bigg|_{\lambda=0} \nonumber \\
   &\approx \beta_k(n) (-i\partial_\lambda)^k \ln\left(1 + \frac{1}{2} \exp(i\lambda)\right)\bigg|_{\lambda=0}.
\end{align}
This analysis establishes a scaling law applicable across different values of $k$:
\begin{align}
   \frac{ h_{n;k}(m \in (-2,0); k_x=0)}{\max  h_{n;k}(m \in (-2,0); k_x=0)} = \frac{h_{n;k}(m \in (0,2); k_x=\pi) }{\max h_{n;k}(m \in (0,2); k_x=\pi) }\approx 1.
\end{align}
This scaling law is also valid in the topological domain of $m \in (0,2)$, demonstrating a consistent pattern across the specified ranges.

Then we vary the parameter $m$ across different phases using the partition scheme shown in the upper-right of Fig.~\ref{fig:edgestate}(a), where the $y$-axis is an open boundary and the $x$-axis is periodic. We compute $h_{n;k}$ from the edge of region $A$ as a function of $m$, denoted by $ h_{n;k}(m;{k}_x) = \,  h_{n;k}(m;{k}_x,y\in \partial A)$, where ${k}_x$ is the momentum quantum number along the $x$-axis and $\partial A$ are the sites near the boundary cut between $A$ and $\bar A$. To ensure a fair comparison and mitigate variations in magnitude, we normalize boundary $h_{n;k}$ by its maximum as $h_{n;k}(m;{k}_x) / \max h_{n;k}(m;{k}_x)$.   This normalization process is applied uniformly to all ${k}_x$.

We observe that most of the normalized $ h_{n;k}(m;{k}_x)$ are randomly distributed, as shown in Fig.~\ref{fig:edgestate}(c) and (d) for $k_x=\pi/3$ and $k_x=4\pi/3$, except for those with ${k}_x=0$ and ${k}_x=\pi$. For these special cases, as seen in Fig.~\ref{fig:edgestate}(a) and (b), $h_{n;k}(m;{k}_x)$ with different $k$ converge to a single curve and reach their maximum values for $m$ values in the intervals of $(-2,0)$ and $(0,2)$.
The emergence of this scaling law within the topological mass gap regions, namely $m\in(-2,0)\cup (0,2)$, indicates the existence of critical edge states and a fundamental $1/2$ mode in the entanglement spectrum. This $1/2$ mode represents the most entangled and correlated mode for $h_{n;k}$ across all values of $k$ and corresponds to the edge states in lattice fermion models. It is essentially the Einstein-Podolsky-Rosen (EPR) pair that contributes the maximum entanglement.

In summary, these distribution properties highlight the different features of a mass gap, a critical Dirac cone, and a Fermi surface,  and they reveal a universal scaling behavior in the presence of topological edge states. These discussions can be generalized to a wider class of systems, including other Chern insulators and those with symmetry-protected topological order.

%%%%%%%%%%%%%%%%%%%%%%%%%%%%%%%%%%%%%%%%%%%%%%%%%%%%%%%%%%%%%%%%%%%%%%%%
\section{Experiment}
\label{experiment}
%%%%%%%%%%%%%%%%%%%%%%%%%%%%%%%%%%%%%%%%%%%%%%%%%%%%%%%%%%%%%%%%%%%%%%%%
In this section, we will first introduce an alternative experiment protocol based on quantum point contact to measure the hyperfine structure. Then we will introduce the relationship between the entanglement of pre-measurement states and post-measurement states through the hyperfine structure.

%%%%%%%%%%%%%%%%%%%%%%%%%%%%%%%%%%%%%%%%%%%%%%%%%%%%%%%%%%%%%%%%%%%%%%%%
\subsection{Protocol based on quantum point contact}
\label{Ap:QPC}
%%%%%%%%%%%%%%%%%%%%%%%%%%%%%%%%%%%%%%%%%%%%%%%%%%%%%%%%%%%%%%%%%%%%%%%%
In this part, we will provide a promising experimental setup to measure the hyperfine structure of the entanglement. Before the introduction of the experimental proposal, we will firstly demonstrate the Eq.~(\ref{Renyiexpe}), which is useful for the experiment realization.

Using the equation shown in Eq.~(\ref{dms2}), which is $\langle e^{i\lambda(\sum_{i=1}^{\mathcal{N}_A}c_i^{\dagger}c_i)}c_l^{\dagger}c_l\rangle=e^{i\lambda}\langle e^{i\lambda(\sum_{i=1,i\neq l}^{\mathcal{N}_A}c_i^{\dagger}c_i)}c_l^{\dagger}c_l\rangle$, the function $ \langle e^{i\lambda \hat N_A}\hat n_j\rangle \equiv G(\lambda,j)$ can be simplified as
\begin{align*}
    G(\lambda,j) &=\left \langle \left (\prod_{i=1,i\neq j}^{\mathcal{N}_A}\left [1+   (e^{i\lambda}-1)\hat c_i^{\dagger}\hat c_i\right ]\right )e^{i\lambda}\hat c_j^{\dagger}\hat c_j\right \rangle\nonumber\\
        &=\left \langle  \left (\prod_{i=1,i\neq j}^{\mathcal{N}_A}\left [1+   (e^{i\lambda}-1)\hat c_i^{\dagger}\hat c_i\right ]\right )\left [1+(e^{i\lambda}-1)\hat c_j^{\dagger}\hat c_j\right ]\right \rangle-\left \langle  \prod_{i=1,i\neq j}^{\mathcal{N}_A}\left [1+   (e^{i\lambda}-1)\hat c_i^{\dagger}\hat c_i\right ] \right \rangle\nonumber\\
        &\,\, \, \, \, +\left \langle  \left (\prod_{i=1,i\neq j}^{\mathcal{N}_A}\left [1+   (e^{i\lambda}-1)\hat c_i^{\dagger}\hat c_i\right ]\right )\hat c_j^{\dagger}\hat c_j\right \rangle\nonumber\\
         &=\left \langle \prod_{i=1}^{\mathcal{N}_A}\left [1+   (e^{i\lambda}-1)\hat c_i^{\dagger}\hat c_i\right ]\right \rangle-\left \langle \prod_{i=1,i\neq j}^{\mathcal{N}_A}\left [1+(e^{i\lambda}-1)\hat c_i^{\dagger}\hat c_i\right ]\right \rangle\nonumber\\
        &\,\, \, \, \, +\left \langle  \left (\prod_{i=1,i\neq j}^{\mathcal{N}_A}\left [1+   (e^{i\lambda}-1)\hat c_i^{\dagger}\hat c_i\right ]\right )\hat c_j^{\dagger}\hat c_j\right \rangle\nonumber\\
        &= \chi(\lambda,\mathcal{N}_A)-\chi(\lambda,\mathcal{N}_{A}-1;j)+\left \langle  \left (\prod_{i=1,i\neq j}^{\mathcal{N}_A}\left [1+   (e^{i\lambda}-1)\hat c_i^{\dagger}\hat c_i\right ]\right )\hat c_j^{\dagger}\hat c_j\right \rangle,
\end{align*}where $\chi(\lambda,\mathcal{N}_{A}-1;j)$ is the cumulant generating function of region $A'$ with the $\mathcal{N}_A$ sites, which is a region obtained by removing site $j$ from the original region A. Hence,  we can obtain the following relation
\begin{align}
    G(\lambda,j)  &=\left \langle \left (\prod_{i=1,i\neq j}^{\mathcal{N}_A}\left [1+   (e^{i\lambda}-1)\hat c_i^{\dagger}\hat c_i\right ]\right )e^{i\lambda}\hat c_j^{\dagger}\hat c_j\right \rangle\nonumber\\
&=\left [\chi(\lambda, \mathcal{N}_A)-\chi(\lambda,\mathcal{N}_{A'};j)\right ]\frac{e^{i\lambda}}{e^{i\lambda}-1}.
\end{align}
\begin{figure}[t]
\centering
\includegraphics[width=0.7\linewidth]{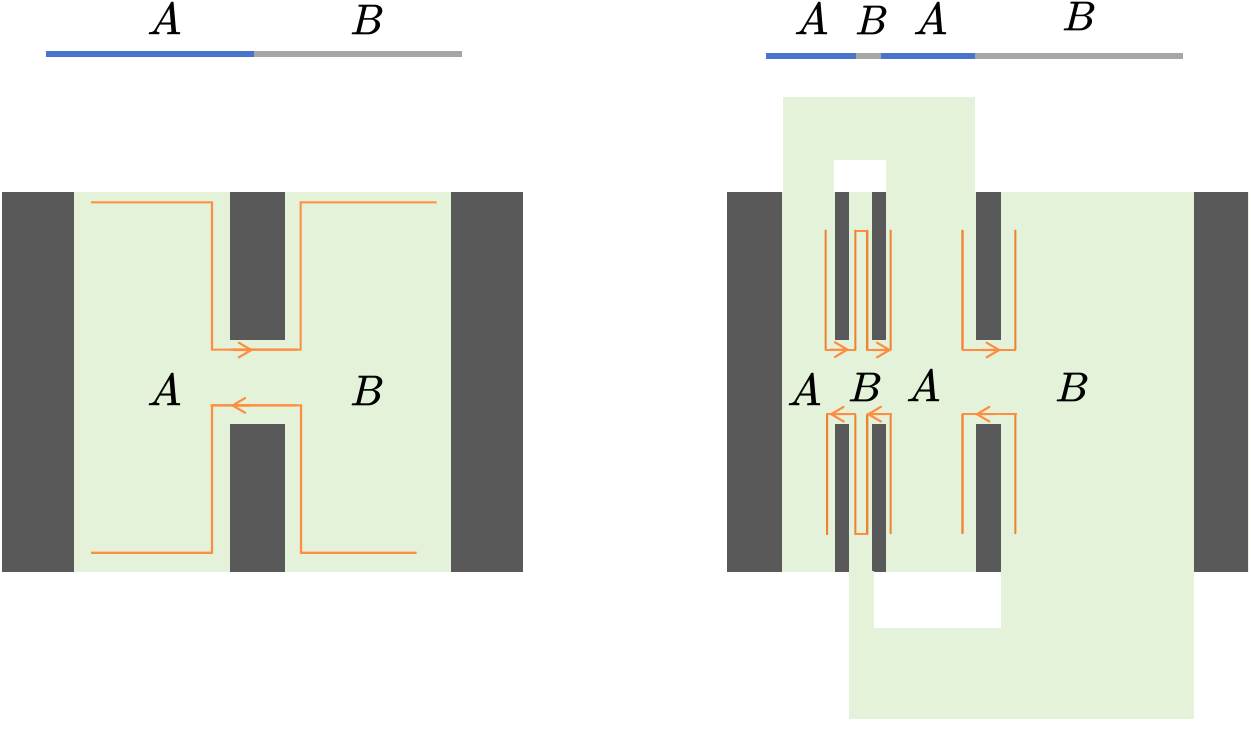}
\caption{Quantum Point Contact settings of two types of partition. The left figure corresponds to the bi-partition of $\chi(\lambda,\mathcal{N}_A)$ and the right one corresponds to the partition way of $\chi(\lambda,\mathcal{N}_{A'};l)$.}
\label{QPC}
\end{figure}

Therefore, the density of particle-number cumulant in Eq.~(\ref{densitycorr})  can be simplified as a function of two different cumulant generating functions with different ways of partition.
\begin{align}
    C_k(l) &=(-i\partial_{\lambda})^{k-1} \bigg \{\frac{[\chi(\lambda,\mathcal{N}_A)-\chi(\lambda,\mathcal{N}_{A}-1;l)]}{\chi(\lambda,\mathcal{N}_A)}\frac{e^{i\lambda}}{e^{i\lambda}-1}\bigg\}\bigg{|}_{\lambda=0}.\label{Renyiexpe}
\end{align}

The key observation is that the cumulant generating function $\chi(\lambda,\mathcal{N}_A)$ and $\chi(\lambda,\mathcal{N}_{A}-1;l)$ can be obtained by measuring the cumulants in specific partition configurations based on the quamtum point contact (QPC) method.  $\chi(\lambda,\mathcal{N}_A)$ corresponds to the bi-partition shown in the left figure of Fig.~\ref{QPC}  and  $\chi(\lambda,\mathcal{N}_{A}-1;l)$ corresponds to the partition arrangement illustrated in the right figure of Fig.~\ref{QPC}, where $A$ and $B$ are not simply connected spaces. The subregion A in the right figure corresponds to the outcome of removing site $j$ from the left subregion $A$, with site j now becoming part of subregion $B$. To mitigate interference within the different parts of $A/B$, we establish additional channels connecting them, positioned above and below the setting. Based on these QPC settings, we can reconstruct the cumulant generating function by measuring the former a few dominant cumulants. To be more precise,  considering that the Taylor expansion$\chi(\lambda,\mathcal{N}_A)=\sum_{k=0}^{\infty} C_k\frac{\lambda^k}{k!}$ centering at $\lambda=0$, we can reasonably approximate $\chi(\lambda,\mathcal{N}_A)$ by determining the first a few terms of the cumulants while the higher-order terms make negligible contributions. These cumulants are experimentally observable in the scheme shown in the left figure of Fig.~\ref{QPC} as demonstrated in previous studies~\cite{2009PhRvL.102j0502K,2009PNAS..10610116F}. These methods can also naturally be applied to the second scheme shown in the right figure of Fig.~\ref{QPC}. Hence,  $\chi(\lambda,\mathcal{N}_A)$ and $\chi(\lambda,\mathcal{N}_{A}-1;l)$ can be approximately determined experimentally, which means that $C_k(l)$ in Eq.~(\ref{Renyiexpe})
 can also be obtained using previous experiment methods.

%%%%%%%%%%%%%%%%%%%%%%%%%%%%%%%%%%%%%%%%%%%%%%%%%%%%%%%%%%%%%%%%%%%%%%%%
\subsection{Local operation and classical communication (LOCC)}
\label{Ap:prepostme}
%%%%%%%%%%%%%%%%%%%%%%%%%%%%%%%%%%%%%%%%%%%%%%%%%%%%%%%%%%%%%%%%%%%%%%%%
In the following, we primarily focus on the behavior of the hyperfine structure under LOCC (Local Operations and Classical Communication). Local operations encompass two types of transformations: unitary transformations and measurements.

First, we consider unitary transformations. As discussed in Appendix~\ref{Ap:property}, the hyperfine structure remains invariant under local unitary transformations.

Second, we demonstrate that the contribution of hyperfine structure on a specific site  does not increase under local measurements. To illustrate this, we consider the single-particle density matrix for a free fermion system and the measurement as a projection operation in real space.

For convenience, we replace the $L$ creation and annihilation operators by $2L$ Majorana operators
\begin{align}
    \psi_{i,1}=\frac{c_i+c_i^{\dagger}}{\sqrt{2}},\psi_{i,2}=\frac{c_i-c_i^{\dagger}}{i\sqrt{2}}.
\end{align}
The single-particle density matrix of free-fermion system is defined as
\begin{align}
\hat \varrho_{ij;\mu,\nu} =\langle \psi_{i,\mu}\psi_{j,\nu}\rangle,
\end{align}
which is exactly the correlation matrix $M$ in the Majorana Basis.

For a subsystem \( A \) comprising less than half of the entire system, the reduced density matrix can be expressed as
\begin{align}
  \hat \varrho_A = \hat P_A \, O \left( \bigoplus_{k=1}^{\mathcal{N}} \begin{bmatrix}
 \eta_k & 0 \\
 0 & 1 - \eta_k
\end{bmatrix} \right) O^{\dagger} \, \hat P_A,
\end{align}
where \(\{\eta_k\} \cup \{1 - \eta_k\}\) represent the eigenvalues of \(\hat \varrho_A\), and \(O\) denotes the diagonalization matrix.

Suppose we perform a projective measurement on the system described by the position projectors $\hat P_i=|i\rangle \langle i|$.  The quantum state after measurement is
\begin{align}
    \tilde \varrho_{\text{post}}=\hat P_i \hat \varrho\hat P_i.
\end{align}
he reduced density matrix of subsystem A after the measurement is obtained by tracing out the degrees of freedom outside A. In terms of the single particle density matrix,
\begin{align}
        \tilde \varrho_{A,\text{post}}=\hat P_i \hat \varrho_A\hat P_i.
\end{align}

It is straightforward to verify that the entanglement entropy of the system post-measurement is directly linked to the entanglement contour of the pre-measurement state \(\hat \varrho\),
\begin{align}
    S_1(\tilde \varrho_{A,\text{post}})
    &= \tr(f(\hat P_i \hat \varrho_A \hat P_i)), \quad f(x) = -x \ln x \nonumber\\
    &= \tr(f(\hat P_i \hat \varrho \hat P_i)) \nonumber\\
    &= S_1(\hat \varrho_ i) \le S_1(\hat \varrho_{A}), \label{Eq:postmeasure}
\end{align}
where site \(i\) is a single site comprised in region \(A\).

The result can be further generalized to the fine structure by utilizing the lower bound established in Ref.~\cite{2014JSMTE..10..011C},
\begin{align}
    s_1(i|\hat \rho_A) \ge S_1(\hat \varrho_i) = s_1(i|\hat \varrho_i),
\end{align}
and
\begin{align}
    \quad s_1(j|\hat \rho_A) = 0, j \neq i,
\end{align}
where \(|...\rangle\) represents the corresponding quantum state. This relation demonstrates that the fine structure is non-increasing under LOCC, implying that under LOCC, the total contribution of hyperfine structure is non-increasing. For site \( j \neq i \), the hyperfine structure vanishes after the measurement, indicating a decrease; for site \( i \), the total contribution of the hyperfine structure is reduced compared to its value before the measurement.

%%%%%%%%%%%%%%%%%%%%%%%%%%%%%%%%%%%%%%%%%%%%%%%%%%%%%%%%%%%%%%%%%%%%%%%%
\section{Conclusions and discussions}
\label{conclusion}
%%%%%%%%%%%%%%%%%%%%%%%%%%%%%%%%%%%%%%%%%%%%%%%%%%%%%%%%%%%%%%%%%%%%%%%%
In this work, we proposed a novel entanglement quantity, termed the hyperfine structure of entanglement $h_{n;k}(j)$, by decomposing the fine structure and R\'{e}nyi contour into contributions from different particle number cumulants. The hyperfine structure exhibits several universal properties, including additivity, normalization, exchange symmetry, and local unitary invariance, making the hyperfine structure physical and meaningful just like other quantum informative measures. We provided some preliminary and prototypical examples of applications. In a Fermi gas, the hyperfine structure can be analytically expressed as a simple formula, revealing its direct relationship with mutual information. Furthermore, in (1+1)D CFT, the hyperfine structure was
shown to be independent of the UV cutoff, demonstrating its robustness as a measurement in field theory. In the context of the AdS/CFT correspondence, we derived the holographic extremal surface for the refined R\'{e}nyi entropy and established its connection to the hyperfine structure by slicing the extremal surface. This bridges the gap between the holographic duality of R\'{e}nyi entropy and the cosmic brane description. In addition, we introduced the notion of R\'{e}nyi entanglement wedge, which can extend beyond the entanglement wedge, and might provide new framework for bulk reconstruction in the AdS/CFT correspondence. For Chern insulators, we demonstrated that the hyperfine structure can distinguish different parameter regions. Specifically, it can differentiate between topological gapped states and trivial gapped states, which are inaccessible using either the R\'{e}nyi entropy or the fine structure alone. This capability highlights the hyperfine structure as a powerful tool for probing topological phases. Experimentally, we proposed an alternative experimental protocol based on quantum point contacts, similar to previous measurements~\cite{2009PhRvL.102j0502K}. The density-density correlator in Eq.~(\ref{RenyiContour}) can also be measured using atomic quantum gases~\cite{2011NJPh...13g5019H} and trapped ions~\cite{2019PhRvA..99e2323E}. These experimental setups also allow us to access the entanglement spectrum and other information about quantum systems.

Our results not only advanced the understanding of entanglement structures in quantum systems but also connected concepts across condensed matter physics, quantum information, and holography. By combining the hyperfine structure of entanglement with other quantum information concepts, such as von Neumann entropy, mutual information, negativity, relative entropy and more, the study of the hyperfine structure will enrich our understanding of quantum entanglement, quantum gravity, and quantum computation. Theoretically, we identify several promising theoretical directions for future study:  (i) discuss more general application of the hyperfine structure in the domain of quantum information; (ii) extend the discussions on the hyperfine structure to correlated models, i.e., emergent fermionic spinons; (iii) explore the connection between the inclined extremal surface and the emergent tension in the holographic dual of $\tilde h_{n;2}(j)$, and study bulk reconstruction and surface growth in terms of the R\'{e}nyi entanglement wedge in the AdS/CFT correspondence; (iv) investigate how the R\'{e}nyi contour changes under measurement-based operations in quantum computations.

%%%%%%%%%%%%%%%%%%%%%%%%%%%%%%%%%%%%%%%%%%%%%%%%%%%%%%%%%%%%%%%%%%%%%%%%
\section{Acknowledgements}
%%%%%%%%%%%%%%%%%%%%%%%%%%%%%%%%%%%%%%%%%%%%%%%%%%%%%%%%%%%%%%%%%%%%%%%%
This work was supported in part by the National Natural Science Foundation of China (No.~12074438), (No.~11675272) and (No.~12475069), the Guangdong Provincial Key Laboratory of Magnetoelectric Physics and Devices under Grant No.~2022B1212010008, and the (national) college students innovation and entrepreneurship training program, Sun Yat-sen University.

%%%%%%%%%%%%%%%%%%%%%%%%%%%%%%%%%%%%%%%%%%%%%%%%%%%%%%%%%%%%%%%%%%%%%%
\begin{appendix}
%%%%%%%%%%%%%%%%%%%%%%%%%%%%%%%%%%%%%%%%%%%%%%%%%%%%%%%%%%%%%%%%%%%%%%

%%%%%%%%%%%%%%%%%%%%%%%%%%%%%%%%%%%%%%%%%%%%%%%%%%%%%%%%%%%%%%%%%%%%%%
\end{appendix}
%%%%%%%%%%%%%%%%%%%%%%%%%%%%%%%%%%%%%%%%%%%%%%%%%%%%%%%%%%%%%%%%%%%%%%

\omits{
\newpage
\thispagestyle{empty}
\begin{center}
    \Large \textbf{Supplementary Materials}
\end{center}

% \tableofcontents

The Supplemental Material (SM) provides further discussions on the technical details and theoretical demonstrations that support the main text. The organization of the SM is as follows: Section~\ref{SM:I} provides the derivation and properties of the hyperfine structure; Section~\ref{SM:II} presents a detailed demonstration and discussion on the hyperfine structure in Fermi gas; Section~\ref{SM:III} provides the derivation of the holographic duality of the hyperfine structure; Section~\ref{SM:IV} discusses the distribution function of the hyperfine structure across different parameter regions of Chern insulator and demonstrates the scaling behavior due to the presence of topological edge states; Finally, Section~\ref{SM:V} proposes a promising experimental setup for measuring the hyperfine structure and discusses the relation between pre-measurement states and post-measurement states through hyperfine structure.
}

\omits{
%%%%%%%%%%%%%%%%%%%%%%%%%%%%%%%%%%%%%%%%%%%%%%%%%%%%%%%%%%%%%%%%%%%%%%
\section{Definition and properties of the hyperfine structure}
\label{SM:I}
%%%%%%%%%%%%%%%%%%%%%%%%%%%%%%%%%%%%%%%%%%%%%%%%%%%%%%%%%%%%%%%%%%%%%%
In this section, we first provide a brief review of the entanglement in free-fermion models, including some entanglement quantities like entanglement entropy, R\'{e}nyi  entropy and the \textit{fine structure} of entanglement . We then introduce the \textit{hyperfine structure} of entanglement to expand the zoo of entanglement quantities within the model.

% In free-fermion models, both Hamiltonian, entanglement Hamiltonian have quadratic forms and its entanglement spectrum and entropy are determined by the single-particle Green’s function.
%%%%%%%%%%%%%%%%%%%%%%%%%%%%%%%%%%%%%%%%%%%%%%%%%%%%%%%%%%%%%%%%%%%%%%
\subsection{Entanglement of free-fermion Models}
%%%%%%%%%%%%%%%%%%%%%%%%%%%%%%%%%%%%%%%%%%%%%%%%%%%%%%%%%%%%%%%%%%%%%%
The Hamiltonian of free-fermion system has a generic quadratic form written as
\begin{align}
    H=\sum_{i,j}\hat c_i^{\dagger}\mathcal{H}_{i,j}\hat c_j,
\end{align}where $\hat c_j^{\dagger}$  and $\hat c_j$ are the creation and annihilation operator, respectively. Denote the vacuum state as $|0\rangle$, the density matrix  of the ground state $|G\rangle$can be written as\begin{align}
    \hat \rho=|G\rangle\langle G|.
\end{align}
By tracing out the remaining region $\bar A$, we can obtain the reduced denstiy matrix
\begin{align}
   \hat  \rho_A=\frac{1}{Z}e^{- \hat K_A}, \hat K_A=\sum_{i,j\in A}\hat c_i^{\dagger}H_{Ai,j} \hat c_j\label{modularhamilt},
\end{align}
where $Z$ is a normalization constant. Based on Ref.~\cite{2003JPhA...36L.205P},  the matrix inside the entanglement Hamiltonian can be written as
\begin{align}
H_A =\ln[M^{-1}-\mathbf{I}],
\end{align}
where $\mathbf{I}$ is the identity matrix and $M$ is the correlation function matrix defined as
\begin{align}
    M_{ij}\equiv\langle G|\hat c_i^{\dagger}\hat c_j|G\rangle.
\end{align}

The entanglement entropy of the subregion $A$ is given by
\begin{align}
    S_1&=-\tr\hat \rho_A\ln\hat  \rho_A=-\sum_k[\xi_k\ln\xi_k+(1-\xi_k)\ln(1-\xi_k)],
\end{align}and R\'{e}nyi  entropy is expressed as
\begin{align}
    S_n&=\frac{1}{1-n}\log \tr\hat \rho_A^n=\frac{1}{1-n}\sum_k\ln[\xi_k^n+(1-\xi_k)^n],
\end{align}
where $\{\xi_k\}$ are the eigenvalues of correlation matrix $M$ and the eigenvectors of $M$ is $\{\psi_k\}$.

Besides entanglement entropy and R\'{e}nyi entropy, Y.~Chen and G.~Vidal have introduced another measure of entanglement known as the \textit{entanglement contour}~\cite{2014JSMTE..10..011C}, the summation of which equates to the entanglement entropy. The entanglement contour represents the \textit{fine structure }of the entanglement ,reflecting its real-space distribution, and can be extended to the R\'{e}nyi contour.

Before discussing the derivation of the contour function, we first review the concept of the projection operator. In the single-particle representation of free-fermion systems, the projection operator is defined as $\hat{P}_l = |l\rangle \langle l|$, satisfying $(\hat P_l)^2=\hat P_l$.  We have
\begin{align}
\hat P_l M   \hat P_l=  \sum_{i,j}\hat P_l\langle \psi| \hat c_i^{\dagger}\hat c_j|\psi\rangle   \hat P_l=\langle \hat\psi| \hat c_l^{\dagger}\hat c_l|\psi\rangle=M_{ll},
\end{align}where $i,j,l$ are real space indices.
Since the correlation matrix $M_{ij}=\langle \hat c_i^{\dagger}\hat c_j\rangle $ can be diagonalized as $ \Lambda $ by a unitary matrix $O$, this above relation can also be written as
\begin{align}
  ( \hat P_l M\hat P_l)_{ip}&=   (\hat P_l O \Lambda O^{\dagger} \hat P_l)_{ip}\nonumber\\
    &=  \sum_{jmn}\delta_{il}\delta_{ij}O_{jm}\Lambda_{mm} (O^{\dagger})_{mn}\delta_{np}\delta_{pl}\nonumber\\
     &=   \sum_{m}\delta_{il}O_{im}\Lambda_{mm} (O^{\dagger})_{mp}\delta_{pl}\nonumber\\
     &= M_{ip}\delta_{il}\delta_{pl}.
\end{align}
Besides, the following relation is also useful:
\begin{align}
  ( \hat P_l M^2\hat P_l)_{ip}&= ( \hat P_l)_{ij} \langle \hat c_j^{\dagger}\hat c_m\rangle  \langle \hat c_m^{\dagger}\hat c_n\rangle (\hat P_l)_{np}\nonumber\\
  &=\sum_{jmn}\delta_{il}\delta_{ij} M_{jm}M_{mn}\delta_{np}\delta_{pl}\nonumber\\
  &=\sum_{m}\delta_{il} M_{im}M_{mp}\delta_{pl}\nonumber\\
  &=(M^2)_{ip}\delta_{il}\delta_{pl}.
\end{align}
We can also insert some arbitrary operators, such as $D$ into the equation:
\begin{align}
  ( \hat P_l MDM\hat P_l)_{ip}&= ( \hat P_l)_{ij} \langle \hat c_j^{\dagger}\hat c_m\rangle D_{mq} \langle \hat c_q^{\dagger}\hat c_n\rangle (\hat P_l)_{np}\nonumber\\
  &=\sum_{jmnq}\delta_{il}\delta_{ij} M_{jm}D_{mq}M_{qn}\delta_{np}\delta_{pl}\nonumber\\
  &=\sum_{mq}\delta_{il} M_{im}D_{mq}M_{qp}\delta_{pl}\nonumber\\
  &=(MDM)_{ip}\delta_{il}\delta_{pl}.
\end{align}
The entanglement entropy can be written as
\begin{align}
   S(A)  &=-\tr[M\ln M+(\mathbf{I}-M)\ln(\mathbf{I}-M)]\label{renenr}\nonumber\\
   &=\sum_{ml}O_{ml}(f(\Lambda))_{ll} (O^{\dagger})_{lm},f(x)=-x\ln x-(1-x)\ln (1-x).
\end{align}
Applying the projection operator onto the above equation, we have
\begin{align}
  s(j) &= \sum_{mli}\delta_{im}O_{ml}(f(\Lambda))_{ll} (O^{\dagger})_{lm}\delta_{mi}\delta_{ij}\nonumber\\
   &=\sum_{l}O_{jl}(f(\Lambda))_{ll} (O^{\dagger})_{lj}\nonumber\\
    &=\sum_{l}(-\xi_l\ln \xi_l-(1-\xi_l)\ln (1-\xi_l))|\psi_l(j)|^2,
\end{align}which is exactly the entanglement contour put forward in Ref.~\cite{2014JSMTE..10..011C}.
Therefore, the entanglement contour $s(j)$ and R\'{e}nyi contour $s_n(j)$ between subregion $A$ and $\bar A$  can be written as
\begin{align}
 s(j)&\equiv-\tr \{\hat P_j[M\ln M+(\mathbf{I}-M)\ln(\mathbf{I}-M)]\} ,
\end{align}
and
\begin{align}
 s_n(j)&\equiv\frac{1}{1-n}\tr \{\hat P_j\ln[M^n+(1-M)^n)]\} \label{one}.
\end{align}
One can easily checked that the summation of the contour function turns out to be the entropy
\begin{align}
    S_n(A)=\sum_{j\in A}s_n(j).
\end{align}

}

\omits{
%%%%%%%%%%%%%%%%%%%%%%%%%%%%%%%%%%%%%%%%%%%%%%%%%%%%%%%%%%%%%%%%%%%%%%
\subsection{Derivation of the hyperfine structure (Eq.~(1)-(2)of the main text) }
\label{Ap:EQ1}
%%%%%%%%%%%%%%%%%%%%%%%%%%%%%%%%%%%%%%%%%%%%%%%%%%%%%%%%%%%%%%%%%%%%%%
In the following, we will introduce the hyperfine structure of entanglement by decomposing the fine structure $s_n(j)$.
Denoted the particle-number operator in  $A$ as $\hat N_A\equiv\sum_{i\in A}\hat c_i^{\dagger}\hat c_i$ and its maximal expectation value, or equivalently, the sites of subregion $A$ as $\mathcal{N}_A$.
The R\'{e}nyi contour $s_n(j)$ can be decomposed into the contribution from different particle-number cumulants as presented in Eq.~(1) of the main text:
\begin{align}
   s_n(j)&\equiv\sum_{k=1}^\infty h_{n;k}(j)=\sum_{k=1}^\infty\bigg(\beta_k(n)C_k(j)\bigg)
\end{align}where
\begin{align}
 C_k(j)\equiv(-i\partial_{\lambda})^{k-1} \frac{G(\lambda,j)}{\chi(\lambda,\mathcal{N}_A)}\bigg{|}_{\lambda=0}.\label{ckj}
\end{align}
The denominator in Eq.~(\ref{ckj}) is the cumulant generating function $\chi(\lambda,\mathcal{N} _A)\equiv\langle \exp(i\lambda  \hat N_A)\rangle$ and we also introduce a function defined as follows:
\begin{align}
    G(\lambda,j)\equiv \langle  \exp(i\lambda \hat  N_A)\hat n_j\rangle.\label{G}
\end{align}

Therefore, we demonstrated that fine structure in Eq.~(\ref{one} ) can be decomposed into the hyperfine structure.

\subsection{Information reconstruction from the fine structure $s_{n}(j)$ }\label{Ap:complete}

From the above sections, we know that the complete set of $h_{n;k}(j)$ with even $k$ can be utilized to obtain the complete set of $s_n(j)$. In the following, we will demonstrate that the fine structure $s_n(j)$ can be used to reconstruct the sets of $C_k(j)$ when a cutoff is applied. Furthermore, based on the fine structure, we can derive the R\'{e}nyi entropy, which allows us to reconstruct the entanglement spectra.

\subsubsection{Particle-number correlation}

Using R\'{e}nyi Contour as the building block, we can reconstruct the information of the particle-number correlation. To illustrate this, consider a free-fermion system with $\hat N_A$ fermions in subsystem A. Now, let's introduce  a \textit{cumulant coefficient matrix} denoted as $B$
\begin{align}
    B=\begin{pmatrix}
 \beta_2(1) &   \beta_4(1) &...  & \beta_{2\mathcal{N}_A}(1)  \nonumber\\
 \beta_2(2) &   \beta_4(2) &...  & \beta_{2\mathcal{N}_A}(2) \nonumber\\
 \vdots  &  & \ddots  & \nonumber\\
\beta_2(\mathcal{N}_A) &   \beta_4(\mathcal{N}_A) &...  & \beta_{\mathcal{N}_A}(\mathcal{N}_A)
\end{pmatrix},\label{pmatrix}
\end{align}
where $\beta_k(n)$ is non-zero only when $k$ is even and $\beta_k(n)=\frac{2}{n-1}\frac{1}{k!}\left(\frac{2\pi i}{n}\right)^k\zeta\left(-k,\frac{n+1}{2}\right)$, then Eq.~(1) of the main text can be written as
\begin{align}
    [s_1(j),
 s_2(j) ,
... ,
s_{\mathcal{N}_A}(j)]^T=B[C_2(j),
C_4(j) ,
... ,
C_{2\mathcal{N}_A}(j)]^T.
\end{align} By solving this linear algebra Eq.~(\ref{pmatrix})  inversely, we can reconstruct the even-rank particle-number fluctuation cumulants $C_k$ with the cut off $k_c=2\mathcal{N}_A.$ Since $C_k$ are the coefficient of the  Taylor expansion  $\chi(\lambda,\mathcal{N}_A)=\sum_{k=0}^{\infty} C_k\frac{\lambda^k}{k!}$,  centered around $\lambda=0$, the dominant $C_k$ terms correspond to smaller values of $k$, such as the former $k_c$ terms.

\subsubsection{Entanglement spectrum}
Besides the set of particle-number cumulant, R\'{e}nyi contour can also be utilized to reconstruct the R\'{e}nyi entropy and entanglement spectrum.  The summation of R\'{e}nyi contour over all sites yields the R\'{e}nyi entropy, from which the entanglement spectrum can be derived.

Let $T_n$ be the trace of the $n$-th power of the reduced density matrix~\cite{2012PhRvB..85c5409S}:
\begin{align}
T_n = \tr(\hat{\rho}_A^n) = e^{(1-n)S_n}.
\end{align}
Let us define \(T_1 = 1\) for a normalized density matrix and introduce a \(D \times D\) matrix as follows:
\begin{align}
U = \begin{pmatrix}
  1 & 1 & 0 & \dots  & \\
  T_2 & 1 & 2 & 0 & \dots  \\
  \vdots & \ddots & \ddots & \ddots & 0 \\
  T_{D-1} & T_{D-2} & \dots & T_2 & D-1 \\
  T_D & T_{D-1} & \dots & T_2 & 1
\end{pmatrix}.
\end{align}

Let's further denote the matrix obtained by taking the first \(n \times n\) submatrix of \(U\) as \(U_n\). Then, we can construct the following polynomial:
\begin{align}
P(x) = \sum_{n=0}^D \frac{(-1)^n}{n!} (\det U_n) x^{D-n},
\end{align}
where $\det U_0 = 1$. This polynomial is actually the characteristic polynomial of the density matrix $\hat{\rho}_A$: $P(x) = \det(xI - \hat{\rho}_A)$. Therefore, the roots of $P(x)$ are the entanglement spectrum.
}

%%%%%%%%%%%%%%%%%%%%%%%%%%%%%%%%%%%%%%%%%%%%%%%%%%%%%%%%%%%%%%%%%%%%%%
\section{Proof of the properties for the hyperfine structure}\label{Ap:property}
%%%%%%%%%%%%%%%%%%%%%%%%%%%%%%%%%%%%%%%%%%%%%%%%%%%%%%%%%%%%%%%%%%%%%%
In the following, we will demonstrate some universal properties of the hyperfine structure $h_{n;k}$ as a quantum information quantity, including additivity, normalization, exchange symmetry, and local unitary invariance.

1. Additivity. For $i,j\in A$ , $h_{n;k}(i) + h_{n;k}(j) = h_{n;k}(i \cup j)$.
\begin{proof}
\begin{align}
    C_k(i)+C_k(j)&=\sum_{l\in \{i,j\}}(-i\partial_{\lambda})^{k-1}\frac{\langle  \exp(i\lambda \hat N_A)\hat n_l\rangle}{\langle \exp(i\lambda \hat N_A)\rangle}\bigg{|}_{\lambda=0}\nonumber\\
    &=C_k(i\cup j)
\end{align}

Correspondingly,  $h_{n;k}(i)+h_{n;k}(j)=h_{n;k}(i\cup j)$.
\end{proof}
2. Normalization. The sum of $h_{n;k}(j)$ over all sites in $A$ equals $C_k$, specifically,  $C_k= \sum_j h_{n;k}(j)/\beta_k(n)$.
\begin{proof}
\begin{align}
    \sum_j C_k(j)&=\sum_j(-i\partial_{\lambda})^{k-1}\frac{\langle  \exp(i\lambda \hat N_A)\hat n_j\rangle}{\langle \exp(i\lambda \hat N_A)\rangle}\bigg{|}_{\lambda=0}\nonumber\\
    &=(-i\partial_{\lambda})^{k-1}\frac{\langle  \exp(i\lambda \hat N_A)\hat N_A\rangle}{\langle \exp(i\lambda \hat N_A)\rangle}\bigg{|}_{\lambda=0}\nonumber\\
    &=(-i\partial_{\lambda})^{k}\ln \langle  \exp(i\lambda \hat N_A)\rangle\bigg{|}_{\lambda=0}\nonumber\\
    &=C_k.
\end{align}
Correspondingly,
\begin{align}
    \sum_j h_{n;k}(j)&=\sum_j \frac{1}{1-n}\sum_{k=1}^{\infty}\beta_k(n)C_k(j)\nonumber\\
    &=\frac{1}{1-n}\beta_k(n)\sum_{k=1}^{\infty}C_k\nonumber\\
    &=   h_{n;k}.
\end{align}
\end{proof}
3.Exchange symmetry. If  $\hat T\hat \rho_A \hat T^{\dagger}=\hat \rho_A $ and $\hat T$ exchanges the site $i$ with $j$, then $h_{n;k}(i)=h_{n;k}(j)$.
\begin{proof}
Notice that
\begin{align}
    [\hat T,\hat N_A]&=[\hat T,\sum_l \hat n_l]\nonumber\\
    &=[\hat T,\sum_{l\neq i,j} \hat n_l]+[\hat T, \hat n_i+\hat n_j]\nonumber\\
    &=\hat T(\hat n_i+\hat n_j)-(\hat n_i+\hat n_j)\hat T\\
    &=\hat T(\hat n_i+\hat n_j)\hat T^{\dagger}\hat T-(\hat n_i+\hat n_j)\hat T\\
    &=(\hat n_j+\hat n_i)\hat T-(\hat n_i+\hat n_j)\hat T\\
    &=0.
\end{align}
Therefore, it is easy to demonstrate that
\begin{align}
    \langle  \exp(i\lambda \hat N_A)\hat n_j\rangle'
    &=\tr(\hat T\hat \rho_A \hat T^{\dagger} \hat T\exp(i\lambda \hat N_A)\hat T^{\dagger} \hat T \hat n_j\hat T^{\dagger} )\nonumber\\
    &=\tr(\hat \rho_A \exp(i\lambda \hat N_A)\hat n_i)\nonumber\\
    &=\langle  \exp(i\lambda \hat N_A)\hat n_i\rangle.
\end{align}
Hence, under the symmetry $\hat T$,  we have
\begin{align}
    C_k(j)=C_k(i),
\end{align}
which implies that
\begin{align*}
   h_{n;k}(j)
   &=h_{n;k}(i).
\end{align*}
\end{proof}

4. Local unitary Invariance. If the quantum state transforms from $|\psi\rangle$ to $|\psi'\rangle$ using a unitary transformation $\hat U^X$ acting on a subset $X \subset A$, then $h_{n;k}(j\in X)$ remains the same for both states $|\psi\rangle$ and $|\psi'\rangle$.

\begin{proof}
  According to the definition in Eq.~(2) of the main text,  the hyperfine structure for state $|\psi\rangle$ and $|\psi\rangle'$ are given by
\begin{align}
    h_{n;k}(j\in X)&=\beta_k(n)(-i\partial_{\lambda})^{k-1} \frac{\langle \exp(i\lambda \hat N_A)\hat n_{j\in X}\rangle}{\langle \exp(i\lambda \hat N_A)\rangle}\bigg{|}_{\lambda=0},\\
    h'_{n;k}(j\in X)&=\beta_k(n)(-i\partial_{\lambda})^{k-1} \frac{\langle \exp(i\lambda \hat N_A)\hat n_{j\in X}\rangle'}{\langle \exp(i\lambda \hat N_A)\rangle'}\bigg{|}_{\lambda=0},\\
\end{align}
respectively.

The correlation functions in the numerator of the above equations are equivalent by noting that
\begin{align}
         \langle  \exp(i\lambda \hat N_A)\hat n_{j\in X}\rangle'
    & =\tr(\hat \rho_A' \exp(i\lambda \hat N_A)\hat n_{j\in X})\nonumber\\
    &=\tr(\hat U^X\hat \rho_A( \hat U^X )^{\dagger}\exp(i\lambda \hat N_A)\hat n_{j\in X})\nonumber\\
     &=\tr(\hat \rho_A( \hat U^X )^{\dagger}\exp(i\lambda \hat N_A)\hat U^X( \hat U^X )^{\dagger}\hat n_{j\in X}\hat U^X)\nonumber\\
      &=\tr(\hat \rho_A( \hat U^X )^{\dagger}\exp(i\lambda \hat n_{j\in X})\hat U^X I^{\bar X}\exp(i\lambda \hat N_{\bar X})\hat I^{\bar X}\hat n_{j\in X})\nonumber\\
      & =\tr(\hat \rho_A \exp(i\lambda \hat N_A)\hat n_{j\in X})\nonumber\\
      &=\langle  \exp(i\lambda \hat N_A)\hat n_{j\in X}\rangle.
   \end{align}  The correlation functions in the denominator are also the same
\begin{align}
         \langle  \exp(i\lambda \hat N_A)\rangle'&=\tr(\hat \rho_A' \exp(i\lambda \hat N_A))\nonumber\\
         &=\tr( \hat U^X \hat \rho_A( \hat U^X )^{\dagger} \exp(i\lambda \hat N_A))\nonumber\\
          &=\tr(\hat \rho_A( \hat U^X )^{\dagger}\exp(i\lambda \hat N_X)\hat U^X I^{\bar X}\exp(i\lambda \hat N_{\bar X})\hat I^{\bar X})\nonumber\\
        & =\tr(\hat \rho_A \exp(i\lambda \hat N_A))\nonumber\\
        &= \langle  \exp(i\lambda \hat N_A)\rangle.
    \end{align}
where $\bar X=A\backslash X$ and $I^{\bar X}$ is the identity operator that act on $\bar X$.
Therefore,
 \begin{align}
         h'_{n;k}(j\in X)= h_{n;k}(j\in X).
    \end{align}
\end{proof}

\omits{
\section{Hyperfine structure in Fermi gas}~\label{SM:II}

In this section, we will first introduce a simplified decomposition  of R\'{e}nyi  contour in Fermi gas. Then we will discuss the hyperfine structure in (1+1)D CFT as an example, which will be useful in the next section.

\subsection{Hyperfine structure in Fermi gas~(Eq.~(3)-(5) of the main text)}\label{Ap:EQ5}

In the $(d+1)$D Fermi gas, we have the following asymptotic relation
\begin{align}
    \frac{s_n(x)}{h_{n;2}(x)}= 1+o(1),
\end{align}
which shows that the hyperfine structure $h_{n;k>2}$ are suppressed in this case.
\begin{proof}

According to the following exact relation between particle fluctuation and R\'{e}nyi entropy~in Fermi gas\cite{2012EL.....9820003C}
\begin{align}
  \frac{S_n}{C_2}&= \frac{(1+n^{-1})\pi^2}{6} +o(1),
 \end{align}
we notice that the contribution from higher cumulant ($C_{k>2}$)  become negligible when compared to $C_2$. Since the elements of projection operator is finite, by applying the projection operator,
\begin{align}
\hat P_j  S_n  \hat P_j=\frac{(1+n^{-1})\pi^2}{6}\hat P_j  C_2  \hat P_j+o(1)(\frac{(1+n^{-1})\pi^2}{6}\hat P_j  C_2  \hat P_j),
\end{align}
the second part($\sum_k h_{n;k>2}$) is also of a magnitude less than $h_{n;2}$.  Hence,
\begin{align}
    \frac{s_n(x)}{h_{n;2}(x)}= 1+o(1).
\end{align}

\end{proof}
Based on the relationship between $h_{n;2}$ and the particle-number cumulant density, we can further derive an important property that relates $h_{n;2}(j)$ or$s_n(j)$ to the mutual information $I_2(A,B)$ between regions $A$ and $B$. This relationship is mathematically expressed as:
\begin{align}
   I_2(A_1,A_2) = S(A_1) + S(A_2) - S(A_1\cup A_2=A) = -2 \sum_{i \in A_1, j \in A_2} [\langle \hat n_i \hat n_j \rangle - \langle \hat n_i \rangle \langle \hat n_j \rangle].
\end{align}
We derive that:
\begin{align}
    S_A - h_{1;2}(i\in A_1) &= \sum_{i \in A_1, j \in A_1} [\langle n_i n_j \rangle - \langle n_i \rangle \langle n_j \rangle] - \sum_{i \in A_1, j \in A} [\langle n_i n_j \rangle - \langle n_i \rangle \langle n_j \rangle] \\
    &= -\sum_{i \in A_1, j \in A_2} [\langle n_i n_j \rangle - \langle n_i \rangle \langle n_j \rangle].
\end{align}
Therefore, the mutual information between regions $A_1$ and $A_2$ inside $A$ can be reformulated as:
\begin{align}
I_2(A_1, A_2)&= 2(S(A_1) - h_{1;2}(i\in A_1)),
\end{align}
which can also be written in a continuous form as
\begin{align}
I_2(A_1, A_2)&= 2(S(A_1) - \int_{x\in {A_1}}h_{1;2}(x)dx),
\end{align}
where  $h_{1;2}(x)$ is the hyperfine structure of region $A_1\cup A_2=A$.

\subsection{Hyperfine structure in (1+1)D CFT~(Eq.~(6) of the main text)}\label{Ap:renyi}
Here we present an example to calculate the hyperfine structure in (1+1)D CFT, which will be useful in the section of holographic duality. In (1+1)D CFT, the leading term of the hyperfine structure $h_{n;2}$ in region A is given by
\begin{align}
    h_{n;2}(x)
    &=\frac{(1+n^{-1})\pi^2}{6}\int dy[\langle \hat n(x) \hat n(y)\rangle-\langle \hat n(x)\rangle\langle \hat n(y)\rangle]\label{eq:renyi222}.
\end{align}
Specifically, we consider the Euclidean action of the Dirac fermion as follows:
\begin{align}
    S=\frac{1}{2} \int d^2x \bar \psi\gamma_{\mu}\partial^{\mu}\psi,
\end{align}where $\psi=(\psi_R,\psi_L)^T.$ The Dirac fermions  $\psi_R$ and $\psi_L$ have conformal weights $(1/2,0)$ and $(0,1/2)$, respectively.  Their corresponding correlation functions are
\begin{align}
    \langle \psi_R^{\dagger}(z)\psi_R(\omega)\rangle=-\frac{1}{2\pi }\frac{1}{z-\omega},\quad \langle \psi_L^{\dagger}(\bar z)\psi_L(\bar \omega)\rangle=-\frac{1}{2\pi }\frac{1}{\bar z-\bar \omega}.
\end{align}

The particle-number operator in this field is
\begin{align}
    \hat n(x)=\psi_R^{\dagger}(x)\psi_R(x)+\psi_L^{\dagger}(x)\psi_L(x).
\end{align}
Inserting the above expression inside Eq.~(\ref{eq:renyi222}), we have
\begin{align}
    \langle \hat n(x)\hat n(y)\rangle=&\langle \psi_R^{\dagger}(x)\psi_R(x)\psi_R^{\dagger}(y)\psi_R(y)\rangle+\langle \psi_L^{\dagger}(x)\psi_L(x)\psi_L^{\dagger}(y)\psi_L(y)\rangle\nonumber\\
    &+\langle \psi_R^{\dagger}(x)\psi_R(x)\psi_L^{\dagger}(y)\psi_L^{\dagger}(y)\rangle+\langle \psi_L^{\dagger}(x)\psi_L(x)\psi_R^{\dagger}(y)\psi_R^{\dagger}(y)\rangle,
\end{align}
and
\begin{align}
    \langle \hat n(x)\rangle\langle \hat n(y)\rangle=&\langle \psi_R^{\dagger}(x)\psi_R(x)\rangle\langle \psi_R^{\dagger}(y)\psi_R(y)\rangle+\langle \psi_R^{\dagger}(x)\psi_R(x)\rangle\langle \psi_L^{\dagger}(y)\psi_L(y)\rangle\nonumber\\
&  + \langle \psi_L^{\dagger}(x)\psi_L(x)\rangle\langle \psi_R^{\dagger}(y)\psi_R(y)\rangle+\langle \psi_L^{\dagger}(x)\psi_L(x)\rangle\langle \psi_L^{\dagger}(y)\psi_L(y)\rangle.
\end{align}
Based on Wick's theorem and the chiral symmetry condition \(\langle \psi_R^{\dagger}\psi_L \rangle = \langle \psi_L^{\dagger}\psi_R \rangle = 0\), we have
\begin{align}
    \langle \hat n(x)\hat n(y)\rangle-\langle \hat n(x)\rangle\langle \hat n(y)\rangle=&-\langle \psi_R^{\dagger}(x)\psi_R(y)\rangle\langle \psi_R^{\dagger}(y)\psi_R(x)\rangle-\langle \psi_L^{\dagger}(x)\psi_L(y)\rangle\langle \psi_L^{\dagger}(y)\psi_L(y)\rangle\nonumber\\
    =&\frac{1}{2\pi^2 }\frac{1}{(x-y)^2}.
\end{align}
Therefore, the leading hyperfine structure $h_{n;2}$ can be written as
\begin{align}
     h_{n;2}(x)&=\frac{(1+n^{-1})\pi^2}{6}\int_{-R+\epsilon}^{R-\epsilon} \frac{1}{2\pi^2 }\frac{1}{(x-y)^2}dy\nonumber\\
     &=\frac{(1+n^{-1})}{12}(\frac{1}{R-x}+\frac{1}{R+x}).
\end{align}

By integrating $s_{n}(x)$ over $A$, we obtain the R\'{e}nyi entropy as
\begin{align}
    S_n=\int_{-R+\epsilon}^{R-\epsilon} dx s_n(x)
\approx[\left(1+n^{-1}\right)/6]\ln\frac{2R}{\epsilon}.
\end{align}
Comparing the known equality $S_n=[\left(1+n^{-1}\right)c/6]\ln\frac{2R}{\epsilon}$,  we end up with the value of the central charge $c=1$, which agrees with the known fact of Dirac fermions models.

Here, we observe that the contour function in a Fermi gas, given by $s_{n} \approx h_{n;2} = \frac{c}{12} (1 + n^{-1}) \left(\frac{1}{R-x} + \frac{1}{R+x}\right)$, aligns with the contour function with conformal symmetry as defined in Ref.~\cite{2016JSMTE..12.3103C}. Remarkably, this expression holds true regardless of the presence of interaction or the specifics of the Hamiltonian.

Consequently, it is natural to ask whether the relationship between the contour function and the density of the particle-number cumulant still persists in general interacting (1+1)D CFT. In the following, we will demonstrate that this relationship does indeed hold, up to a constant. Specifically, we find that
\begin{align}
    \frac{s_{n}(x)}{c_2(x)}=\text{constant}+o(1).
\end{align}
\begin{proof}
According to the definition of the density of particle-number cumulant , we have the following expression
\begin{align}
    c_2(i)&=\sum_{j\in A}[\langle n_i n_j\rangle-\langle n_i\rangle\langle n_j\rangle].
\end{align}
By dividing the subsystem $A$ into $A_1, A_2, A_3$ .It is easy to check the linear combination relation between the density of particle-number cumulant contributed from region $A_2$ and the particle-number cumulant
\begin{align}
    c_2(A_2)=\frac{1}{2}(C_2(A_1\cup A_2)+C_2(A_2\cup A_3)-C_2(A_1)-C_2(A_3))\label{eq:ALC}.
\end{align}
The left side of the equation is
\begin{align}
      c_2(A_2)=\sum_{i\in A_2}\sum_{j\in A}[\langle n_i n_j\rangle-\langle n_i\rangle \langle n_j\rangle],
\end{align}
which is the same as the right side of the equation.

The continuous form of Eq.~(\ref{eq:ALC})  in 1+1d CFT is
   \begin{align}
   c_2(x)=\frac{1}{2}(\frac{\partial   C_2(x_1,x)}{\partial x}-\frac{\partial   C_2(x,x_2)}{\partial x})\label{continuous},
    \end{align}
where $x_1,x_2$ are the are the coordinates of the boundary of region A, and $C(x_1,x)$ is a function of coordinates of the boundary $x_1, x$.

In (1+1)D CFT, the particle-number fluctuation is ~\cite{2010PhRvB..82a2405S}
\begin{align}
  C_2(A)=\frac{g}{\pi^2}\ln(\frac{2R}{\varepsilon})\label{eq:def},
\end{align}
where the prefactor $g$ can be fixed by considering the conserved charge and $\varepsilon$ is the cutoff.

Based on Eq.~(\ref{continuous}) and Eq.~(\ref{eq:def}), we have
\begin{align}
    c_2(x)=\frac{g}{2\pi^2}(\frac{1}{R-x}+\frac{1}{R+x}).
\end{align}

Therefore, the ratio between $s_2(x)$ and $c_2(x)$ is a constant
\begin{align}
    \frac{s_n(x)}{c_2(x)}=\frac{\frac{c}{12}(1+n^{-1})(\frac{1}{R-x}+\frac{1}{R+x})}{\frac{g}{2\pi^2}(\frac{1}{R-x}+\frac{1}{R+x})}=\frac{\pi^2 c(1+n^{-1})}{6g}.
\end{align}

This relation can be generalized to $h_{h;2}$ and $\tilde h_{n;2}$ in CFT directly.
\end{proof}
}

\omits{
\section{Hyperfine structure in AdS/CFT correspondence} \label{SM:III}
In this section, we will introduce the holographic duality of the hyperfine structure in the context of AdS$_3$/CFT$_2$.

\subsection{Refined R\'{e}nyi entropy and its hyperfine structure~(Eq.~(7) of the main text)}\label{Ap:refined}
To explore the holographic duality of the R\'{e}nyi contour $s_n$, it is advantageous to utilize the concept of refined R\'{e}nyi entropy, defined as $\tilde{S}_n \equiv n^2 \partial_n \left(\frac{n-1}{n} S_n\right)$. The contour function and the hyperfine structure can be extended to the refined R\'{e}nyi entropy as indicated in Eq.~(7) of the main text.

In the following, we will first establish that the refined R\'{e}nyi entropy represents the entanglement entropy of a new density matrix, $\tilde{\rho}_A = \frac{(\hat{\rho}_A)^n}{\tr(\hat{\rho}_A)^n}$. Subsequently, we will derive the relationship described in Eq.~(7) of the main text.

\begin{proof}

Considering a pure state $|\psi\rangle$ in Hilbert space $\mathcal{H}=\mathcal{H}_A\otimes \mathcal{H}_B$,  the Schmidt decomposition is as follows,

\begin{align}
|\psi\rangle=\sum_{q}a_q|\psi_A^q\rangle\otimes |\psi_B^q\rangle,
\end{align}
where $\{|\psi_A^q\rangle\}$ is an orthogonal basis for $\mathcal{H}_A$ and $\{|\psi_B^q\rangle\}$ is  an orthogonal basis for $\mathcal{H}_B$ and the Schmidt numbers sum up to one,  $\sum_q a_q^2=1$.
The reduced density matrix of region A is
\begin{align}
\hat \rho_A
&=\sum_q a_q^2(|\psi_A^q\rangle\langle \psi_A^q|).
\end{align}
The refined R\'{e}nyi contour $\tilde S_n(\hat \rho_A)$ can be written as the von Neumann entropy of a new density matrix $\tilde{\rho}_A=\frac{\hat \rho_A^n}{\tr \hat \rho_A^n}$,

\begin{align}
\tilde S_n(\hat \rho_A) &= -n^2\partial_n(\frac{1}{n}\ln \tr \hat \rho_A^n) \nonumber\\
&= \ln\tr\hat \rho_A^n - n\partial_n\ln\tr\hat \rho_A^n \nonumber\\
&= \ln\tr\hat \rho_A^n - \frac{\tr (\hat \rho_A^n\ln\hat \rho_A^n)}{\tr(\hat \rho^n_A)} \nonumber\\
&= -\tr\left(\frac{\hat\rho_A^n}{\tr \hat\rho_A^n}\right)\ln \left(\frac{\hat\rho_A^n}{\tr \hat\rho_A^n}\right) \nonumber\\
&= S\left(\frac{\hat \rho_A^n}{\alpha_n}\right)\label{newrho}
\end{align}

where we denote $\alpha_n=\tr \hat \rho_A^n$.

\end{proof}

Here we further discuss the entanglement Hamiltonian of the $n$-replica reduced density matrix $\tilde{\rho}_A=\frac{(\hat \rho_A)^n}{\tr (\hat \rho_A)^n}$. The original density density matrix is
\begin{align}
    \hat \rho_A&=\sum_p a_p^2 |\psi_A^p\rangle\langle\psi_A^p|\nonumber\\
    &=\frac{e^{- \hat K_A}}{Z},
\end{align}where  the corresponding  entanglement Hamiltonian is
\begin{align}
\hat K_A=\sum_p- \ln a_p^2 |\psi_A^p\rangle\langle\psi_A^p|,
\end{align}and $Z$ is the normalization constant. The density matrix is a Gibbs state with temperature $T=1$. The $n$-replica density matrix is defined as
\begin{align}
    \tilde\rho_A^{(n)}&\equiv\sum_p \frac{a_p^{2n} }{\alpha_n}|\psi_A^p\rangle\langle\psi_A^p|\nonumber\\
    &=\frac{e^{- \hat K_A^{(n)}}}{Z^{(n)}},
\end{align}
where $\alpha_n=\sum_q a_q^{2n}$  and
\begin{align}
\hat K_A^{(n)}
&=n\hat K_A\label{KAn}.
\end{align}Here, $Z^{(n)}$ represents a new normalization constant, and the temperature of the Gibbs state is now set to $T = \frac{1}{n}$. It is important to emphasize that the two entanglement Hamiltonians commute, $[\hat{K}_A^{(n)}, \hat{K}_A] = 0$, yet their spectra differ.

In the following, we will present two different ways to derive the Eq.~(7) of the main text.

Firstly,  Eq.~(7) of the main text can be easily determined from the definition of the refined R\'{e}nyi contour in Eq.~(6) of the main text in CFT.

Secondly, we can use the analytical form of entanglement Hamiltonian $\hat K_A$ and $\hat K_A^{(n)}$ to verify the Eq.~(7) of the main text. In the Fermi gas, the entanglement Hamiltonian takes an analytical form resembling the following tridiagonal matrix, as described in Ref.~\cite{2017JPhA...50B4003E}:
\begin{align}
  K = \begin{pmatrix}
  d_1 & t_1 &  &  & \nonumber\\
  t_1 & d_2 & t_2 &  & \nonumber\\
  & t_2 & d_3 &  & \nonumber\\
  & & & \ddots & t_{\mathcal{N}_A-1} \nonumber\\
  & & & t_{\mathcal{N}_A-1} & d_{\mathcal{N}_A}
\end{pmatrix}, \label{modularham}
\end{align}
with the matrix elements defined as:
\begin{align}
    t_i = \frac{i}{\mathcal{N}_A} \left(1 - \frac{i}{\mathcal{N}_A}\right), \quad d_i = -2 \cos q_F \frac{2i-1}{2\mathcal{N}_A} \left(1 - \frac{2i-1}{2\mathcal{N}_A}\right),
\end{align}
where $q_F = \frac{\pi}{2}$ represents the Fermi momentum at the critical point.
Based on Eq. ~\ref{KAn} and Eq.~\ref{modularham} as well as exact diagonalization,  it is easy to confirm that the refined R\'{e}nyi contour $\tilde s_n(x)$ derived from $nT$ and the R\'{e}nyi contour derived from $T$ satisfy the relationship:
\begin{align}
    \tilde s_n(x) = \frac{1}{n} s_1(x), \label{refinedrenyia}
\end{align}
which aligns with Eq.~(7) of the main text. Equivalently, we have
\begin{align}
    \tilde S_n=\frac{S}{n}.
\end{align}

\begin{align}
    \tilde{S}_n &\equiv n^2 \partial_n \left(\frac{n-1}{n} S_n\right)\\
    &=-n^2 \partial_n  \left(\frac{n-1}{n}\frac{1}{1-n}(n-\frac{1}{n})\log L\right)\frac{c}{6}\\
     &=-n^2 \partial_n  \left((1-\frac{1}{n^2})\log L\right)\\
      &=-n^2 \partial_n  \left((1-\frac{1}{n^2})\right)\log L\\
      &=\frac{c}{3n}\log L
\end{align}

\begin{align}
S_n = (1+ \frac{1}{n})\frac{c}{6}\log L
\end{align}

}
\omits{
\subsection{Rindler transformation in AdS$_3$/CFT$_2$}\label{Ap:Rindler}

\begin{figure}[t]
\centering
\includegraphics[width=0.7\linewidth]{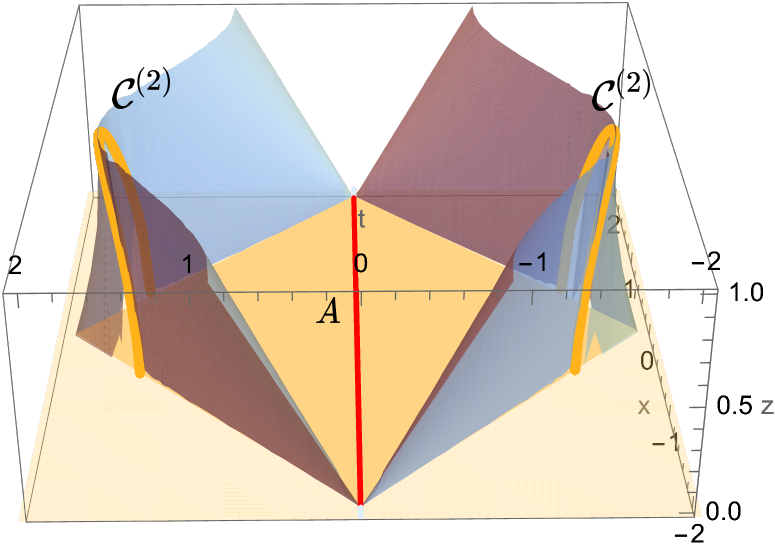}
 \caption{The red line represents subregion A. The two transparent surfaces are $\mathcal{N}^{(n)}_{\pm}$ with $n=2$, and their intersections are highlighted by yellow lines, which correspond to the extremal surfaces for $n=2$. These extremal surfaces, denoted $\mathcal{C}^{(2)}$, are anchored at and traverse along the entangling surface. They also exhibit tension and conical defects that cause backreactions to the bulk geometry, such as the discontinuity in $\mathcal{N}^{(n)}_{\pm}$.
We set $l_u = l_v = 2$ in the figure. }
            \label{n=2}
\end{figure}

In the following, we will introduce the holographic duality of hyperfine structure $h_{n;2}$ using Rindler transformation. The basic idea of Rindler transformation is to convert the entanglement entropy of a region into the thermal entropy of another region which can be mapped to the horizon entropy of a hyperbolic black hole. It turns out that the horizon corresponds to the null hypersurfaces in the original  Poinc\'{a}re AdS spacetime.  The intersection of null hypersurfaces $\mathcal{N}^{(1)}_{\pm}$ in Poinc\'{a}re $\mathrm {AdS_3}$ forms the extremal surface $\mathcal{E}_A$.
Furthermore, the  intersection of the $n$-dependent null hypersurfaces $\mathcal{N}^{(n)}_{\pm}$ in the  $n$-dependent Poinc\'{a}re $\mathrm {AdS_3}$ represents the holographic dual of $\tilde{S}_n$.

Suppose $A$ is the interval on the infinite line in the ground state in (1+1)D CFT, which is  a straight interval with the two boundaries on $(u_1,v_1)=(-\frac{l_u}{2},-\frac{l_v}{2})$ and $(u_2,v_2)=(\frac{l_u}{2},\frac{l_v}{2})$ in the Poinc\'{a}re  ${\mathrm {AdS_3}}$ with the metric
\begin{align}
ds^2=2rdudv+\frac{dr^2}{4r^2}\label{PoincareAdS}.
\end{align}Here we consider the AdS radius $l=1$. We can go back to the original Poinc\'{a}re ${\mathrm {AdS_3}}$ $ds^2=\frac{dx^2+dz^2-dt^2}{z^2}$  by setting
\begin{align}
u=\frac{x+t}{2},\quad v=\frac{x-t}{2},\quad r=\frac{2}{z^2}.
\end{align}
In the original Poinc\'{a}re ${\mathrm {AdS_3}}$, the two boundaries of subregion A are $(x_1,t_1)=(-R,0)$ and $(x_2,t_2)=(R,0)$, where $R=\frac{l_u+l_v}{2}$. The causal development of $A$ is the diamond-shape subregion
\begin{align}
    -\frac{l_u}{2}\le u\le\frac{l_u}{2},  -\frac{l_v}{2}\le v\le\frac{l_v}{2}.
\end{align}

In the case of  refined R\'{e}nyi entropy,  we consider the following metric of the $n$-dependent Poinc\'{a}re $\mathrm {AdS_3}$ ~\cite{2016NatCo...712472D}
\begin{align}
ds^2=\rho^2  d\tilde y^2+(\rho^2-\rho_h^2)d\tilde \tau^2+\frac{d\rho^2}{\rho^2-\rho_h^2},
\end{align}
with $\rho_h=\frac{1}{n}$.

In the following, we will utilize the Rindler transformation to derive the extremal surface associated with this metric, which corresponds to the refined R\'{e}nyi entropy. Before proceeding further, we include some remarks regarding the application of this metric. As shown in Ref.~\cite{2011JHEP...12..047H}, the entanglement entropy is equivalent to the thermal entropy, or the horizon entropy, of the dual black hole at a temperature $T_0 = \frac{1}{2\pi R}$.

\begin{align}
  &  S= S_{\text{thermal}}(T_0).
\end{align}
The R\'{e}nyi entropy can also be derived from the thermal entropy through
\begin{align}
     S_n= \frac{n}{n-1}\frac{1}{T_0} \int_{T_0/n}^{T_0}S_{\text{thermal}}(T)dT.
\end{align}
We are going to show that the refined R\'{e}nyi entropy is nothing but the thermal entropy at temperature $T_n=T_0/n$ by noticing that
\begin{align}
    \tilde S_n&=n^2\frac{d}{dn}\left ( \frac{n-1}{n}S_n\right )\\
    &=n^2\frac{d}{dn}\left ( \frac{n-1}{n}\frac{n}{n-1}\frac{1}{T_0} \int_{T_0/n}^{T_0}S_{\text{thermal}}(T)dT\right )\\
    &=-\frac{n^2}{T_0}S_{\text{thermal}}(\frac{T_0}{n}) \frac{d}{dn}(T_0/n)\\
    &=S_{\text{thermal}}(T_0/n).
\end{align}
Therefore, we can use Rindler transformation to derive the refined R\'{e}nyi entropy.
% \[
% \frac{d}{dt}\int^{a(t)}_{b(t)} f(x)dx=f(a(t))a'(t)-f(b(t))b'(t)
% \]

This  $n$-dependent  $\mathrm {AdS_3}$ can also be written as
\begin{align*}
ds^2=&\frac{d\tilde r^2}{4(\tilde r+1)(\tilde r+1-2\rho_h^2)}
+\rho_h^2d\tilde u^2+\rho_h^2d\tilde v^2+2(\frac{\tilde r+1}{\rho_h^2}-1)\rho_h^2d\tilde u d\tilde v,
\end{align*}
by setting
\begin{align}
\tilde r+1=2\rho^2,\tilde u=\frac{\tilde y+\tau}{2},\tilde v=\frac{\tilde y- \tau}{2},\tau =i\tilde \tau.
\end{align}
The above form can be easily transformed into the Rindler $\mathrm {AdS_3}$
\begin{align}
ds^2=du^{*2}+2r^{*}  du^{*}dv^{*}  +dv^{*2}+\frac{dr^{*2}}{4(r^{*2}-1)},
\end{align}using
\begin{align}
u^*=\rho_h \tilde u,v^*=\rho_h \tilde v,r^*=\frac{\tilde r+1}{\rho_h^2}-1.
\end{align}
Therefore, the Rindler transformation from coordinate $(u,v,r)$ to $(u^*,v^*,r^*)$ is
\begin{align*}
&u^*=\frac{\rho_h}{4}\ln (\frac{4(rv(l_u+2u)+1)^2-l_v^22 r^2(l_u+2u)^2}{4(rv(l_u-2u)-1)^2-l_v^2 r^2(l_u-2u)^2}),\nonumber\\
&v^*=\frac{\rho_h}{4}\ln(\frac{l_u^2r^2(l_v+2v)^2-4(l_vru+2ruv+1)^2}{l_u^2 r^2(l_v-2v)^2-4(-l_v ru+2ruv+1)^2}),\nonumber\\
&(r^*+1)\rho_h^2-1=\frac{r^2(l_u^2(l_v^2-4v^2)-4l_v^2 u^2)+4(2ruv+1)^2}{4l_ul_v r}.
\end{align*}

By setting $r^*=1$,  the horizon of the $n$-Rindler $\mathrm {AdS_3}$ maps to the two null hypersurfaces $\mathcal{N}_{\pm}^{(n)}$ in the original space $(u,v,r)$

\begin{align}
&\mathcal{N}^{(n)}_+:
\quad r=\frac{-2 n^2}{l_u^2(n^2-2)+4 n^2 u v+2l_u \sqrt{l_u^2(1-n^2)-4 n^2 uv+n^4(u+v)^2)}},
\end{align}
\begin{align}
&\mathcal{N}^{(n)}_-:
\quad r=\frac{-2 n^2}{l_u^2(n^2-2)+4 n^2 u v-2l_u \sqrt{l_u^2(1-n^2)-4 n^2 uv+n^4(u+v)^2)}}.
\end{align}

The horizon of the $n$-Rindler $\mathrm {AdS_3}$ in Poinc\'{a}re coordinates  can be written as
\begin{align}
x_n^+&=    \frac{2 n^2 u - 2 \sqrt{-n^4 (1 - n^2) (1 - u^2)}}{n^4}\\
t_n^+&=2 \left(u - \frac{u}{n^2} + \frac{\sqrt{-n^4 (1 - n^2) (1 - u^2)}}{n^4}\right)\\
z_n^+&=\sqrt{l_u^2 \left(1 - \frac{2}{n^2}\right) + \frac{4 u \left(n^2 (2 - n^2) u + \sqrt{n^4 (1 - n^2) (l_u^2 - 4 u^2)}\right)}{n^4}}
\end{align}
and
\begin{align}
    x_n^-&=\frac{2 n^2 u + \sqrt{n^4 (1 - n^2) (l u^2 - 4 u^2)}}{n^4}\\
t_n^-&=-\frac{2 n^2 u - 2 n^4 u + \sqrt{n^4 (1 - n^2) (l u^2 - 4 u^2)}}{n^4}
\\
z_n^-&=\sqrt{l u^2 \left(1 - \frac{2}{n^2}\right) + 4 u^2 - \frac{4 u \left(2 n^2 u + \sqrt{n^4 (1 - n^2) (l u^2 - 4 u^2)}\right)}{n^4}}
\end{align}
\begin{figure}[t]
\centering
\includegraphics[width=0.5\linewidth]{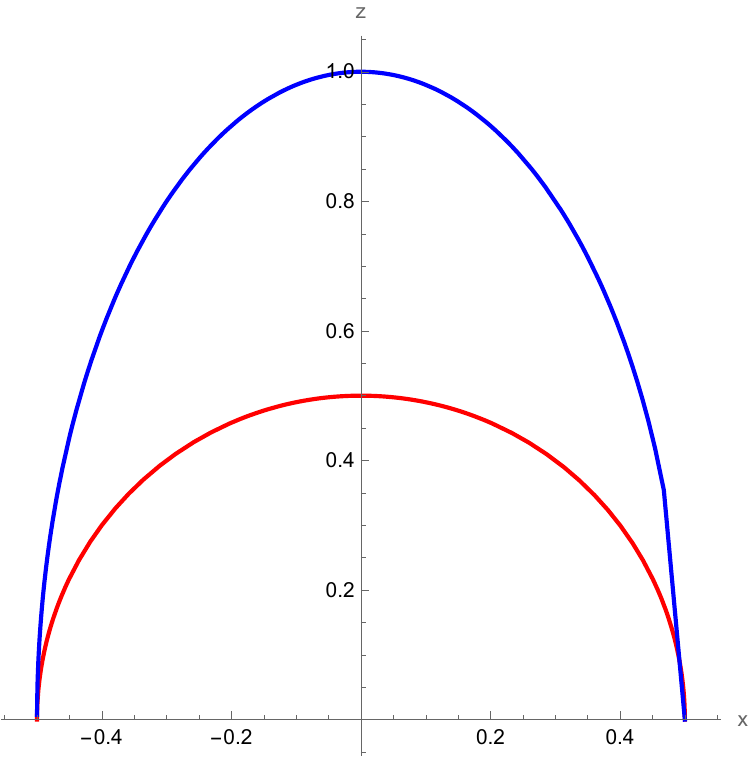}
\caption{The holography duality of refined R\'{e}nyi entropy lies outside the entanglement wedge. Here we take the time slice $t_n=\frac{1}{n^2}$, at which the extremal surface anchored on. The blue line denotes the the projection of $\mathcal{C}^{(2)}$ on the time slice and the red line denotes the time slice of the entanglement wedge.}
\label{Fig:outside}
\end{figure}
As we can see in the Fig.~1(c) of the main text, the intersection of the null hypersurfaces $\mathcal{N}^{(1)}_{\pm}$ provides the RT surface $\mathcal{E}_A$. The entanglement wedge is defined as the  bulk domain of dependence bounded by $A$ and the RT surface $\mathcal{E}_A$. Within the entanglement wedge, the bulk operators can be reconstructed from the boundary CFT, which is the well-known subregion-subregion duality in holographic duality~\cite{2014JHEP...12..162H}.

The intersection of $\mathcal{N}^{(n)}_{\pm}$ provides the extremal surface $\tilde{S}_n$ within the original coordinates $(x,t,z)$, as depicted in Fig.~\ref{n=2} .  The extremal surfaces, denoted as $\mathcal{C}^{(n)}$, for $n > 1$ differ markedly from the $n=1$ RT surface. It is noteworthy that the extremal surfaces $\mathcal{C}^{(n)}$ are inclined and exhibit a conical defect, manifesting as discontinuous regions in the null hypersurface $\mathcal{N}_{\pm}^{(n)}$. However, this singularity disappears when $n=1$.

Interestingly,  $\mathcal{C}^{(n)}$ extend beyond the entanglement wedge(see Fig.~\ref{Fig:outside}), which implies that they cannot be connected to $\mathcal{E}_A$ through time unitary evolution. This aligns with the principle that a unitary transformation does not modify the entanglement spectrum of a density matrix. Therefore, we can further define the R\'{e}nyi entanglement wedge enclosed by $\mathcal{C}^{(n)}$, which may be used to reconstruct more information about the bulk from the boundary.

% The holographic dual of the refined Renyi entropy lies outside the causal wedge because the refined Renyi entropy exceeds the entanglement entropy at time $t_n$.
% \begin{align}
% &S(t_n)\le \tilde S
% \end{align}
% \begin{proof}

% \begin{align}
%     S(t_n)&=2R\log\frac{l_{t_n}}{a}\frac{1}{4G}\\
%     &=2R\log(\frac{l}{a}\frac{x_{t_n}}{x_{t_1}})\frac{1}{4G}\\
%     &=2R\log(\frac{l}{a})\frac{1}{4G}+2R\log(\frac{x_{t_n}}{x_{t_1}})\frac{1}{4G}\\
%     &=\frac{c}{3}\log\frac{l}{a}+\frac{c}{3}\log(\frac{1}{n^2})
% \end{align}

% The result should be non-negative, so we require
% \begin{align}
%     a\le \frac{l}{n^2}.\label{Eq:larger}
% \end{align}
% We know that refined Renyi entropy can be written as
% \begin{align}
%     \tilde S_n=\frac{c}{3n}\log\frac{l}{a}.
% \end{align}
% Based on the above elements, we can check that the following relation holds:

% \begin{align}
%     \tilde S-S(t_n)&=\frac{c}{3}[ \frac{1}{n}\log\frac{l}{a}-\log\frac{l}{a}+\log n^2]\\
%     &=\frac{c}{3}[\log n^2- \frac{n-1}{n}\log \frac{l}{a}]\\

% \end{align}

% \begin{align}
% &    \frac{1}{n}\log\frac{l}{a}\ge \log\frac{l}{a}-\log n^2\\
% &\log n^2\ge \frac{n-1}{n}\log \frac{l}{a},
% \end{align}
% which holds because of the following relation
% \begin{align}
%     \log n^2\ge \frac{n-1}{n}\log \frac{l}{a}\ge \frac{n-1}{n}\log n^2,
% \end{align}
% where we use the inequality~\ref{Eq:larger}.
% \end{proof}

Moreover,  as is shown in Fig.~\ref{Fig:contourplane}, with the help of the hyperfine structure we introduced, we can find out a finer description for the subregion-subregion duality in entanglement wedge and R\'enyi entanglement wedge. We can use the planes that link the boundary sites with the extremal surfaces $\mathcal{C}^{(n)}$ to slice the extremal surface  according to the contour distribution. Each segment derived from the extremal surface represents the holographic dual of the hyperfine structure. Thus, the holographic description of R\'enyi entropy is refined into the duality between the boundary density-density correlator and the segment of bulk extremal surface. This finer correspondence indicates that a bulk local operator can be reconstructed by CFT operators in a segment of the subregion on the boundary.
%%%%%%%%%

\begin{figure}[t]
\centering
\includegraphics[width=0.8\linewidth]{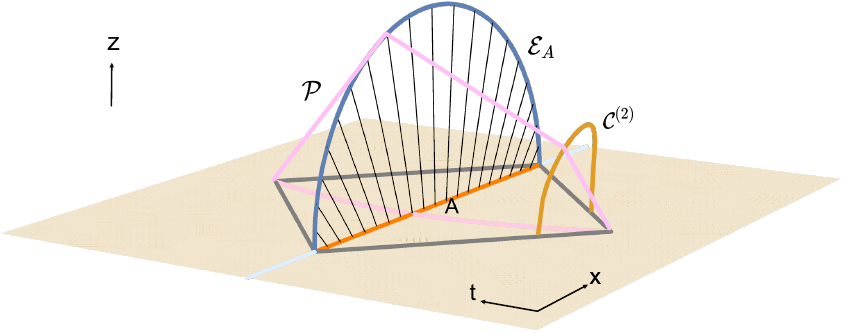}
\caption{Slice the extremal surfaces according to the distribution of the hyperfine structure using the pink plane.}
\label{Fig:contourplane}
\end{figure}

\subsection{Discussions on the R\'{e}nyi entanglement wedge}

Since the holographic duality of the entanglement entropy of the replica state extends beyond the entanglement wedge, one may wonder whether this contradicts the well-known bulk reconstruction principle, which states that the largest region for bulk reconstruction is typically the entanglement wedge. To verify the bulk reconstruction property, we further analyze the distribution of boundary and bulk modular flows associated with subregion \( A \).
\subsubsection{Modular flow on the boundary and bulk}

The modular Hamiltonian is the generator along the thermal circle $k_t\equiv \tilde\beta^i\partial_{\tilde x^i}$ in Rindler space. In order to map it to the original space, we need to solve the following differential equations
\begin{align}
\partial_u&=(\partial_u \tilde u)\partial_{\tilde u}+(\partial_u \tilde v)\partial_{\tilde v}+(\partial_u \tilde r)\partial_{\tilde r}\\
\partial_v&=(\partial_v \tilde u)\partial_{\tilde u}+(\partial_v \tilde v)\partial_{\tilde v}+(\partial_v \tilde r)\partial_{\tilde r}\\
\partial_r&=(\partial_r \tilde u)\partial_{\tilde u}+(\partial_r \tilde v)\partial_{\tilde v}+(\partial_r \tilde r)\partial_{\tilde r}
\end{align}
For the general \( n \) case, these equations become:

\begin{align}
\partial_u&=(\partial_u u^*)\partial_{ u^*}+(\partial_u v^*)\partial_{\tilde v}+(\partial_u r^*)\partial_{r^*}\\
\partial_v&=(\partial_v u^*)\partial_{u^*}+(\partial_v  v^*)\partial_{ v^*}+(\partial_v r^*)\partial_{ r^*}\\
\partial_r&=(\partial_r u^*)\partial_{u^*}+(\partial_r v^*)\partial_{v^*}+(\partial_r r^*)\partial_{r^*}
\end{align}
By plugging the bulk Rindler transformations (from \((u, v, r)\) to \((u^*, v^*, r^*)\)) into the above differential equations and solving them, we obtain the modular flow in both the boundary and bulk. In Rindler \(\widetilde{\mathrm{AdS_3}}\), the generator of the Hamiltonian is simply the translation along the thermal circle. Mapping this to the original space yields the modular flow in the boundary and bulk.

The modular flow in the bulk is given by:
\begin{align*}
k_t^{\text{bulk}} &= \beta^*_{u^*} \partial_{u^*} + \beta^*_{v^*} \partial_{v^*} \\
&= -\pi \partial_{u^*} + \pi \partial_{v^*} \\
&= -n\pi \partial_{\tilde{u}} + n\pi \partial_{\tilde{v}} \\
&= n\left(\frac{2\pi u^2}{l_u} - \frac{\pi l_u}{2} + \frac{\pi}{l_v r}\right) \partial_u + 4n\pi r \left(\frac{v}{l_v} - \frac{u}{l_u}\right) \partial_r \\
&\quad + \frac{n}{2} \pi \left(-\frac{2}{l_u r} - \frac{4v^2}{l_v} + l_v\right) \partial_v.
\end{align*}
Solving the system of differential equations:
\begin{align}
\frac{\partial u(s)}{\partial s} &= n\left(\frac{2\pi u^2}{l_u} - \frac{\pi l_u}{2} + \frac{\pi}{l_v r}\right), \\
\frac{\partial v(s)}{\partial s} &= \frac{n}{2} \pi \left(-\frac{2}{l_u r} - \frac{4v^2}{l_v} + l_v\right), \\
\frac{\partial r(s)}{\partial s} &= 4n\pi r \left(\frac{v}{l_v} - \frac{u}{l_u}\right),
\end{align}
we obtain the modular flow in the bulk. By reparameterizing time as \( s \rightarrow ns \), these equations reduce to the modular flow of the \( n = 1 \) case. This implies that although the R\'{e}nyi entanglement wedge extends beyond the entanglement wedge, the largest bulk region in which local operators can be reconstructed using the modular flow is still the entanglement wedge. However, the extension beyond the entanglement wedge can be used to reconstruct other information about the bulk spacetime, such as the metric~\cite{2020PhRvD.101f6011B}.

The modular flow on the boundary is given by:
\begin{align}
k_t = \left(\frac{2n\pi u^2}{l_u} - \frac{n\pi l_u}{2}\right) \partial_u + \frac{n}{2} \pi \left(-\frac{4v^2}{l_v} + l_v\right) \partial_v.
\end{align}
Solving the differential equations \( \left(\frac{\partial u(s)}{\partial s}, \frac{\partial v(s)}{\partial s}\right) = k_t \), we obtain the boundary solution:
\begin{align}
u(s) &= -\frac{1}{2} l_u \tanh(n\pi s - 2 l_u k), \\
v(s) &= \frac{1}{2} l_v \tanh(n\pi s + 2 l_v k),
\end{align}
which scales the time \( s \) by a factor of \( n \).

In summary, the boundary and bulk modular flows in the R\'{e}nyi entanglement wedge remain consistent with those in the entanglement wedge, confirming that the largest bulk region in which local operators can be reconstructed by the modular flow in the bulk is still the entanglement wedge.

}

\omits{
\section{Hyperfine structure in lattice fermion models}\label{SM:IV}

In Fermi gas systems, higher ranks of the hyperfine structure are suppressed. However, in more general cases, we observe a variety of parameter regions where $h_{n;k>2}$ contributes to both the fine structure and entropy. In the forthcoming sections, we will explore the behavior of all ranks of the hyperfine structure in lattice fermion models. Initially, we will examine the distribution function of $h_{n;2}$ across different parameter regions. Subsequently, we will demonstrate the emergent scaling law of normalized $h_{n;k}(m;k_x=0,\pi)$ in the presence of topological edge states.

\subsection{Distribution function of $h_{n;2}$ in various parameter regions}\label{Ap:scaling}

As illustrated in Fig.~2 in the main text, the behavior of the leading component $h_{n;2}$ is similar across different across parameter regions: it predominantly localizes on the boundary and exhibits a decay from the boundary towards the center. To further explore the distribution function of $h_{n;2}$ across various parameter regions, we present a cross-sectional analysis of the hyperfine structure $h_{n;2}$ in Fig.~\ref{Fig:scal}. For clarity, we set $n = 2$. As shown in Fig.~\ref{Fig:scal}(a), it is apparent that in both trivially gapped and topologically gapped scenarios, $h_{n;2}$ exhibit an exponential decrease. In contrast, at critical points and along Fermi surfaces, $h_{n;2}$ decreases following a power-law decay, which is notably slower than the exponential decline.

Furthermore, it is noteworthy that the power-law exponents at critical points and Fermi surfaces are distinct. The decay of $h_{n;2}$ at critical points is faster than that at Fermi surfaces. Despite the similarities in scaling behavior of entanglement entropy between critical points and trivially gapped cases, the distribution functions of the hyperfine structure $h_{n;2}$ exhibit marked differences.

 \begin{figure}[t]
            \centering
            \subfigure[]{\includegraphics[width=0.4\linewidth]{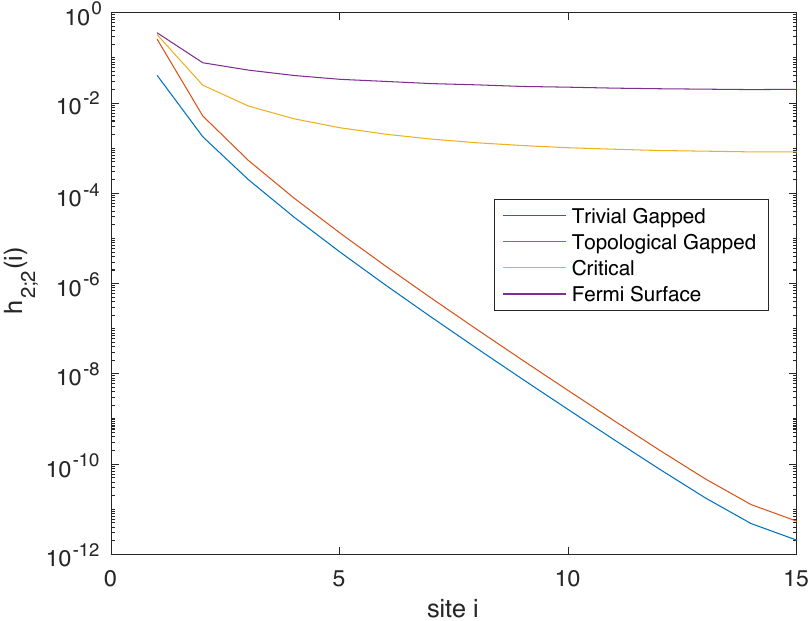}}
            \subfigure[]{\includegraphics[width=0.4\linewidth]{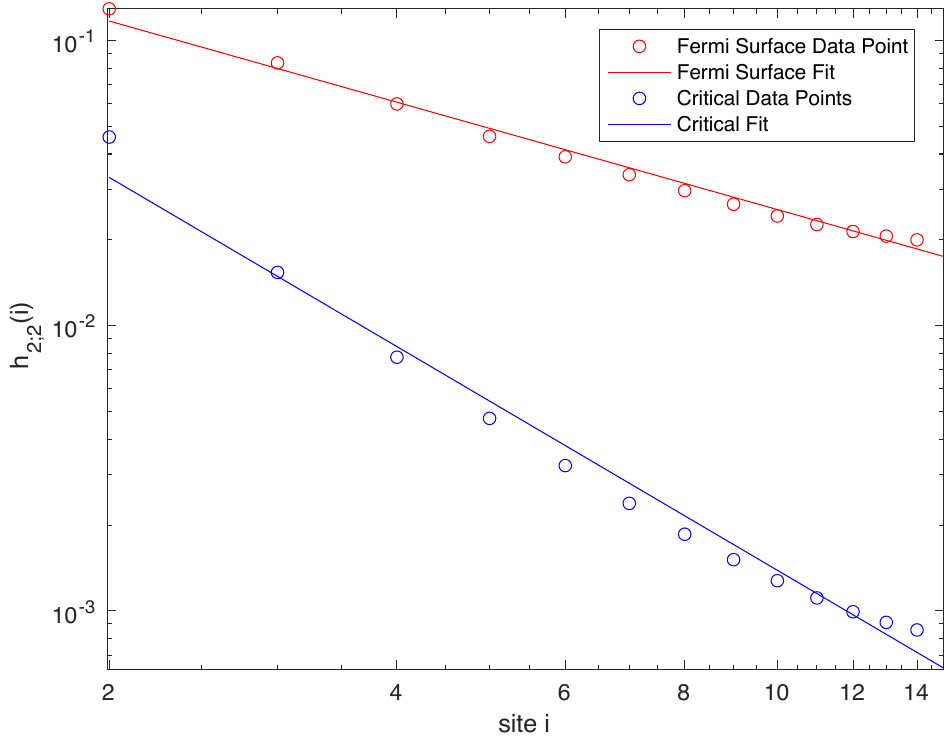}}
            \caption{The cross section of the hyperfine structure $h_{2;2}(x)$ in the four parameter regions. From Fig.(a) we can find that in the cases of both trivial gapped and topological gapped, $h_{2;2}(x)$ decrease exponentially from the boundary to the bulk. From both (a)  and (b) we can see that the $h_{2;2}(x)$ of critical and Fermi surface decrease in a power law. }
            \label{Fig:scal}
      \end{figure}

\subsection{Edge states of Chern insulator}\label{Ap:edge}
In this subsection, we will demonstrate the emergent scaling law of normalized $h_{n;k}(m;k_x=0,\pi)$ shown in Fig.~3 in the presence of topological edge states. Initially, we establish the link between the bulk Chern number and the zero-energy modes in the Chern insulator model, denoted as  $\hat{\mathcal{H}}=\sum_{\mathbf{k}}\hat c_{ \mathbf{k}}^{\dagger}h(\mathbf{k})\cdot \mathbf{\sigma} \hat c_{  \mathbf{k} }$ with $ h(\mathbf{k})=(\sin  {k}_x,\sin  {k}_y,m+\cos  {k}_x+\cos  {k}_y),$where $\sigma$ is the vector of Pauli matrices.  By considering an open boundary along the y-axis and a periodic boundary along the x-axis, we classify different quantum states using the quantum number $k_x$. We demonstrate that within the topological interval $m\in(-2,0)$, the zero-energy edge mode corresponds to $k_x=0$, whereas for $m\in(0,2)$, it aligns with $k_x=\pi$.
Here we only present a torus argument as shown in Fig.~\ref{Fig:bbc} with the rigorus proof demonstrated in Ref.~ \cite{2011PhRvB..83l5109M}. The critical loops of fixed $k_x$ are highlighted  in red. In Fig.~\ref{Fig:bbc}(b) and (d), since the Chern number $Q\neq 0$,  the torus wraps the origin and it is always possible to find two loops that are coplanar with the origin, one of which encloses the origin and the other does not.  Specifically, at $k_x=\pi$ in Fig.\ref{Fig:bbc}(d), the loop that is coplanar with and encircles the origin signifies the presence of zero-energy edge states at this $k_y$ value.

\begin{figure}[t] %
\centering
\includegraphics[width=\linewidth]{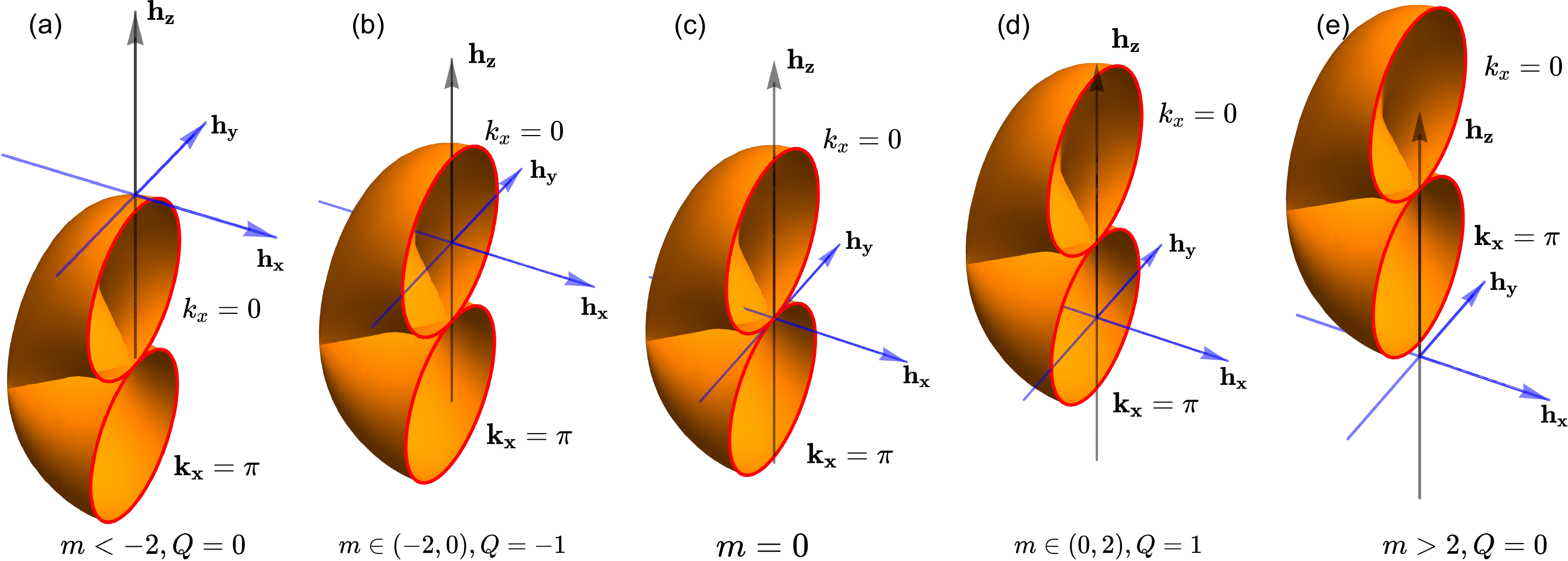}
\caption{The torus $\mathbf{h}(\mathbf{k}) $ of the Chern insulator for different values of m. In (a) ,(c) and (e), the torus do not contain the origin hence Chern number Q=0. In (b), the red circle with $k_x=0$ encloses the origin, meaning that there are zero-energy edge states at this $k_x$.  At $k_x=\pi$, the red loop lies in the plane of the origin without containing it, indicating no edge state at $k_x=\pi$.  (d) is similar to (b) but has zero-energy edge state at $k_x=\pi$.}
\label{Fig:bbc}
\end{figure}
According to Refs.~\cite{2009arXiv0909.3119T,2010PhRvL.104m0502F}, it has been identified that in specific topological zones, zero-energy edge states at $k_x=0$ for $m \in (-2,0)$ and at $k_x=\pi$ for $m \in (0,2)$ correspond to entanglement modes with a fractional value of $\frac{1}{2}$.

In the topological region where $m \in (-2,0)$, we derive the following expression:
\begin{align}
   h_{n;k}(m;k_x=0) &= \beta_k(n) C_k(m;j \in \partial A, k_x=0) \nonumber \\
   &\approx \beta_k(n) C_k(m; k_x=0) \nonumber \\
   &= \beta_k(n) (-i\partial_\lambda)^k \ln \det \left(1 + M(m; k_x=0) \exp(i\lambda)\right)\bigg|_{\lambda=0} \nonumber \\
   &= \beta_k(n) \operatorname{tr}\left[(-i\partial_\lambda)^k \ln \left(1 + M(m; k_x=0) \exp(i\lambda)\right)\right]\bigg|_{\lambda=0} \nonumber \\
   &\approx \beta_k(n) (-i\partial_\lambda)^k \ln\left(1 + \frac{1}{2} \exp(i\lambda)\right)\bigg|_{\lambda=0}.
\end{align}
This analysis establishes a scaling law applicable across different values of $k$:
\begin{align}
   \frac{ h_{n;k}(m \in (-2,0); k_x=0)}{\max  h_{n;k}(m \in (-2,0); k_x=0)} = \frac{h_{n;k}(m \in (0,2); k_x=\pi) }{\max h_{n;k}(m \in (0,2); k_x=\pi) }\approx 1.
\end{align}
This scaling law is also valid in the topological domain of $m \in (0,2)$, demonstrating a consistent pattern across the specified ranges.

}

\bibliography{Reference}
\end{document}